\documentclass[12pt]{article}
\usepackage[dvips]{graphicx}
\usepackage{xspace}
\usepackage{rotating}
\begin{document}
\begin{flushright}
\end{flushright}
\begin{center}
\begin{large}
{\bf Accelerator Neutrino Beams}
\end{large}
\vskip 0.25 cm
Sacha E. Kopp, \\
Department of Physics, University of Texas at Austin
\end{center}          

\begin{abstract}
 Neutrino beams at from high-energy proton accelerators 
have been instrumental discovery tools in particle physics.  Neutrino beams are derived from 
the decays of charged $\pi$ and $K$ mesons, which in turn are created from 
proton beams striking thick nuclear targets.  The precise 
selection and manipulation of the $\pi/K$ beam control the energy spectrum and 
type of neutrino beam.  This article describes the physics of 
particle production in a target and manipulation of the particles to 
derive a neutrino beam, as well as numerous innovations 
achieved at past experimental facilities.

\end{abstract}

\begin{small}
\setcounter{tocdepth}{2}
\tableofcontents
\end{small}


\section{Introduction}
\label{intro}

Neutrino beams at accelerators have served as laboratories for greater understanding of the neutrino itself, but also have harnessed the neutrino as a probe to better understand the weak nuclear force and its unification with the electromagnetic force, the existence of strongly-bound quarks inside the proton and neutron, and the recent revelations that neutrinos undergo quantum mechanical oscillations between flavor types, a strong indication that neutrinos have non-zero mass.  Excellent reviews of these topics are available, for example, in \cite{barish1978b,conrad1998,fisk1982,delellis2004,mckeown2004,musset1978,steinberger1988}.

The present article discusses so-called conventional neutrino beams, those in which a high-energy proton beam is impinged upon a nuclear target to derive a beam of pion and kaon secondaries, whose decays in turn yield a neutrino beam.  Such beams have been operated at Brookhaven, CERN, Fermilab, KEK, Los Alamos and Serpukhov, and new facilities at Fermilab, J-PARC, and CERN are underway.   The present article cannot be taken as a complete catalog of every facility.  Rather, the intent is to discuss some of the basic physical processes in meson production in a nuclear target, the manipulation (focusing) of the secondary beam before its decay to neutrinos, and the measurements which can validate the experimentally-controlled spectrum.  As such, it is useful to refer to earlier papers in which such ideas were first developed, in addition to ``state-of-the art'' papers written about contemporary facilities.  These notes will not cover so-called ``beam-dump'' experiments, for which very thorough reviews are already available \cite{burman2003,wachsmuth1978}.

There are two valuable references on conventional neutrino beams to which readers may refer:  the first is the proceedings of three workshops held at CERN \cite{cern-nu-63}-\cite{cern-nu-69} at a time in which the accelerator neutrino beam concept was in its infancy.  The second is a set of workshops \cite{nbi1999}-\cite{nbi2006} held at KEK, Fermilab, and CERN.  Initiated by Kazuhiro Tanaka of KEK, this workshop series arose at a time of renaissance for the neutrino beam, when ``long-baseline'' neutrino oscillation experiments required new developments in accelerator technology to deliver intense neutrino beams across distances 200-900~km through the Earth.  Both workshop series are valuable documentation of the ingenuities of the experimental teams. 

While much has been written about neutrino interactions and detectors, comparatively little has been written about the facilities to produce these beams.  In as much as these notes collect those references which may not be commonly known, I hope they will be helpful.

\begin{figure}[t]
\vskip -0. cm
  \centering
  \includegraphics[width=5.8in]{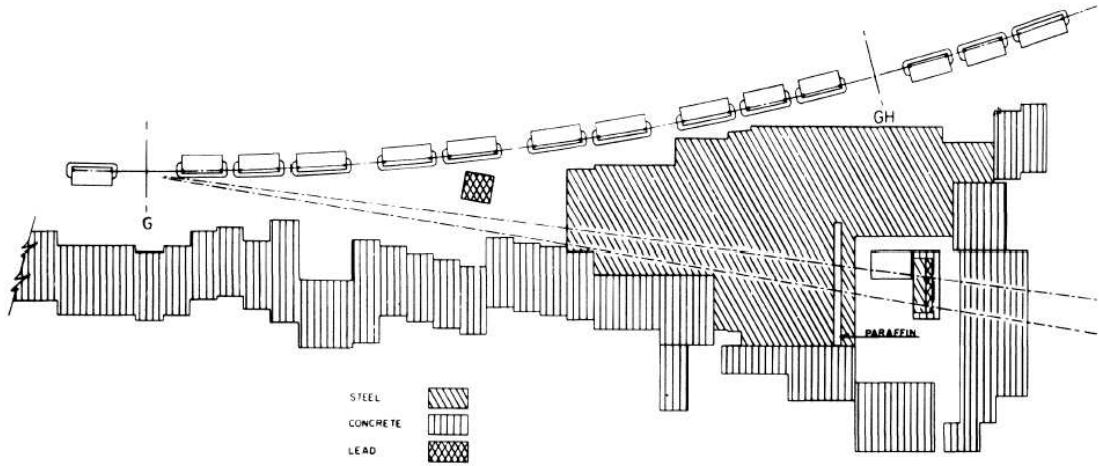}
\vskip -0. cm
  \caption{Plan view of the first accelerator neutrino experiment.  Taken from Ref. \cite{danby1962}.}
  \label{fig:danby1962-fig1}
\end{figure} 

\section{Accelerator Neutrino Beam Concept}

The idea of an accelerator neutrino beam was proposed independently by Schwartz\cite{schwartz1960} and Pontecorvo \cite{pontecorvo1960}.  The experiment, first carried out by Lederman, Schwartz, Steinberger and collaborators \cite{danby1962}, demonstrated the existence of two neutrino flavors.\footnote{It's interesting to note that the first accelerator neutrino beam was sufficiently new that the authors felt a need to put the word ``beam'' in quotations \cite{danby1962}.  Though we perhaps should still do so today (no one has yet focused a neutrino), the current experimental facilities have certainly evolved in 44 years.}  Figure~\ref{fig:danby1962-fig1} shows their apparatus.  In brief, a proton beam strikes a thick nuclear target, producing secondaries, such as pions and kaons.  Those secondaries leave the target, boosted in the forward direction but with some divergence given by production cross sections $d^2\sigma/dp_Tdx_F$ ($p_T$ is the momentum of the secondary transverse to the proton beam axis, $x_F\approx p_L/p_{\mbox{proton}}$ is the ratio of the secondary particle's longitudinal momentum along the beam axis to the proton beam momentum).  The mesons, permitted to drift in free space, decay to neutrino tertiaries.  In the 1962 experiment, the drift space was $\sim21$~m.  Shielding, often referred to as the ``beam stop'' or ``muon filter,'' removes all particles in the beam except for the neutrinos, which continue on to the experiment.  The 1962 experiment was a ``bare target'' beam, meaning that the experiment saw the direct decays of the secondaries, which were not in any way focused prior to their decay.

The decays $\pi^\pm\rightarrow\mu\nu_\mu$ (BR$\sim$100\%), $K^\pm\rightarrow\mu\nu_\mu$ (BR=63.4\%), and $K_L\rightarrow \pi\mu\nu_\mu$ (BR=27.2\%) make the development of muon-neutrino beams the most profitable.  While some muons will decay via $\mu\rightarrow e\nu_e \nu_\mu$ in the drift volume giving rise to electron neutrinos, the long muon lifetime makes this source more of a nuisance background than a source to be exploited.  Proposals have been made to produce an enhanced $\nu_e$ beam from $K_L\rightarrow \pi e\nu_e$ decays (BR=38.8\%) \cite{mori1977,mori1979,ong1985,camerini1980,camerini1980b}, though these have not been realized.  Comparatively few experiments have utilized the residual $\nu_e$ contamination in their beam \cite{ahrens1986,eichten1973}.  Most $\nu_e$ beams are produced from beam dump experiments \cite{wachsmuth1978,burman2003}, as are $\nu_{\tau}$ beams arising from $D_s\rightarrow\tau\nu_\tau$ decays \cite{pontecorvo1975,kodama2001}.  For conventional neutrino beams, the neutrino spectra may be derived from the $\pi/K$ meson spectra and the kinematics of meson decay in flight.  Some useful relations for the kinematics of $\pi/K$ decay in flight are given in Appendix~\ref{kinematics}.

The 1962 neutrino experiment didn't actually extract a proton beam.  The circulating protons in the BNL AGS were brought to strike an internal Be target in a 3~m straight section of the accelerator and those resulting secondaries at 7.5$^\circ$ angle with respect to the proton direction contributed to the neutrino flux.  A deflector sent the protons to strike the target for $25~\mu$sec bursts \cite{danby1962}.  The idea of an extracted proton beam dedicated for a neutrino experiment came from CERN \cite{giesch1963,plass1963}.  There are at least three important motivators for the switch from the internal target to fast-extracted external beams:  
\begin{itemize} 
\item The extraction efficiency onto the internal target was not perfect (about 70\%, according to \cite{danby1962} or 50\% in \cite{kustom1969}, to be compared with nearly 100\% for the fast-extraction\cite{giesch1963}).  
\item The CERN team developed a lens \cite{vandermeer1961} to better collect the pions leaving the target, which was much more efficient than taking those few secondaries at 7.5$^\circ$ to the beam direction, and this lens system is large (couldn't be located in or around the synchrotron).
\item The van der Meer lens is an electromagnet sourced by a pulsed current which required short beam pulses ($<1~$msec) to avoid overheating from the pulsed current.
\end{itemize}
The second BNL neutrino run did use an extracted beam \cite{burns1965,danby1965}, though still no focusing of the secondaries \cite{lederman1965}.  Extracted beams are the norm in today's experiments.

\begin{sidewaystable}[p]
  \centering
  \begin{tabular}{c|c|c|c|c|c|c|c}

     &      & $p_0$  & Protons/      & Secondary& Dec. Pipe &  $\langle E_\nu \rangle$ & \\  
Lab  & Year &(GeV/$c$) & Pulse ($10^{12}$)& Focusing & Length (m)&   (GeV)  &Experiments \\  \hline\hline
ANL  & 1969 & 12.4 & 1.2  & 1 horn WBB   & 30       &  0.5  &     Spark Chamber     \\\hline
ANL  & 1970 & 12.4 & 1.2  & 2-horn WBB   & 30       &  0.5  &   12$^\prime$ BC \\\hline
BNL  & 1962 & 15 & 0.3 & bare target     & 21       &  5  &  Spark Ch. Observation of 2 $\nu$'s \\\hline
BNL  & 1976 & 28   & 8  & 2-horn WBB   & 50      &  1.3 & 7$^\prime$ BC, E605, E613, E734, E776 \\\hline
BNL  & 1980 & 28   &  7   & 2-horn NBB   & 50       &  3  &  7$^\prime$ BC, E776 \\\hline
CERN & 1963 & 20.6& 0.7  & 1 horn WBB    & 60       & 1.5    & HLBC, spark ch. \\\hline
CERN & 1969 & 20.6& 0.63 & 3 horn WBB   & 60       & 1.5    & HLBC, spark ch. \\\hline
CERN & 1972 & 26  & 5    & 2 horn WBB   & 60       & 1.5    & GGM, Aachen-Pad.  \\\hline     
CERN & 1983 & 19 & 5   & bare target & 45    &  1      & CDHS, CHARM  \\\hline
CERN & 1977 & 350 & 10   & dichromatic NBB & 290    &  50,150$^{(a)}$  & CDHS, CHARM, BEBC  \\\hline
CERN & 1977 & 350 & 10   & 2 horn WBB      & 290    &  20 & GGM,CDHS, CHARM, BEBC \\\hline
CERN & 1995 & 450 & 11   & 2 horn WBB   & 290       & 20     & NOMAD, CHORUS \\\hline
CERN & 2006 & 450 & 50   & 2 horn WBB   & 998       & 20     & OPERA, ICARUS \\\hline
FNAL & 1975 & 300, 400 & 10   & bare target   & 350      & 40 & HPWF\\\hline
FNAL & 1975 & 300, 400 & 10   & Quad. Trip., SSBT   & 350      & 50,180$^{(a)}$ & CITF, HPWF\\\hline
FNAL & 1974 & 300 & 10   & dichromatic NBB   & 400      & 50, 180$^{(a)}$ & CITF, HPWF, 15$^\prime$ BC \\\hline
FNAL & 1979& 400& 10    & 2-horn WBB  & 400      & 25     & 15$^\prime$ BC \\\hline
FNAL & 1976 &350& 13     & 1-horn WBB   & 400      & 100    & HPWF, 15$^\prime$ BC \\\hline
FNAL & 1991 & 800 & 10   & Quad Trip.   & 400      & 90, 260 &  15$^\prime$ BC, CCFRR \\\hline 
FNAL & 1998 & 800 & 12   & SSQT  WBB  & 400      & 70, 180 &  NuTeV exp't   \\\hline
FNAL & 2002 &8  & 4.5    & 1-horn WBB   & 50       & 1      & MiniBooNE      \\\hline
FNAL & 2005 &120& 32     & 2-horn WBB   & 675      & 4-15$^{(b)}$      & MINOS, MINER$\nu$A          \\\hline
FNAL & 2009 &120& 70     & 2-horn NBB   & 675      & 2      & NO$\nu$A off-axis          \\\hline
IHEP &  1977 & 70   & 10  & 4 horn WBB & 140 & 4   & SKAT, JINR \\\hline
JPARC&  2009 & 40   & 300 & 3 horn NBB & 140 & 0.8 & Super K off-axis \\\hline
KEK  &  1998 & 12   & 5   & 2 horn WBB & 200 & 0.8 & K2K long baseline osc. \\\hline
  \end{tabular}
$^{(a)}$ pion and kaon peaks in the momentum-selected channel~~~~~~
$^{(b)}$ tunable WBB energy spectrum.
  \caption{Tabulation of neutrino beam lines at high energy proton synchrotrons.}
  \label{tab:nu-beams}
\end{sidewaystable}

The short beam pulses from single-turn extractions are one of the advantages of accelerator neutrino experiments:  with the beam only live for a duty fraction $\sim10^{-6}$, the experiment (provided it has fast enough electronics) has the ability to close its trigger acceptance around a small ``gate'' period around the accelerator pulse, reducing false triggers cosmic ray muons.\footnote{The BNL experiment even reduced the $25~\mu$sec period further by gating on the RF structure of the circulating beam consisting of a train of 20~nsec ``buckets'' of protons separated by 220~nsec \cite{gaillard1963}.}

\begin{figure}[t]
\vskip -2.2 cm
  \centering
  \includegraphics[width=6.5in]{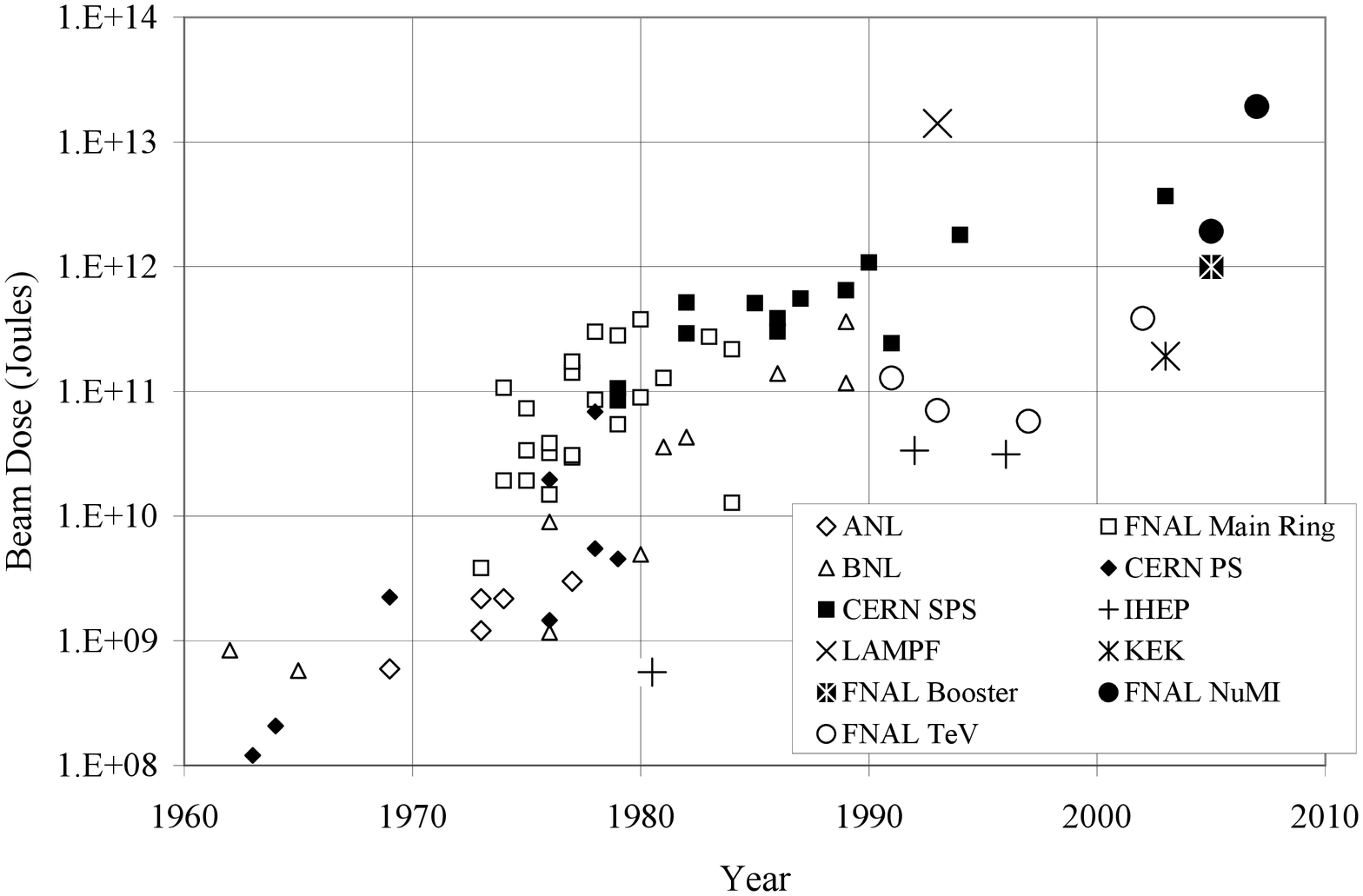}
\vskip -2.2 cm
  \caption{Compilation of total protons-on-target times beam energy per proton delivered to neutrino experiments at various laboratories, by date of publication.  Experiments running concurrently in the same neutrino line are not plotted separately.}
  \label{fig:neutrino-pot-history}
\end{figure} 
Neutrino experiments require expansive numbers of protons delivered to their targets.  The 1962 experiment received $1.6\times10^6$ ``pulses'' at an average of $1.9\times10^{11}$~protons-per-pulse (ppp)\cite{gaillard1963}.  Today's experiments require $10^{20}-10^{21}$ protons on target (POT)\footnote{A LANL experiment \cite{allen1993} received an impressive 10$^{23}$, but at 800~MeV/$c$ momentum.}.  Since the number of pion and kaon secondaries per proton grows with the incident proton beam energy, a good figure of merit is (POT $\times$ Beam Energy). Figure~\ref{fig:neutrino-pot-history} shows Joules per experiment since the first accelerator neutrino experiment.    Forthcoming experiments such as CNGS, JPARC and NO$\nu$A are not shown, but are another order of magnitude in accumulated dose.

\section{Production of Hadrons in the Target}
\label{hadprod}

\subsection{Introduction}

Neutrino experiments require information about the production of $\pi^+$, $\pi^-$, $K^+$, $K^-$, and $K_L$.  Further, production yields, $d^2N/dpd\Omega$, as a function of the secondary's momentum and angle emerging from the target are necessary:  the secondary's momentum is related to the resulting neutrino energy (see Appendix~\ref{kinematics}), and the production angle relates to how well the secondary points along the direction of the desired neutrino beam, or to the degree to which the secondary is captured by the focusing system.  Models of secondary production have been derived by fitting and interpolating experimental data on $p+A\rightarrow\pi^\pm X$ or $p+A\rightarrow K X$.

\begin{figure}[b]
\vskip -0. cm
  \centering
  \includegraphics[width=5.5in]{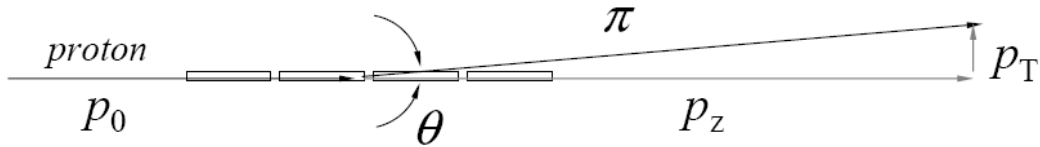}
\vskip -0.5 cm
  \caption{Pion secondary produced by a proton striking segmented target, with definition of momentum components. }
  \label{fig:pt-pz-target}
\end{figure}

The prediction of the neutrino flux starting from the yield of secondary hadrons from a target is the bane of every neutrino experiment.  ANL, for example, performed a ``beam survey'' of the yield of secondaries from 12.5~GeV protons on thick targets of A$\ell$ and Be \cite{lundy1965}, only to be surprised \cite{derrick1969} by their neutrino flux being off by a factor of two compared with subsequent but more limited beam surveys \cite{asbury1969,marmer1969}.  The experiment scaled up the older, more complete $d^2N/dpd\Omega$ results to agree with the normalizations of the newer experiment (such was suggested by Sanford \& Wang, who had tried a fit to all invariant cross section data\cite{sanfordwang1967}) and quoted \cite{kustom1969} 30\% errors on the neutrino flux as a result.  Another round of beam surveys was done \cite{cho1971} which fixed the normalization problem and covered the full phase space, and these results were used in subsequent papers \cite{campbell1973,mann1973}.  As the authors of \cite{campbell1973} put it:  {\it ``The calculation of the $\nu$ flux ... requires a detailed discussion, which we will defer to a subsequent publication.''} These are hard experiments to get right.

\begin{figure}[t]
\vskip -1 cm
  \centering
  \includegraphics[width=5.5in]{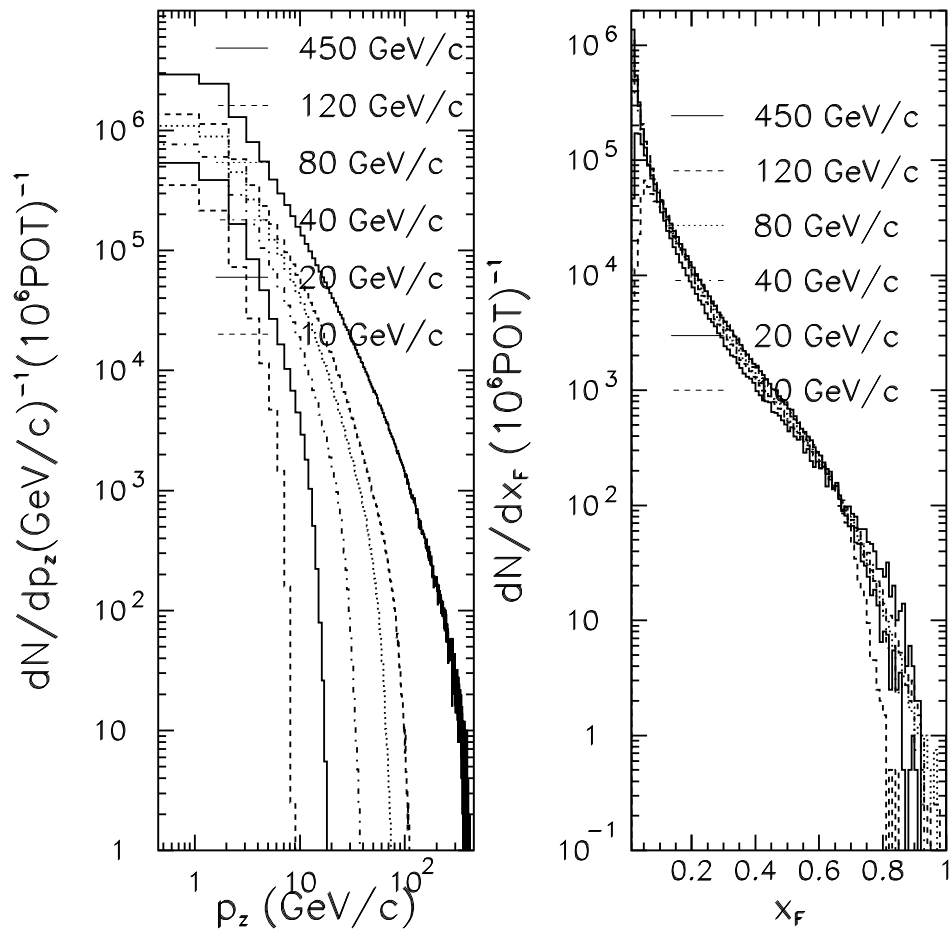}
\hfill
\vskip -0.5 cm
  \caption{Fluka \cite{fluka} calculations of (left) $p_z$ and (right) $x_F$ distributions of $\pi^+$ in $p+$C collisions at incident momenta $p_0=10, 20, 40, 80, 120, 450$~GeV/$c$ and $p_z>0.5$~GeV/$c$.  A 94~cm long target, $6.4\times15$~mm$^2$ transverse size, is assumed.
Taken from \cite{pavlovic2007}.}
  \label{fig:fluka-xf-dists}
\end{figure}

Figure~\ref{fig:fluka-xf-dists} demonstrates one of the aspects of hadron production predicted by Feynman scaling \cite{feynman1969} of relevance for neutrino flux predictions.  Shown is the distribution of $p_z$ and of $x_F \approx p_z/p_0$ for $\pi^+$ produced by protons striking a graphite target as estimated using the Fluka-2005 \cite{fluka} Monte Carlo code, where $p_0$ is the primary proton beam momentum and $p_z$ the longitudinal momentum of the secondary (defined in Figure~\ref{fig:pt-pz-target}).  Distributions are shown for incident proton momenta $p_0$=10, 20, 40, 80, 120, 450~GeV/$c$.  The shapes of the $x_F$ distributions are quite similar, indicating that the pion momenta scale with $p_0$.  It is also of note that the integrals of these curves, {\it i.e.} the mean number of $\pi^+$ produced per proton on target, grows nearly linearly with $p_0$ (see Table~\ref{tab:fluka-table}).

\begin{figure}[t]
\vskip -1 cm
  \centering
  \includegraphics[width=2.5in]{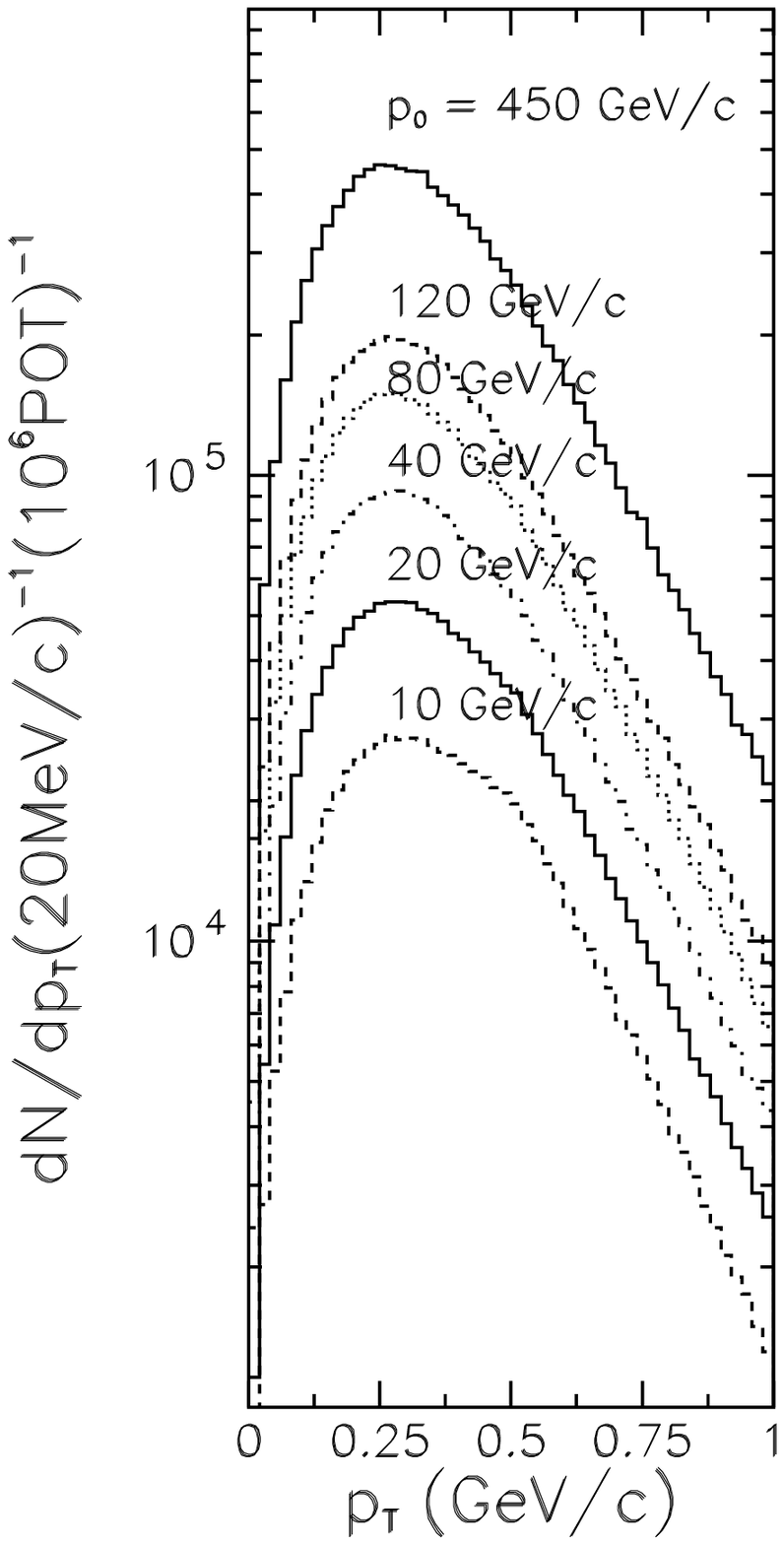}
  \includegraphics[width=2.5in]{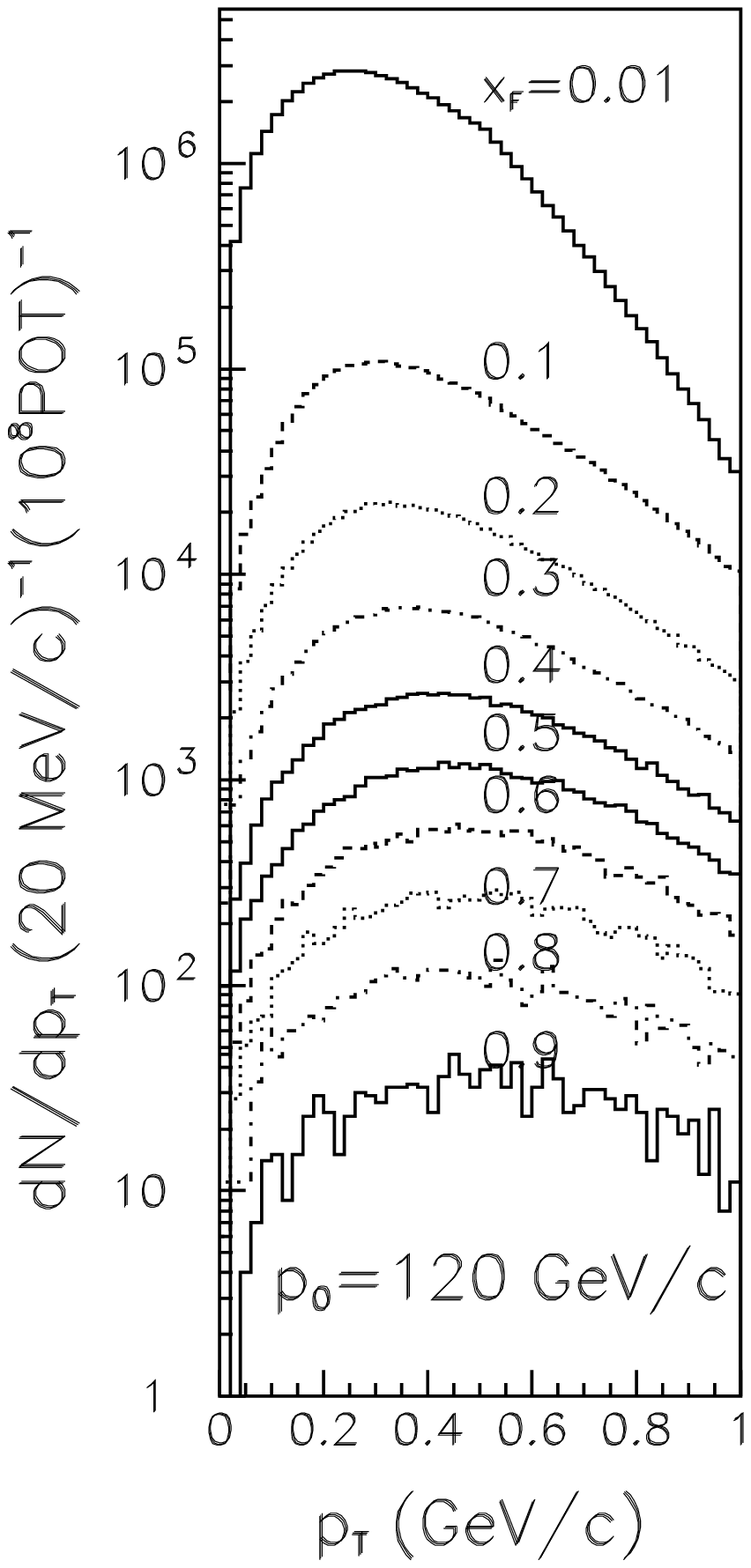}
\hfill
\vskip -0.5 cm
  \caption{Fluka \cite{fluka} calculations of (left) $p_T$ spectra of $\pi^+$ produced in $p+$C collisions at various incident proton momenta $p_0$;  (right) $p_T$ spectra of $\pi^+$ produced in 120~GeV/$c$ $p+$C collisions for various values of $x_F$.  Taken from \cite{pavlovic2007}.}
  \label{fig:fluka-pt-dists}
\end{figure} 

Figure~\ref{fig:fluka-pt-dists} demonstrates another important aspect of hadron production: the Fermi momentum of partons inside the nucleons being $\sim\hbar c$/1~fm$\approx$200~MeV, and the fact that momentum components transverse to the boost direction are invariant, implies that the production spectra in transverse momentum $p_T$ should be independent of $x_F$, {\it i.e.} $$\frac{d^2N}{dx_F dp_T}\approx f(x_F)g(p_T)$$ and the peak transverse momentum is of order 250~MeV for the secondaries.  Figure~\ref{fig:fluka-pt-dists} shows very little evolution of the $p_T$ shape for different incident momenta $p_0$ or exiting pion momenta $p_z$. That $p_T$ does not scale (very much) is important because the transverse momentum is what controls the divergence of the secondary beam:  mesons with $p_T=0$ are directed along the beam line, and their neutrino daughters tend to follow the secondaries' direction.  It is fortunate that the amount of $p_T$ to remove by focusing (see Section~\ref{wbb}) does not grow rapidly with pion momentum. 

The linearly increasing secondary yield with incident beam momentum has an important impact on neutrino beam design.  It is often argued that to produce a lower-momentum neutrino beam one must deliver a lower momentum proton beam at the target, the rationale being that at lower energy machine can be operated at higher repetition rate.  However, a given neutrino beam energy is achieved by focusing a particular secondary beam pion momentum.  As shown in Figure~\ref{fig:fluka-xf-dists} and Table~\ref{tab:fluka-table}, the yield at a fixed momentum appears to drop (approximately linearly) with decreasing proton beam momentum.  Thus, the benefit of a lower-momentum, higher rep-rate, accelerator is cancelled by the lower pion yield per proton on target.  The only reason for changing the accelerator energy might be to achieve higher secondary momenta than accessible at a lower-energy machine.

\begin{table}[t]
  \centering
  \begin{tabular}{c|c|c|c}

  $p_0$ (GeV/$c$)  & $\langle n_\pi \rangle$ & 	$\langle p_T \rangle$ (MeV/$c$) & $K/\pi$ \\  \hline
	10		&	0.68		&	389		&0.061 \\
	20		&	1.29		&	379 		&0.078 \\
	40		&	2.19		&	372 		&0.087 \\
	80		&	3.50		&	370 		&0.091 \\
	120		&	4.60		&	369 		&0.093 \\
	450		&	10.8		&	368 		& 0.098 \\ \hline
  \end{tabular}
  \caption{Fluka \cite{fluka} predictions for $\pi^+$ production above $p_z>0.5$~GeV/$c$ in a $6.4\times15\times94$~mm$^3$ graphite target per incident proton.  Shown are the mean number $\langle n_\pi\rangle$ produced per incident proton, mean transverse momentum $\langle p_T\rangle$ for $\pi^+$, and ratio of $K/\pi$ yields for several incident proton momenta $p_0$.  To good approximation, $\langle n_\pi\rangle \propto  (p_0)^{0.7}$.}
  \label{tab:fluka-table}
\end{table}

While the above discussion of scaling is qualitatively correct, current experimental data indicate that these scaling behaviours are not exact.  In fact, the Fluka Monte Carlo shown in Figures~\ref{fig:fluka-xf-dists} and \ref{fig:fluka-pt-dists}, being tuned to such data, demonstrates such scaling violations.  

\begin{figure}[t]
\vskip -1 cm
  \centering
  \includegraphics[width=5.2in]{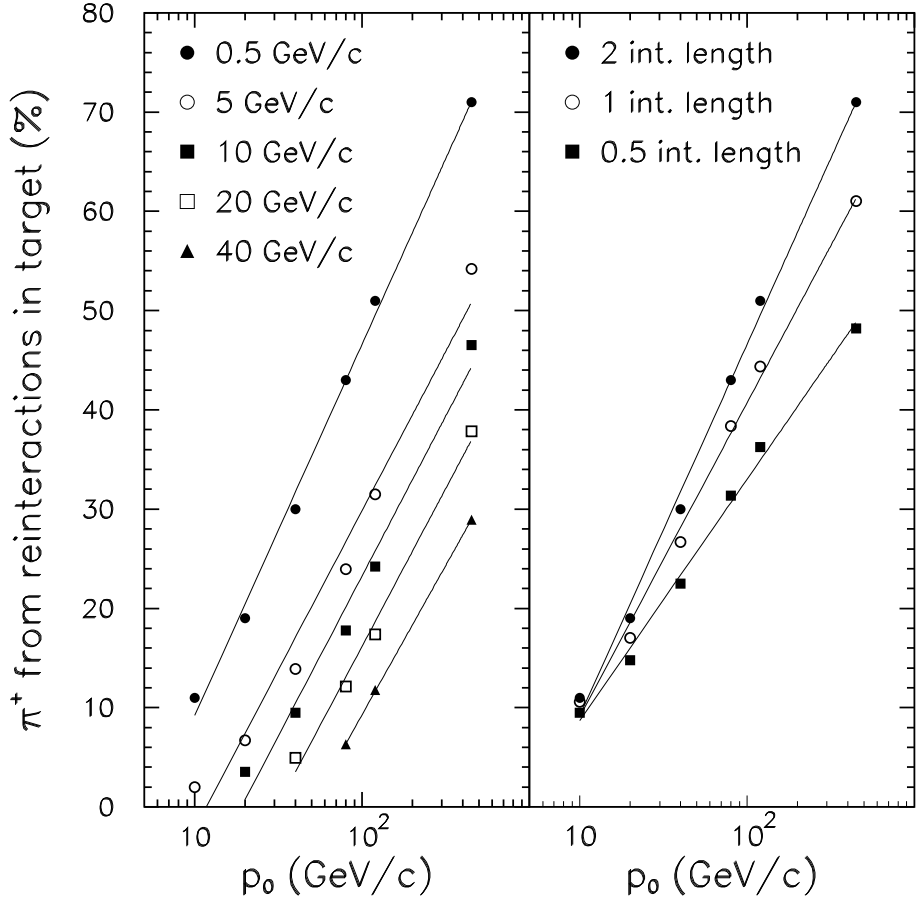}
\hfill
\vskip -0.5 cm
  \caption{Fluka \cite{fluka} calculations of the fraction of tertiary $\pi^+$ production from reinteractions in a graphite target 6.4$\times$15~mm$^2$ in transverse size as a function of primary beam momentum $p_0$. (left)  The reinteraction fraction is plotted for a 2.0 interaction length target for $\pi^+$ with $p_z>0.5, 5.0$, {\it etc.} GeV/$c$ momentum threshold.  (right)  The reinteraction fraction is plotted for targets of 0.5, 1.0, and 2.0 interaction lengths.  Taken from \cite{pavlovic2007}.  }
  \label{fig:reinteractions}
\end{figure} 

The geometry of the target is of particular note for prediction of the neutrino spectrum.  The geometry's significance arises because secondary particles exiting the $p+A$ collision have greater probability of reinteraction in the target material for longer pathlengths.  Secondary interactions are expected to decrease the yield of high-energy particles and increase the yield of low-energy particles, as reflected in the Fluka calculation of Figure~\ref{fig:reinteractions}.  Plotted are the fraction of the $\pi^+$ which are not produced by the primary $p+$C collision, but instead by subsequent reinteractions of the exiting particles.  As also shown in the figure, such reinteractions occur with greater probability in high energy proton beam experiments.  For very high energy neutrino beams, produced from high-momentum secondaries, the target is segmented as shown in Figure~\ref{fig:pt-pz-target}, with $\sim(1-10)$~cm ``slugs'' separated by gaps so as to permit small-angle, high-momentum secondaries to escape the target with less path length for reinteraction.  For low-energy neutrino beams, derived from low-momentum secondaries, such segmentation is not advantageous from the point of pion yield.\footnote{Segmented targets are of benefit for all high-power neutrino beams, however, for reducing longitudinal stress accumulation in the target due to heating from the proton beam.  Solid targets have failed under the shock wave which propagates along the target's length \cite{white1976}.}  At CERN's CNGS line, the target consists 13 slugs of 10~cm graphite separated by 9 cm, appropriate for its focus on collection of 40~GeV/$c$ pions \cite{bruno2002}.

When lacking the hadron production data which reproduces the exact conditions in a neutrino experiment, experimenters must rely on models to extrapolate such data to conditions of relevance for a given accelerator neutrino beam.  Some of the factors which must be accounted for are:
\begin{enumerate}
\item[(a)] interpolating between a sparse set of measurements at fixed values of secondary momenta $p$ or transverse momenta $p_T$, 
\item[(b)] extrapolating from measured yields off one nuclear target material to the one of relevance (Be, A$\ell$, C, W, ...) for the neutrino beam. 
\item[(c)] extrapolating to the correct projectile momentum $p_0$ on the target
\item[(d)] extrapolating to the correct target dimensions from those used in a hadron production experiment.
\end{enumerate}
Neutrino experiments have tried to derive these yields in auxilliary particle production experiments, either controlling or correcting for effects (a)-(d).

\subsection{Hadron Production Experiments}

\begin{figure}[t]
\vskip -0.6 cm
  \centering
  \includegraphics[width=5.2in]{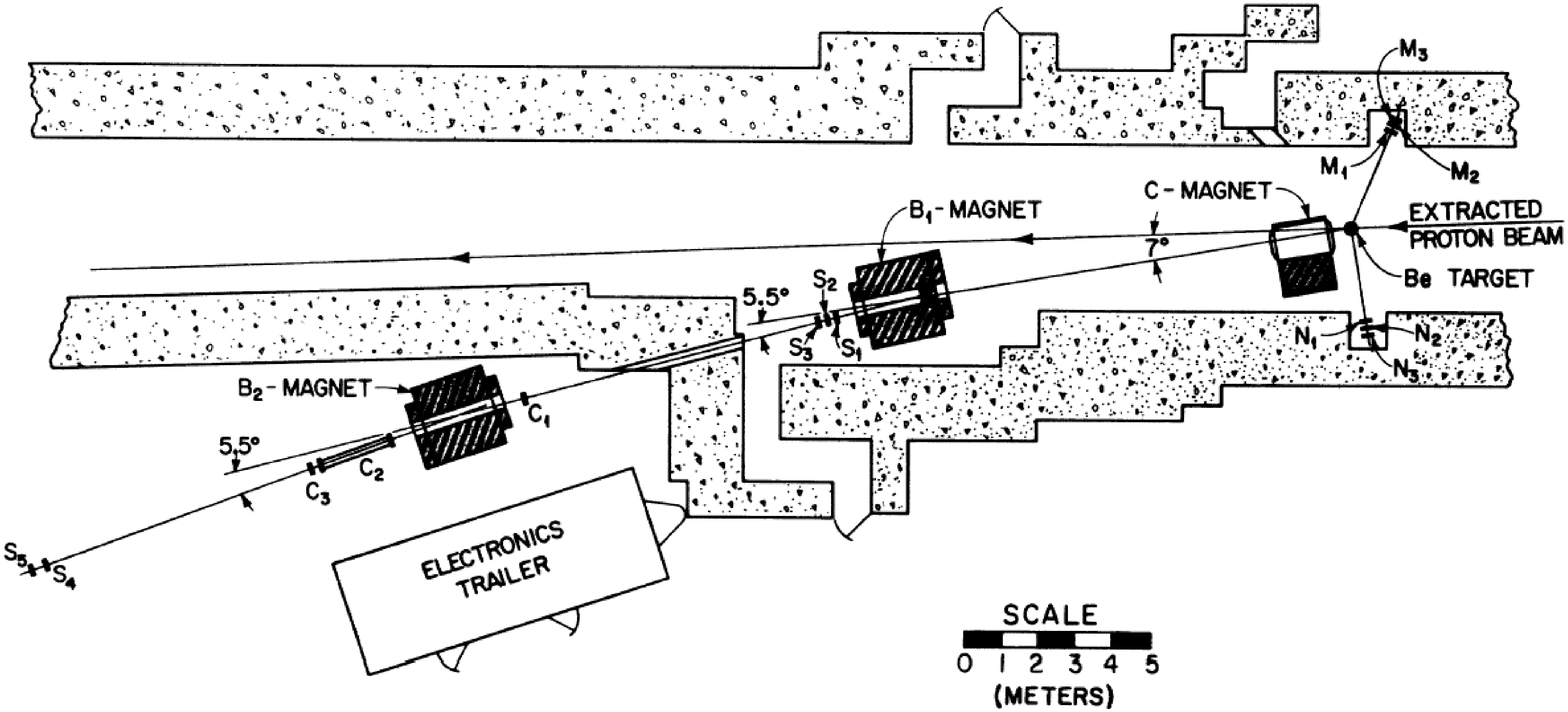}
\hfill
\vskip -0.5 cm
  \caption{Layout of the Cho {\it et al.} \cite{cho1971} spectrometer measurement of particle production at ANL.  The extracted proton beam is directed at a target, and secondaries are bent toward a Cherenkov detector by a set of dipoles.  Quadrupoles keep the secondary beam focused, and slits or collimators aid in the secondary momentum definition. }
  \label{fig:cho1971-fig1}
\end{figure}

\begin{figure}[t]
\vskip -0.6 cm
  \centering
  \includegraphics[width=4.99in]{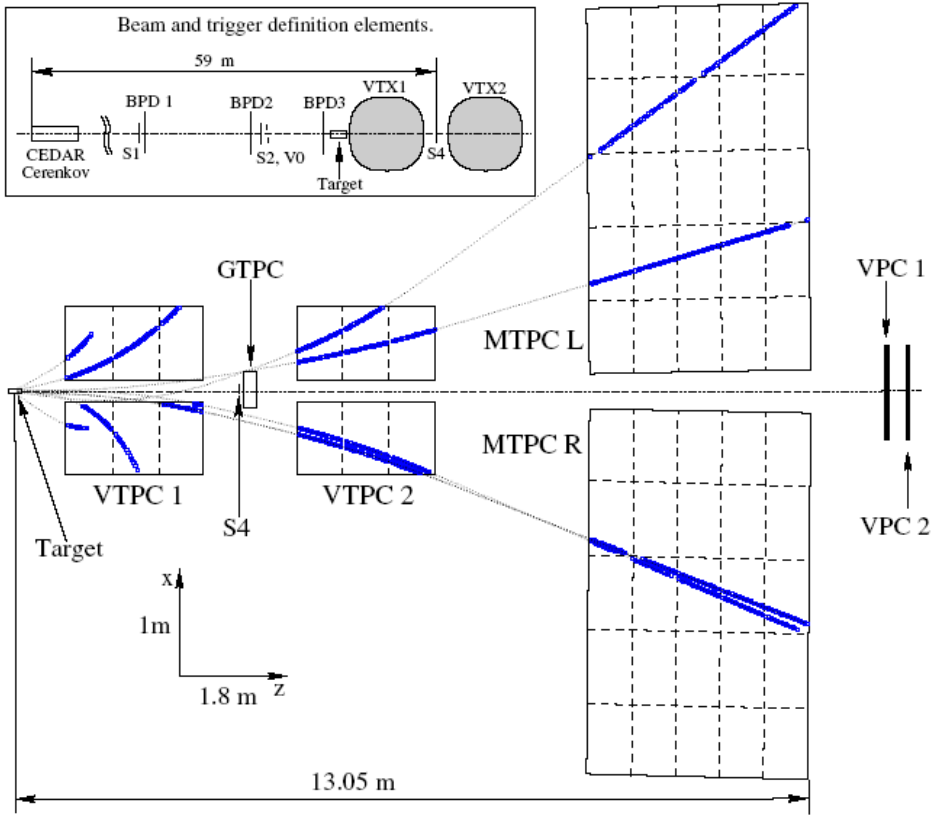}
\hfill
\vskip -0.5 cm
  \caption{Layout of the Alt {\it et al.} (NA49) \cite{alt2006a} full-acceptance spectrometer at CERN.  The extracted proton beam is directed at a target and the secondaries are tracked by time-projection chambers embedded in analyzing magnets. }
  \label{fig:alt2006-fig2}
\end{figure}

Table~\ref{tab:had-prod-expts} summarizes several of the hadron production experiments conducted over a range of incident proton momenta from 10~GeV/$c$ to 450~GeV/$c$ .  As can be seen from the table, many cover only limited ranges of $x_F$ and $p_T$, owing to the geometry of the experiment.  There are two main types of experiments:  single-arm spectrometers (shown schematically in Figure~\ref{fig:cho1971-fig1}) and full-acceptance specrometers (shown schematically in Figure~\ref{fig:alt2006-fig2}).

Single-arm spectrometers direct secondary particles within a small angular acceptance $\Delta\Omega$ into a magnetic channel in which dipoles define a secondary momentum bite $\Delta p$ and quadrupoles are used to focus the secondaries within this momentum bite into the analyzing channel.  Particle identification is accomplished by either TOF or Cherenkov systems or both.  The measurements are conducted with slow-spill beams to enable single secondary particle counting.   Normalization uncertainties on yields range from $\sim(10-25)\%$ due to the difficulty in proton counting:  current-integrating toroids function well only in fast-spill ($\mu$sec) beam pulses, and in slow spill beams the proton intensity must be monitored either by secondary emission monitors (SEMs) or by the induced radioactivation in thin foils placed upstream of the target.  SEMs are difficult to calibrate due to the decreasing secondary-electron yield after prolonged exposure \cite{ambrosini1999,ferioli1997}, and foil-based normalizations require knowledge of the production cross sections for the radionuclides, which are not typically known to better than 10\% \cite{baker1961}.\footnote{Some neutrino experiments, which are fast-extracted and so could use current toroids to measure protons-on-target, would take their normalization from foil activation techniques, to better match what the hadron production experiments did for proton normalization \cite{barish1977}.}  In addition, spectrometer measurements require accurate knowledge of their acceptance.   Ratios such as $\pi^+/\pi^-$ or $K^+/\pi^+$ are often better-measured.

Full-acceptance spectrometers are a relatively recent and quite sophisticated undertaking.  A wide acceptace tracking device, such as a time-projection chamber (TPC) is placed downstream or even surrounding the target.  Analyzing magnets surround the tracking system.  For small-angle particles, downstream drift chamber planes are used.  Particle identification is achieved by $dE/dx$ in the tracking chamber or by downstream TOF or Cherenkov counters.  The first attempt at such a full-acceptance measurement was at CERN, in which a replica Cu target for the CERN-PS neutrino line was placed inside the Ecole Polytechnique heavy liquid bubble chamber \cite{orkin1965}.  More recent examples are the NA49, HARP, and MIPP experiments, all based on TPCs.

\begin{figure}[t]
\vskip -0.6 cm
  \centering
  \includegraphics[width=4.5in]{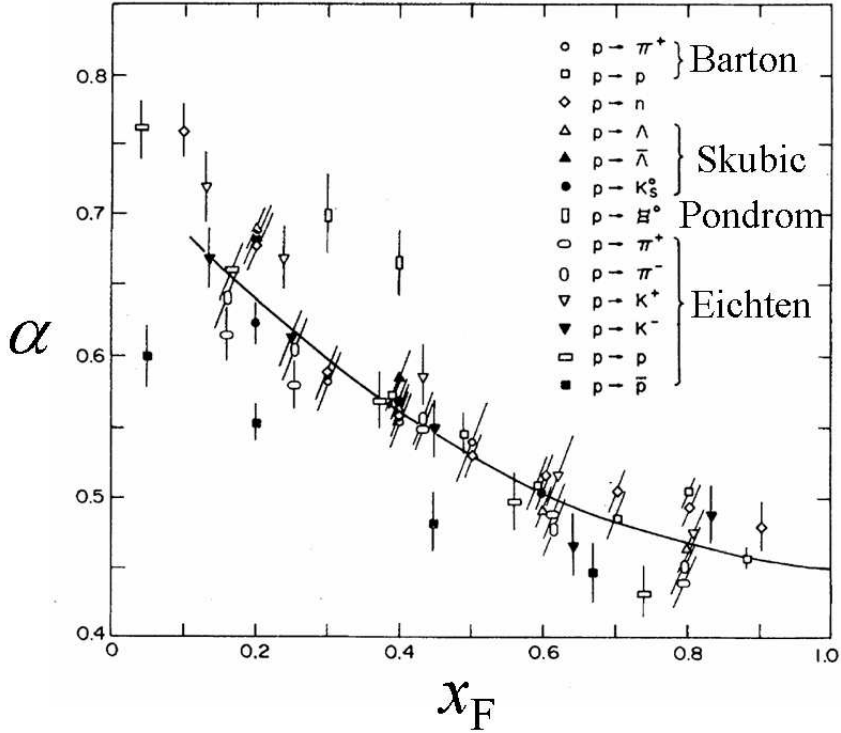}
\hfill
\vskip -0.5 cm
  \caption{The parameter $\alpha$ defined in Equation~\ref{eq:alpha}.  Taken from \cite{barton1983}.}
  \label{fig:barton1983-fig6}
\end{figure} 

$K^0$ production is important for accurate calculation of the $\nu_e$ flux from $K_L\rightarrow \pi e \nu_e$ decays.  While not focused, the $K_L$ do contaminate most beam lines.\footnote{The NuTeV experiment explicitly tried to reduce this background by targeting their proton beam at an angle with respect to the beam line.  Dipole magnets swept the desired $\pi^+$ and $K^+$ secondaries toward the decay tunnel. leaving the $K_L$ to travel in the forward direction off the target \cite{bernstein1994}.}  While $K_L$ production can be approximated as 
\begin{equation}
N(K_S)=N(K_L)=\frac{1}{4}(N_{K^+}+3\cdot N_{K^-})
\end{equation}
from quark-counting arguments\cite{bmpt}, direct data for comparison is limited to \cite{edwards1978} and \cite{skubic1978}.

Extrapolation must sometimes be done from a dataset collected on one nuclear target material to the target material relevant for a neutrino experiment. Data on $pp$ collisions at $p_0=19.2$~GeV/$c$ \cite{allaby1970}, $p_0=100$~GeV/$c$ \cite{brenner1982}, and $p_0=158$~GeV/$c$ \cite{alt2006a} are quite complete in $x_F$ and $p_T$ and are relevant for this purpose.  Additionally, studies of the $A$ dependence of cross sections at 100~GeV/$c$ \cite{barton1983} and 25~GeV/$c$ \cite{eichten1972} were used to show a scaling behaviour 
\begin{equation}
E\frac{d^3\sigma}{dp^3}=\sigma_0A^\alpha,
\label{eq:alpha}
\end{equation}
where $\alpha$ is graphed in Figure~\ref{fig:barton1983-fig6}.  This scaling proscribes how to extrapolate data taken at one target material to another relevant for a particular neutrino experiment.

\begin{sidewaystable}[p]
  \centering
  \begin{tabular}{c|c|c|c|c|c}

          & $p_0$   &                & Target	& $t/\lambda_{\mbox{int}}$ &   \\  
Reference & (GeV/$c$)  & Beam & Material	& (in \%) &  Secondary Coverage \\  \hline\hline

HARP \cite{catanesi2006} & 12    & PS      & Al  & 5     & $0.75<p<6.50$~GeV/$c$, $30<\theta<210$~mrad $^d$\\\hline
Asbury\cite{asbury1969} & 12.5 & ANL     & Be             & 4.9, 12.3 & $p=$3, 4, 5, $\theta=12^\circ,15^\circ$\\\hline
Cho \cite{cho1971}      & 12.4 & ANL     & Be             & 4.9, 12.3 & $2<p<6$~GeV/$c$, $0^\circ<\theta<12^\circ$\\\hline
Lundy\cite{lundy1965}$^a$  & 12.4 & ANL   &Be & 25,50,100 & $1<p<12$~GeV/$c$, $2^\circ<\theta<16^\circ$\\\hline
Marmer\cite{marmer1969} & 12.3 & ANL & Be, Cu & 10 & $p=$0.5, 0.8, 1.0~GeV/$c$, $\theta=0^\circ,5^\circ,10^\circ$\\\hline
Abbot \cite{abbot1992}	& 14.6 & AGS & Be, Al, Cu, Au & 1.0-2.0 & $0<p<8$~GeV/$c$, $\theta=5^\circ, 14^\circ, 
									24^\circ, 34^\circ, 44^\circ$ \\\hline
Allaby \cite{allaby1970}& 19.2 & PS & Be, Al, Cu,  & 1-2 & $p=$6, 7, 10, 12, 14~GeV/$c$,  \\
                        &      &         & Pb, B$_4$C &         & $\theta=$12.5, 20, 30, 40, 50, 60, 70~mrad \\\hline
Dekkers \cite{dekkers1964}$^b$ & 18.8, 23.1& PS & Be, Pb  & ``thin'' & $1<p<12$~GeV/$c$, $\theta=0, 100$~mrad\\\hline
Eichten \cite{eichten1972}& 24 & PS & Be, Al, Cu, & 1-2 & $4<p<18$~GeV/$c$,  \\
                        &      &         & Pb, B$_4$C &         & $17<\theta<127$~mrad \\\hline
Baker \cite{baker1961} &10,20,30& AGS& Be, Al& ?? & $1<p<17$~GeV/$c$, $\theta=4.75^\circ,9^\circ,13^\circ,20^\circ$\\\hline

Barton\cite{barton1983}& 100& FNAL & C,Al,Cu,Ag,Pb&1.6-5.6 & $0.3<x_F<0.88$, $0.18<p_T<0.5$~GeV/$c$\\\hline
NA49 \cite{alt2006b} & 158 & SPS & C & 1.5 & $0.05<p_T<1.8$~GeV/$c$, $-0.1<x_F<0.5^c$\\ \hline

Aubert \cite{aubert1975} & 300 & FNAL & Al & 76 & $\theta=0.8$~mrad, $x_F=$0.083, 0.17, 0.25,\\
                         &     &      &    &    &  0.33, 0.42, 0.5, 0.58, 0.67, 0.0.75\\\hline
Baker \cite{baker1974} & 200, 300 & FNAL & Be & 50 & $\theta=3$~mrad$^d$, $60<p<370$~GeV/$c$\\\hline
Baker \cite{baker1978} & 400 & FNAL & Be & 75 & $\theta=3.6$~mrad, $23<p<197$~GeV/$c$\\\hline
Atherton\cite{atherton1980} & 400 & SPS & Be & 10,25,75,125 & $x_F=$0.15, 0.30, 0.50, 0.75, $p_T=$0, 0.3, 0.5~GeV/$c$\\\hline
NA56/SPY \cite{ambrosini1999} & 450  & SPS& Be & 25,50,75 & $x_F=$0.016, 0.022, 0.033, 0.044, 0.067, 0.089, 0.15, 0.30,\\
                              &      &         &    &      & $p_T$=0, 75, 150, 225, 375, 450, 600~MeV/$c$ \\\hline\hline
  \end{tabular}

\begin{tabular}{ll}
\multicolumn{2}{l}{$^a$ Possible normalization discrepency with \cite{marmer1969,asbury1969,cho1971}.}\\
\multicolumn{2}{l}{$^b$ Possible normalization discrepency with Allaby {\it et al.}\cite{allaby1970}.}\\
$^c$ Full-acceptance spectrometer. \\
\multicolumn{2}{l}{$^d$ They report angular variation between 2-3.5~mrad, consistent with then-running FNAL neutrino experiments.}\\

\end{tabular}
  \caption{Tabulation of published $p+A$ hadron production experiments.  }
  \label{tab:had-prod-expts}
\end{sidewaystable}

\begin{figure}[t]
\vskip -0.0 cm
  \centering
  \includegraphics[width=5.99in]{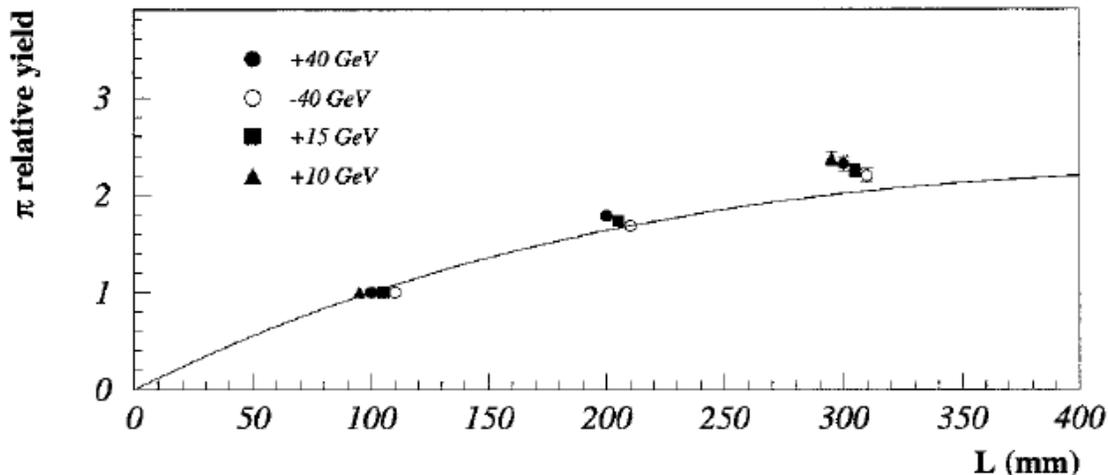}
\hfill
\vskip -0.5 cm
  \caption{Measurement of the $\pi^+$ yield from Be targets as a function of the target length.  Taken from \cite{ambrosini1999}.}
  \label{fig:ambrosini1999-fig18}
\end{figure} 

Neutrino targets are 1-2 nuclear interaction lengths so as to increase the fraction of the proton beam reacting in the target, hence the yield of secondaries.  Many particle production experiments, however, by measuring invariant cross sections, must perform their experiments on thin (1-5)\% interaction length targets.  In so doing such experiments do not have any sensitivity to the effect of reinteractions of particles produced in the primary $p+A$ as these secondary particles traverse the target.  A measurement of particle production in ``thick targets'' is shown in Figure~\ref{fig:ambrosini1999-fig18}.  The data are compared to a ``naive absorption model'' \cite{bmpt} 
\begin{equation}
f(\theta,L)=\int^L_0\exp(-z/\lambda_p)\exp(-t_{res}(z)/\lambda_s)\frac{dz}{\lambda_p}
\label{eq:absorption}
\end{equation}
where $\theta$ is the production angle, $L$ the target length, $z$ the longitudinal position along the target, $t_{res}(z)$ the residual target thickness to be crossed by the secondary particle to escape the target.  The three terms in the integral represent the probability that the proton does not interact up to $z$, the secondary is not reabsorbed, and the primary proton does interact between $z$ and $z+dz$ \cite{ambrosini1999}.  The data show excess particle production over such a naive model.

New measurements with full-acceptance spectrometers are forthcoming from  BNL E910 \cite{link2004,kirk2000} which took thin-target data, from HARP at CERN which studied a replica of the MiniBooNE Be target \cite{schmitz2005} and of the K2K Al target, and from Fermilab E907 which studied 120~GeV/$c$ protons incident on a replica of the NuMI target \cite{messier2005}.

\subsection{Some Parameterizations and Models}

Without going into a complete list of all models, here are mentioned some models which have been employed in neutrino flux calculations.  We shall not discuss some older models/parameterizations, such as the Von Dardel \cite{vondardel1962} used at CERN, Stefanski-White \cite{stefanski1976} used at Fermilab, or thermodynamic models \cite{hagedorn1970} used at CERN, Fermilab.  

The merit of the shower cascade models is that they (claim to) contain all the necessary physics.  They tend to be ``black boxes,'' however, in that one cannot modify them to suit one's neutrino data.  Such is the merit of parametric models.  Comparing models to one's neutrino data is an unsatisfying way to evaluate systematic uncertainties, and in Sections~\ref{fluxmonitor} and \ref{two-det} other techniques are discussed to adapt one's models to neutrino data.

\subsubsection{Shower Cascade Models}

Shower cascade models offer physics-driven descriptions of the cascade of particles initiated by a proton interaction in a nuclear target.  These codes allow the user to describe a complex geometry of a nuclear target, impinge a beam into the target, and follow the progeny of the interactions through the target, allowing them to subsequently escape the target, or further scatter/interact to produce other particles.  Such models therefore are critical to extrapolating data with respect to the beam momentum, target material $A$, and understanding thick target effects.  The state-of-the-art models include MARS-v.15 \cite{mars}, Fluka-2005\cite{fluka}, and DPMJET-III \cite{dpmjet}.  Other models, such as GHEISHA \cite{gheisha}, GCALOR \cite{gcalor}, Geant/Fluka \cite{gfluka}, or Geant4 \cite{geant4} appear to have discrepencies with published hadron production data in certain kinematic regimes \cite{jaffe2006}.  MARS-v.15 and Fluka-2005 have been tuned to accomodate the SPY data, but not measurements from HARP, BNL-E910, or NA49.

\subsubsection{Parametric Models}

\vskip .1 cm
\noindent\begin{large}{\it Malensek}\end{large}
\vskip .25 cm

Malensek \cite{malensek} parameterized the Atherton {\it et al} \cite{atherton1980} data and included an extrapolation for different beam energies:
\begin{equation}
\frac{d^2N}{dpd\Omega} = Kp(1-x_F)^A\frac{(1+ 5e^{-Dx_F})}{(1+p_T^2/m^2)^4}
\label{eq:malensek}
\end{equation}
with separate parameter sets $A, B, m, D$ for $\pi, K, p$.  Scaling to target lengths other than 1.25$\lambda_{int}$ is done by the naive absorption model, and scaling to different nuclear targets is done using the data of Eichten {\it et al} \cite{eichten1972}.  The formula maintains scaling, and the $p_T^{-8}$ at large $p_T$ was suggested by experimental data \cite{antreasyan1977}.  This formula fails to replicate the evolution of $\langle p_T \rangle$ with $x_F$ found by NA56/SPY \cite{ambrosini1999}.

\vskip .5 cm
\clearpage
\noindent\begin{large}{\it BMPT}\end{large}
\vskip .25 cm

The authors of \cite{bmpt} developed a new parameterization that fit not only the Atherton \cite{atherton1980} but also the NA56/SPY \cite{ambrosini1999} data, the latter of which indicated the evolution of $p_T$ with $x_F$.  The functional form of their parameterization for for the invariant cross section is:
\begin{equation}
(E\times\frac{d^3\sigma}{dp^3})=A(1-x_R)^\alpha(1+Bx_R)x_R^{-\beta}(1+a^\prime(x_R)p_T+b^\prime(x_R)p_T^2)e^{-a^\prime(x_R)p_T}
\label{eq:bmpt}
\end{equation}
where $x_R\equiv \overline{E}/\overline{E}_{max}$ is the ratio of the particle's energy to its maximum possible energy in the C.M. frame, and the functions $a^\prime(x_R)\equiv a/x_R^\gamma$ and $b^\prime(x_R)\equiv a^2/x_R^\delta$ control the scale-breaking of $p_T$. Separate parameters were fitted to $\pi^+$, $\pi^-$, $K^+$, $K^-$ data, subject to constraints on ratios of positives and negatives $\propto(1+ax_R)^b$ which have been well-measured previously.  For application to other nuclear targets, the scaling Equation~\ref{eq:alpha} from Barton \cite{barton1983} is applied.  An improved version of the naive absorption model was developed for thicker targets.

\vskip .5 cm
\noindent\begin{large}{\it Sanford-Wang}\end{large}
\vskip .25 cm

The Sanford-Wang parameterization \cite{sanfordwang1967,wang1970} was used by the CERN PS \cite{wachsmuth1969}, FNAL-MiniBooNE \cite{schmitz2005}, K2K \cite{ahn2006}, and BNL beams:
\begin{equation}
\frac{d^2N}{dpd\Omega}= Ap^B(1-p/p_0)\exp[-\frac{cp^D}{p_0^E}-F\theta(p-Gp_0\cos^H\theta)]
\label{eq:sanfwang}
\end{equation}
where $p_0$ is the proton momentum, $p$ is the pion or kaon momentum, $\theta$ is the pion or kaon production angle, and the parameters $A$-$H$ are fitted to experimental data, with separate parameters derived for $\pi^+$, $\pi^-$, $K^+$, and $K^-$.  Such is a thin-target parameterization.  Fits to Cho {\it et al} \cite{cho1971} and Allaby {\it et al} \cite{allaby1970} are found to be consistent with new data from HARP \cite{catanesi2006} within 10\%, so this model appears quite satisfactory for ``low-energy'' thin target data.

Wang\cite{wang1974} also published a variation of this parameterization suitable for extrapolation to higher energies upon the publication of Baker {\it et al.} \cite{baker1974}:
\begin{equation}
\frac{d^2N}{dpd\Omega}= ap_0x_F(1-x_F)\exp[-bx_F^c-dp_T]
\label{eq:wang-highenergy}
\end{equation}
which is quite similar to the original Sanford-Wang with the omission of the last term in the exponential and a new scaling for beam momentum.  This function fit well to \cite{baker1974} with a factor 0.37 to account for the thick target, though it conflicts with NuMI data.

\vskip .5 cm
\noindent\begin{large}{\it CKP}\end{large}
\vskip .25 cm

The CKP model \cite{ckp} apparently dates back to cosmic ray work, and was used in neutrino beam simulation for the CERN-PS beam \cite{vandermeer1963} (adapted by \cite{vondardel1962}) and by BNL \cite{burns1965}:
\begin{equation}
\frac{d^2\sigma}{dpd\Omega} = Ap^2(p_0-p) e^{-(p-a)(b + c\theta)}
\label{eq:ckp}
\end{equation}
which was said \cite{vandermeer1963} to be in good agreement with previous data \cite{baker1961,diddens1961}.  The effective $e^{-ap_T}$ dependence comes from cosmic ray data \cite{burns1965}.  This is a thin-target model.


\clearpage
\newpage

\section{Focusing of Wide Band Beams}
\label{wbb}

The first accelerator neutrino experiment \cite{danby1962,burns1965} was a ``bare target beam,'' meaning that the proton beam was delivered to the target, and the meson secondaries emanating from the target were permitted to drift freely away from the target.  The only collimation or increase of flux is achieved by the relativistic boost of the secondaries in the forward direction.  The first neutrino experiment at Fermilab \cite{benvenuti1973,benvenuti1974} was likewise supplied by a bare-target beam.

Focusing of the secondaries from the target is essential for increasing the neutrino flux to the detectors on axis with the beam line.  In pion decay, the flux of neutrinos at a given decay angle $\theta$ with respect to the pion direction is (see Appendix~\ref{kinematics}):
\begin{equation}
\phi_{\nu}=\frac{A}{4\pi z^2}\left(\frac{2\gamma}{1+\gamma^2\theta^2}\right)^2,
\label{eq:flux-angle}
\end{equation}
where $A$ is the size of the detector, $z$ is its distance from the pion decay point, and $\gamma$ is the pion boost factor.  If no focusing is employed, the pions diverge from the target with a typical angle 
\begin{equation}
\theta_\pi\approx p_T/p_{\pi}\approx\langle p_T\rangle/p=280\hbox{MeV}/p_{\pi}=2/\gamma,
\label{eq:pi-divergence}
\end{equation}
where a typical $p_T\approx280$~MeV/$c$ off the target was assumed (see Section~\ref{hadprod}), and $p_\pi\approx E_\pi=\gamma m_\pi$.  This angle of the pions off the target is larger than the typical angle of neutrinos from pion decay, $\sim1/\gamma$, so is important to correct.  Perfect focusing of pions should, in this simple model, improve the flux of neutrinos by $\sim25$.

\subsection{Horn Focusing}
\label{horns}

Simon van der Meer developed the idea of the ``magnetic horn,'' \cite{vandermeer1961} a focusing device to collect the secondary pions and kaons from the target and directing them toward the downstream experiments, thereby increasing the neutrino flux.\footnote{The name is said to be given by the similarity of the horn's geometric shape to a Swiss alpenhorn.  Panofsky \cite{panofsky1962}, however, called van der Meer's device the ``Horn of Plenty.''  The name in the U.S. almost stuck \cite{kustom1969}.}

\begin{figure}[t]
\vskip 0.cm
  \centering
  \includegraphics[height=1.2in]{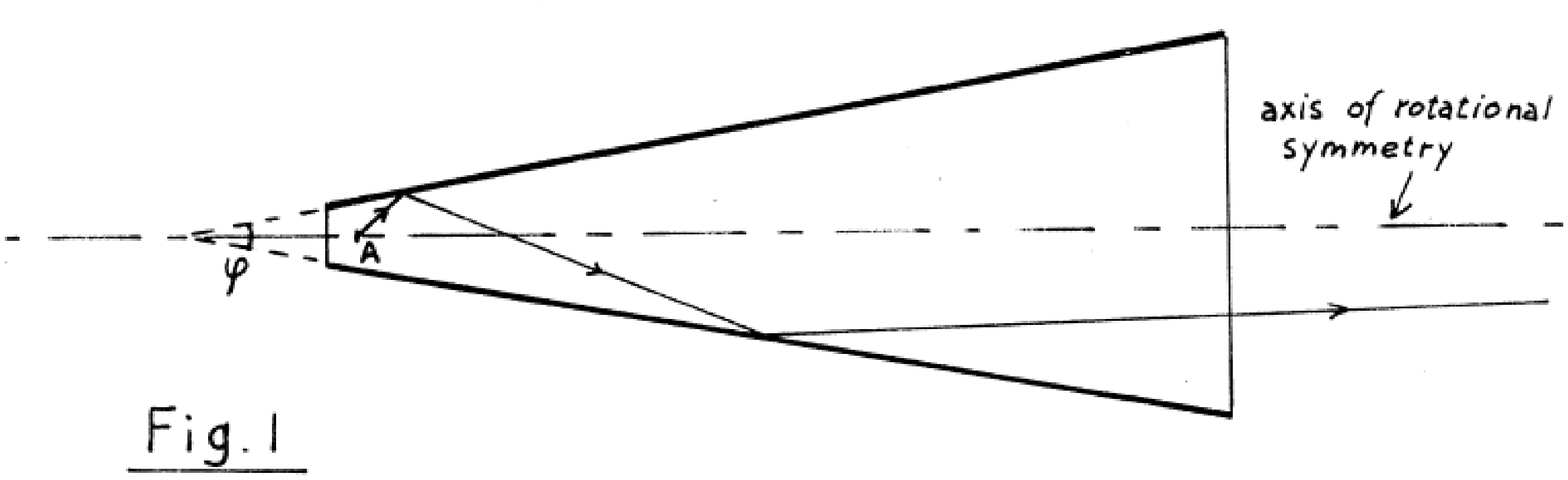}
  \includegraphics[height=1.3in]{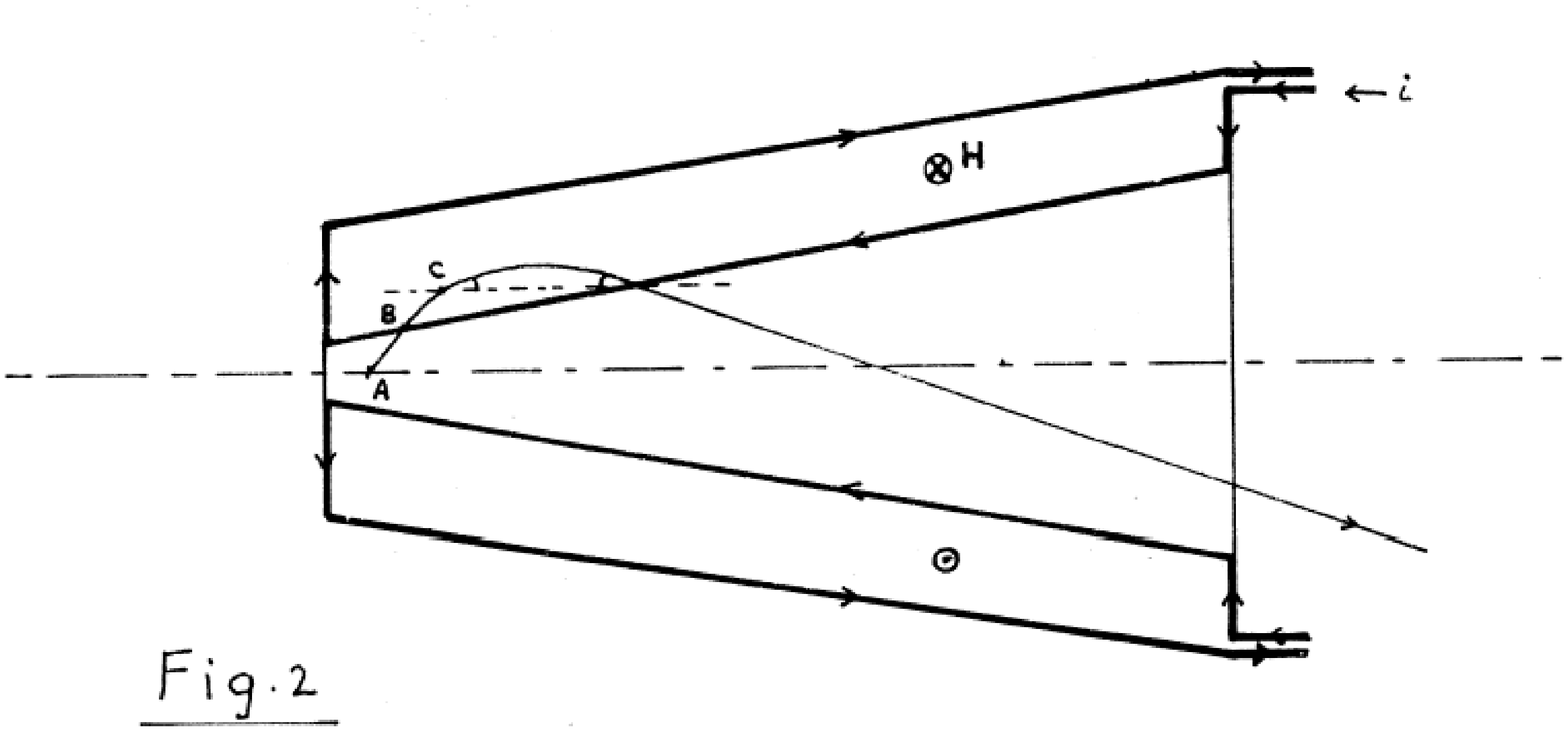}
\vskip 0.cm
  \caption{Van der Meer's schematic diagram of the neutrino horn, taken from \cite{vandermeer1961}.  (left) An optical source placed inside a reflective cone will result in exiting rays closer to the cone axis after several reflections.  (right) A negatively charged particle emanating from a source is deflected by a toroidal magnetic field. }
  \label{fig:vandermeer1961-horn}
\end{figure}

The magnetic horn consists of two axially-symmetric conductors with a current sheet running down the inner conductor and returning on the outer conductor, as shown in Figure~\ref{fig:vandermeer1961-horn}.  Between the conductors is produced a toroidal magnetic field whose $q{\bf v}\times {\bf B}$ force provides a restoring force for particles of one sign ($\pi^+$ or $\pi^-$), and defocuses particles of the other sign, thus enhancing a $\nu_\mu$ beam while reducing $\overline{\nu}_\mu$ background, for example.  The focusing device is unusual in accelerator physics in so far as the particles must traverse the lens conductors, causing some loss and scattering of particles.  Ref \cite{vandermeer1963} is a thorough consideration of a various trajectories of particles through such a lens and the angles and momenta that can be focused by a particular horn geometry.

Horns must withstand magnetic forces and the thermal load from the pulsed current and beam energy deposition in the horn conductors.  Since the early 1970's, beam intensities were high enough that these components become quite radioactive following extended running.  Systems for remotely-handling any failed components are necessary \cite{dusseux1972,fermi-handling,walker1987}.  Designs of horns now are quite refined and employ full analyses of the vibrations and strains on the horn (see, {\it e.g.} \cite{danilov1967,baratov1974} and the many contributions to \cite{nbi2000}-\cite{nbi2006}).  Further, current-delivery systems have gone from large coaxial cables over to metallic transmission lines\cite{nezrick1975,baratov1977d} able to better withstand intense radiation fields and magnetic forces.  
The following sections consider various geometries of horns and their focusing properties.  

\begin{figure}[t]
  \centering
  \includegraphics[width=4.5in]{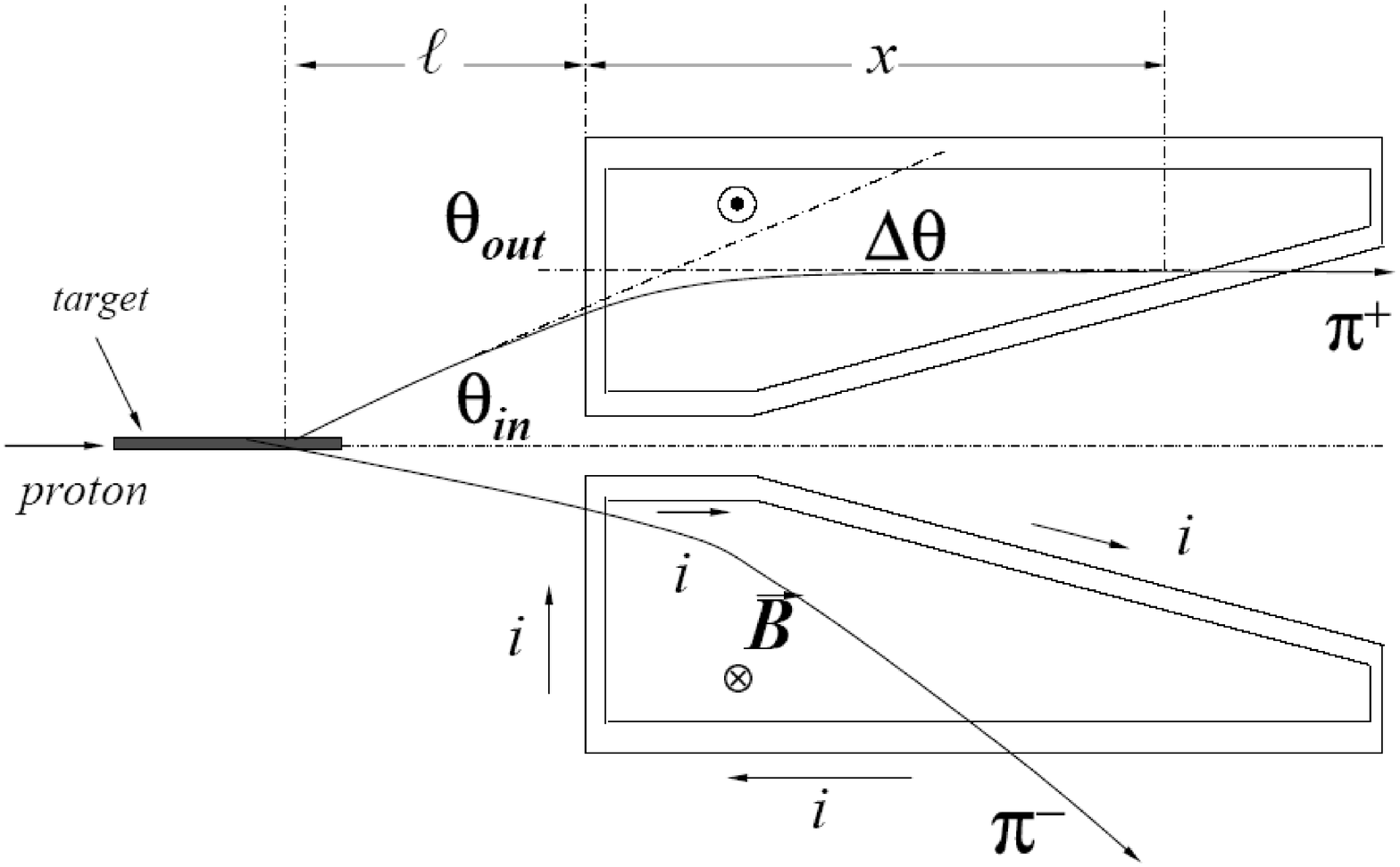}
\vskip -.5 cm
  \caption{Trajectory of a $\pi^+$ and a $\pi^-$ through a single conical horn focusing element.  An axially-symmetric current sheet down the inner conducting shell produces a toroidal magnetic field between the inner and outer conductors of the horn, providing a restoring force for one sign of particles.  }
  \label{fig:single-horn-focusing}
\end{figure}

\subsubsection{Conical Horns}

Van der Meer's original horn was a conical surface for the inner conductor \cite{vandermeer1961,vandermeer1963}.  Such a device, indicated in Figure~\ref{fig:single-horn-focusing}, does a good job at focusing all momenta for a given angle of pion into the horn, $\theta_{\mbox{in}}=r/\ell$.  To see this, note that the magnetic field of the device varies inversely with radius, $B=\mu_0 I/2\pi r,$ and the angular deflection of the pion in the magnetic field (the ``$p_T$ kick''), in the ``thin-lens approximation,'' is: $$\Delta\theta = \frac{Bx}{p}=\frac{\mu_0 I}{2\pi r}\frac{x}{p}$$ where $I$ is the horn current, $p$ is the pion momentum, and $x$ is the pathlength of the pion through the horn magnetic field region (see Figure~\ref{fig:single-horn-focusing}).

Recalling that the incident pion angle and momentum are inversely related (c.f. Equation~\ref{eq:pi-divergence}), we have that the average incident angle for pions into the horn is $\overline{\theta}_{\mbox{in}}\approx \langle p_T\rangle/p$.  A focused pion is one in which $\theta_{\mbox{out}}=0$, or in other words the $p_T$ kick cancels the incident angle of the pion into the horn.  One sets this $p_T$ kick to the average incident angle:
$$\Delta \theta=\overline{\theta}_{\mbox{in}}$$
$$\frac{\mu_0 I}{2\pi}\frac{x}{pr}=\frac{\langle p_T\rangle}{p}$$
\begin{equation}x = \langle p_T\rangle \frac{2\pi}{\mu_0I}r\label{eq:horn-cone}\end{equation}
This says that the pathlength in the horn should grow linearly with the radius of entrance into the horn, in other words a cone-shaped horn.  The momentum cancels out of the final equation, implying this is a broad-band beam.

Equation~\ref{eq:horn-cone} is derived in the limit of large source distance compared to the horn size, $\ell>>x$, and the small angle approximation for the pion angle $\theta_{\mbox{in}}$.  For many horns, these approximations are not valid.  Further, not all pions have the average $\langle p_T \rangle$ (not all incident angles $\theta_{\mbox{in}}$ are at the average or most likely angle $\overline{\theta}_{\mbox{in}}$).  

\subsubsection{Parabolic Horns}
\label{parabolic}

\begin{figure}[t]
\vskip -.5 cm
  \centering
  \includegraphics[width=4.2in]{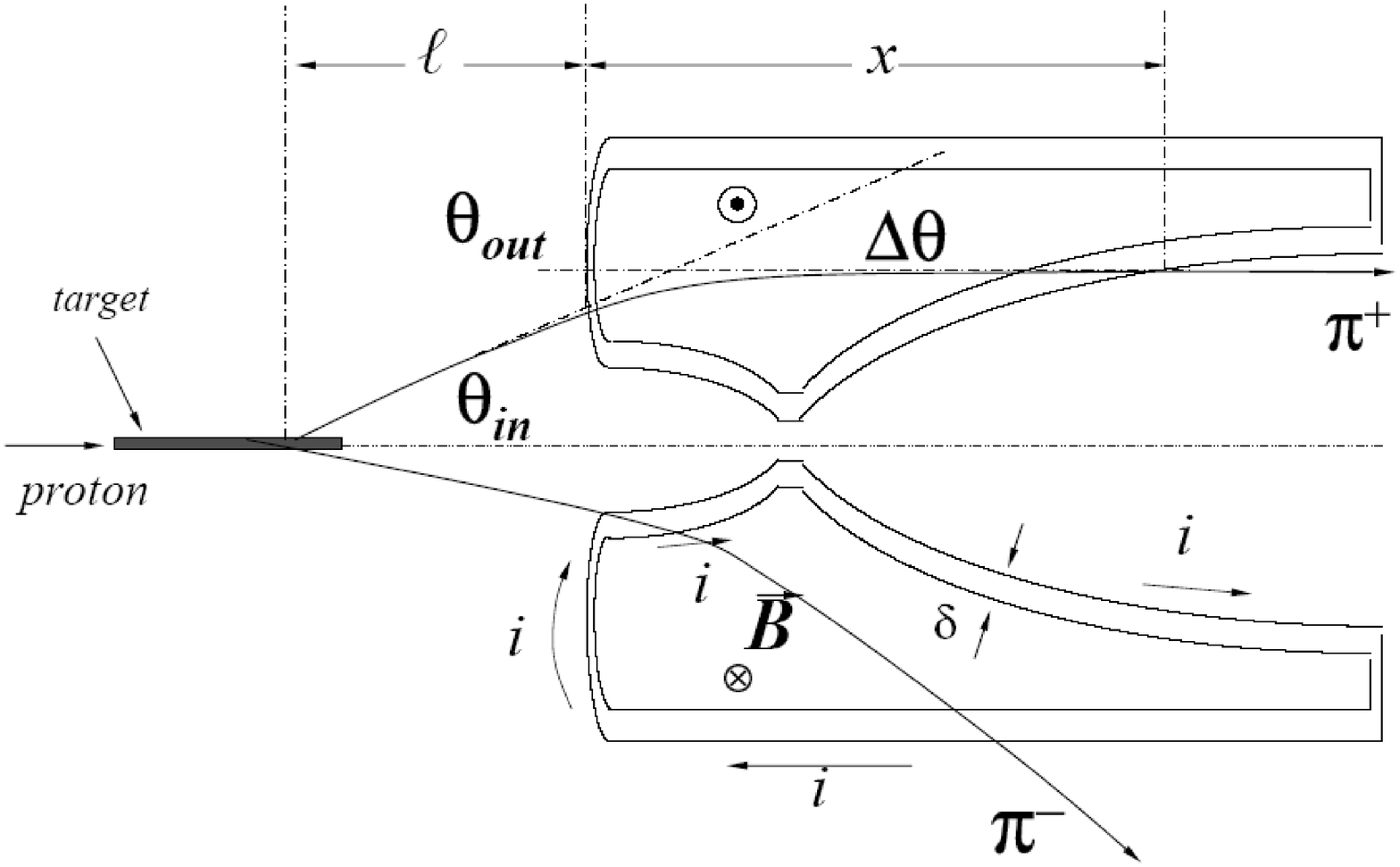}
\vskip -.5 cm
  \caption{Trajectory of a $\pi^+$ and a $\pi^-$ through a single horn focusing element.  An axially-symmetric current sheet down the inner conducting shell produces a toroidal magnetic field between the inner and outer conductors of the horn, providing a restoring force for one sign of particles.  }
  \label{fig:single-horn-parabola}
\end{figure}

It was apparently Budker who first conceived of a magnetic horn with parabolic-shaped inner conductors \cite{budker1961}.  Such a device focuses a given momentum for all possible angles of entry into the horn.  It appears that such was conceived in 1961, and first attempted by his group at Novosibirsk to improve the collection of positrons from a target for an $e^+e^-$ collider \cite{danilov1967}.  The parabolic lens was studied for its efficiency in collecting mesons for a neutrino beam by a Serpukhov group \cite{danilchenko1972}, and first implemented in a neutrino beam at the IHEP accelerator \cite{baratov1977a,baratov1977d}.

A parabolic horn, like that shown in Figure~\ref{fig:single-horn-focusing}, is one whose inner conductor follows a curve $z=a r^2$, with the parabolic parameter $a$ in cm$^{-1}$.  The $p_T$ kick of any horn results in a change in angle of 
$$\Delta\theta = \frac{Bx}{p} = \frac{\mu_0 I}{2\pi r}\frac{x}{p},$$ 
where $x = 2ar^2$ is the pathlength through the horn (for a parabolic conductor on either side of the neck).  Setting $\Delta\theta=\theta_{\mbox{out}}-\theta_{\mbox{in}}=\theta_{\mbox{out}}-r/\ell$, a point source located a distance $\ell=f$ (focal length) upstream of the target is focused like a lens if $\theta_{\mbox{out}}=0$, or 
\begin{equation}f=\frac{\pi}{\mu_0 a I} p.\label{eq:para-horn-f}\end{equation}
There are two differences with the conical horn: (1) the parabolic horn works for all angles (within the limit of the small angle approximation), not just the ``most likely angle'' $\overline{\theta}_{\mbox{in}}=\langle p_T\rangle/p$, and (2) a single parabolic horn has a strong chromatic dependence (its focal length depends directly on particle momentum $p$).

For the parabolic horn, the Coulomb scattering of particles through the horn conductors does not degrade the focusing quality for any pion momentum:  considering a parallel beam incident on the horn, the spot size, $S$, at the focal point of the horn will be due to Coulomb scattering in the horn material:
$$S = f \theta_Z$$
where 
$$\theta_Z = \frac{13.6~\mbox{mrad}}{p}\sqrt{\frac{t}{X_0}}$$
is the typical scattering angle in the horn conductor, $t$ the conductor thickness, and $X_0$ the conductor material radation length.  Thus 
$$S \propto \sqrt{\frac{t}{X_0}}\frac{1}{a I}$$
Thus, the quality of the focus is independent of the momentum, and improves with larger horn current, thinner conductors, lighter-weight materials with longer radiation lengths $X_0$, or longer horns with larger parameter $a$.  The fact that the focusing quality is independent of $p$ means one can almost calculate a spectrum with simple ray tracing and require no Monte Carlo calculation \cite{garkusha2006}.  To compensate the fact that particles entering the horn at larger radii traverse greater thickness of material $t\approx \delta\sqrt{1+4a^2r^2}$, horns are often designed with tapered conductor thicknesses, the neck region being the thickest.\footnote{This greater neck thickness is also beneficial for its greater strength.  The neck is the location of greatest mechanical strain due to the magnetic force of the pulsed current.}

\begin{figure}[t]
\vskip -0 cm
  \centering
  \includegraphics[width=6in]{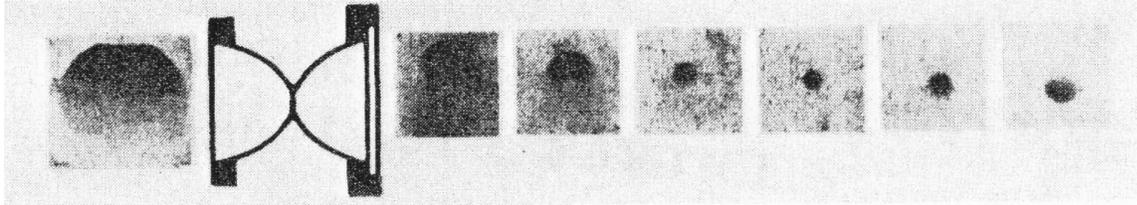}
\vskip -0 cm
  \caption{Demonstration of the focusing of a parabolic horn, taken from \cite{danilov1967}.}
  \label{fig:danilov1967-fig7}
\end{figure}

Figure~\ref{fig:danilov1967-fig7} is a demonstration performed by the Serpukhov group \cite{danilov1967} of the momentum-focusing properties of the parabolic horn.  A 130~MeV/$c$ electron beam is injected into the parabolic horn off-axis from the left.  After passing through the horn, the focusing causes a convergence of the electron rays at a distance from the horn equal to the focal length, after which the electron beam enlarges in size again.  The beam size before the horn and at several locations after the horn is measured using photographic film.  The circular spot indicates no aberrations despite off-axis injection and the measured focal length agreed with predictions.

\subsubsection{Ellipsoidal Lenses}

The authors of \cite{danilov1967} (from Budker's Novosibirsk group) show, in addition to the proposed parabolic surface, a slightly-less tapered inner conductor shape which they term the ``aberrationless'' surface.  The nature of such an alternative inner conductor shape is better-elucidated in Ref.~\cite{dohm1975}, in which is shown that an ellipsoidal inner conductor surface is a better focusing device across wider angles of entrance to the horn.  Such also appears to have been understood by Budker \cite{budker1977}.  

\begin{figure}[t]
\vskip -1.cm
  \centering
  \includegraphics[height=4.in]{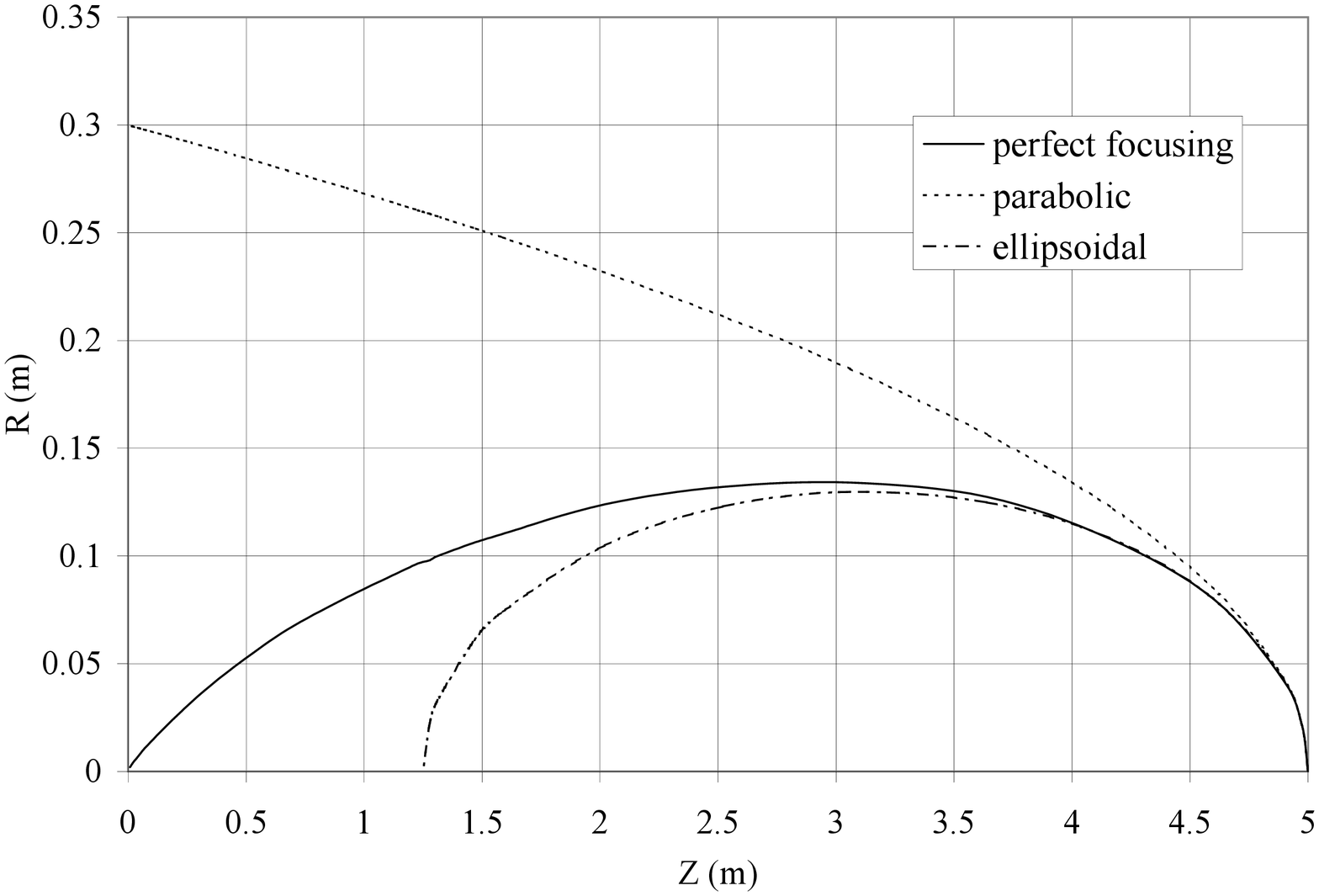}
\vskip -1.5cm
  \caption{Comparison of horn inner-conductor shapes required to focus particles assuming $f=5$~m, $I/p=60$~kA/GeV.  The three curves show the required horn shape for perfect focusing and for the thin lens approximations of parabolic or ellipsoidal horns.  The particle source is located at $(z,r)=(0,0)$.  The horns compared are ``half-lenses'' with vertical current sheet at $z=5$~m.}
  \label{fig:dohm1975-fig3}
\end{figure} 

The ellipsoidal lens is again one in which the focal length $f$ is a linear function of momentum:
\begin{equation}
f=\frac{4\pi b^2}{\mu_0 a I}p,
\label{eq:focal-ellipse}
\end{equation}
where the $I$ is again the horn current, and $a$ and $b$ are the major and minor half-axes (in cm) of the ellipsoid.  

As noted in \cite{dohm1975}, the parabolic lens is derived in the ``thin lens'' approximation, and further requires a small-angle approximation for the particles' incident angles into the horns: 
\begin{equation}
\theta_{\mbox{in}}^2<<1, ~~\mbox{and}~~ A_0\theta_{\mbox{in}}^2<<1,~~~~~\mbox{(parabolic~approx.)}
\end{equation}
where $A_0\equiv\frac{2\pi}{\mu_0}\frac{p}{I}=p(\mbox{GeV}/c)/(6\times10^{-5} I\mbox{(kA)})$.  Given that $A_0\sim200-400$, such is more restrictive than the small-angle approximation required for the ellipsoidal lens:
\begin{equation}
\theta_{\mbox{in}}^2<<1,~~\mbox{and}~~A_0^2\theta_{\mbox{in}}^4<<1,~~~~~\mbox{(ellipsoidal~approx.)}
\end{equation}
so that the ellipsoidal lens achieves an exact momentum-focus across a wider angular spread.  As can be seen in Figure~\ref{fig:dohm1975-fig3}, the parabolic lens is an approximation of the ellipsoid surface for small-angle particles.  

\subsubsection{Magnetic Fingers}

\begin{figure}[t]
\vskip -0.cm
  \centering
  \includegraphics[height=2.1in]{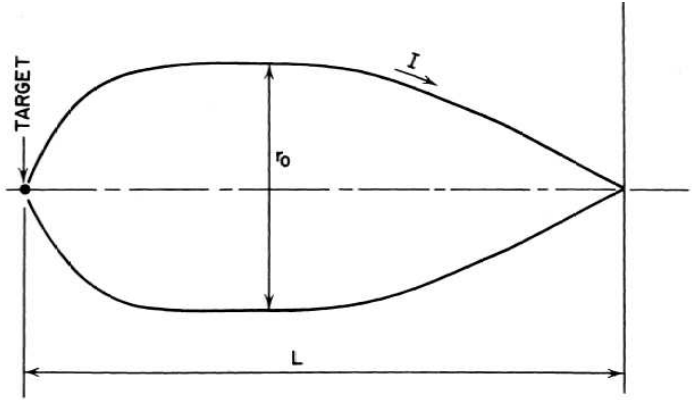}
\vskip -0cm
  \caption{Inner conductor of the ideal-focusing ``magnetic finger'' of Palmer \cite{palmer1965}.}
  \label{fig:palmer1965-fig6}
\end{figure}

Palmer \cite{palmer1965} proposed a variant of the magnetic horn which he dubbed ``magnetic fingers.''  His variation required an axially symmetric pair of pulsed conductors, but considered inner conductor shapes other than conical surfaces.  Following numerical calculations, his inner conductor shape reminded him of a human digit, shown in Figure~\ref{fig:palmer1965-fig6}.  Such shapes were adopted for two-horn beams at BNL \cite{carroll1985,carroll1987}, and the BNL horns subsequently informed the designs for KEK \cite{yamanoi1997}, MiniBooNE\cite{kourbanis2002}, and JPARC \cite{itow2001}.  

The numerical calculation of ideal focusing for a particle of momentum $p$ is detailed thoroughly in \cite{dohm1975}, and dispenses with both the small-angle and thin-lens approximations, computing the curvature of a particle through the lens itself to obtain the required incident coordinates $(z,r)$ at which a particle of momentum $p$ should enter the horn in order to be focused ($\theta_{\mbox{out}}=0$ at the horn exit).  Figure~\ref{fig:dohm1975-fig3} shows such a horn shape in comparison to the ellipsoid and parabolic approximations, in which it is assumed that the horn is a ``half-lens,'' {\it i.e.:} one in which the conductor is tapered upstream of its neck, but whose current sheet becomes vertical at $z=f$, where $f$ is the focal length (similar assumption to Figure~\ref{fig:palmer1965-fig6}).

The ideal surface in Figure~\ref{fig:dohm1975-fig3} has the visual appearance of a lopsided ellipse, similar to lenses described by Budker's group \cite{budker1977}.

\subsection{Multi-horn Systems}
\label{multi-lens}

Palmer \cite{palmer1965} noted that multiple focusing elements can improve the neutrino flux because subsequent focusing elements can be used to ``rescue'' pion trajectories improperly focused by the first focusing element.  Such a multi-lens system was adopted at CERN PS neutrino beam \cite{asner1965,asner1966,pattison1969} and nearly every WBB since (see Table~\ref{tab:nu-beams}).  A double horn system was also implemented for the CERN Antiproton Accumulator \cite{vandermeer1980}.  

Palmer \cite{palmer1965} gives a clear motivation for the multiple lenses:  a lens provides a definite ``$p_T$ kick'' given by $\Delta\theta$ whose value can be calculated given the horn shape, current, and the particle momentum $p$.  The horn is tuned to give a $p_T$ kick equal to this most probable entrance angle $\overline{\theta}_{\mbox{in}}=\langle p_T\rangle/p$ into the horn:$$\Delta\theta=\overline{\theta}_{\mbox{in}}.$$Many particles emerging from the target will have a $p_T$ not equal to the mean $\langle p_T \rangle$, resulting in particles, at the same momentum $p$, entering the horn at a variety of angles.  Assume we would like to focus all particles between $\theta_{\mbox{in}}=0$ and $\theta_{\mbox{in}}=2\overline{\theta}_{\mbox{in}}$.  A particle entering the horn at $\theta_{\mbox{in}}$ will thus emerge from the horn with outgoing angle $$\theta_{\mbox{out}}=|\theta_{\mbox{in}}-\Delta\theta|.$$  A particle entering the horn with $\theta_{\mbox{in}}=\overline{\theta}_{\mbox{in}}$ will exit at $\theta_{\mbox{out}}=0$, while a particle entering the horn at either $\theta_{\mbox{in}}\approx0$ or $\theta_{\mbox{in}}=2\overline{\theta}_{\mbox{in}}$ will emerge with an angle $\theta_{\mbox{out}}=\overline{\theta}_{\mbox{in}}$.  A particle beam entering the horn with angular divergence 2$\overline{\theta}_{\mbox{in}}$ will emerge with divergence $\overline{\theta}_{\mbox{in}}$.

\begin{figure}[t]
\vskip -1. cm
  \centering
  \includegraphics[width=5.8in]{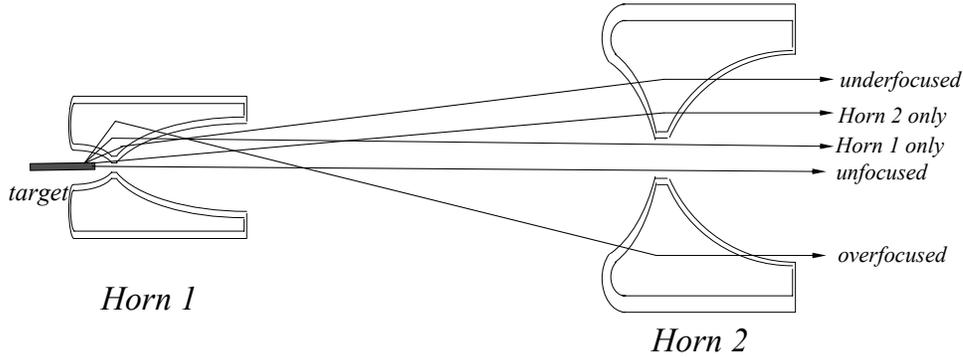}
\vskip -6. cm
  \caption{Two-lens focusing system:  a second lens, significantly further from the target than the first, improves the collection efficiency of particles over-or underfocused by the first lens.  The horns shown are for the Fermilab NuMI line \cite{abramov2002}.  The scale transverse to the beam axis is 4$\times$ the scale along the beam axis.}
  \label{fig:two-horn-focusing}
\end{figure} 

\begin{figure}[b]
\vskip -.5 cm
  \centering
  \includegraphics[width=5.5in]{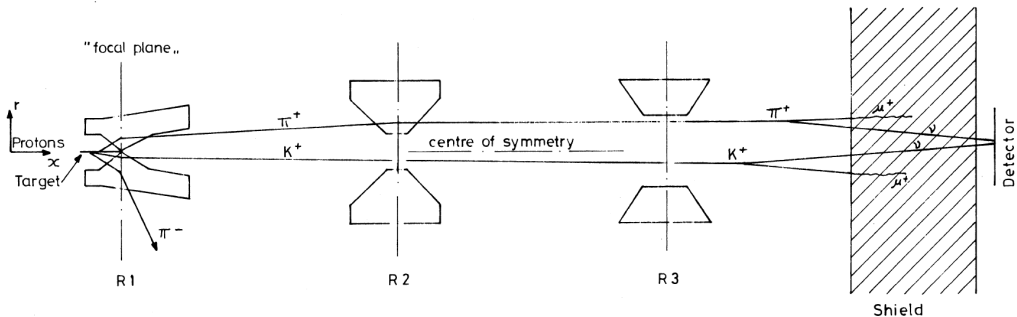}
\vskip -.5 cm
  \caption{Three-horn focusing system employed at CERN for the 1967 neutrino run, taken from \cite{venus1969}.  Each successive downstream horn is larger to capture errant particle trajectories, and each has a larger inner aperture to leave un-perturbed those particles well-focused by the upstream horns.}
  \label{fig:venus1969-fig1}
\end{figure}

A second lens far from the first will see a point source of particles with a span of angles 0 to $\overline{\theta}_{\mbox{in}}$.  It would be likewise expected to halve the divergence of the beam.  Its inner aperture should be larger so as to leave unperturbed those particles already well-focused by the first lens.  A third lens could similarly be expected to bring the overall divergence down a factor of 8, but must be located even further downstream to continue the point source approximation for the incoming particles.  Techniques for design of multiple lens systems, including lens sizes, focal lengths, and inter-lens distances, based upon transfer matrices have been developed in \cite{danilchenko1972}.  

A three-lens system was adopted for the 1967 CERN run\cite{asner1965,asner1966,pattison1969}, with the second horn 15~m from the horn-1 and the third $\sim35$~m from the target, more than half-way down the 60~m decay path (see Figure~\ref{fig:venus1969-fig1}).  

\begin{figure}[t]
\vskip 0.cm
  \centering
  \includegraphics[width=5.in]{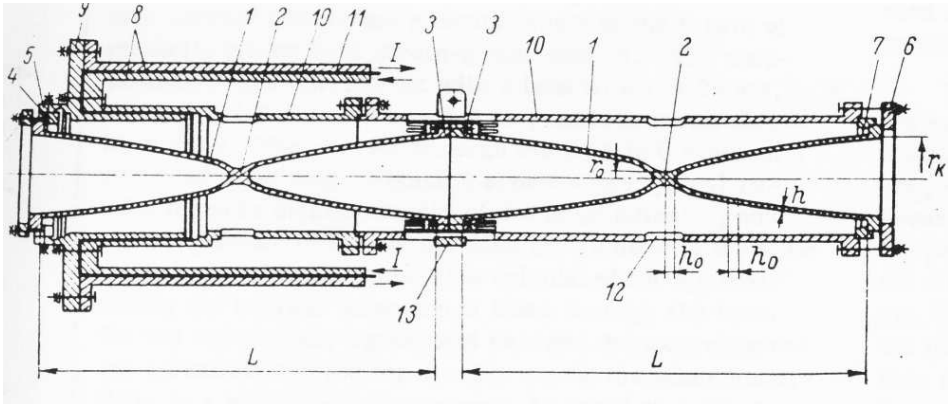}
\vskip 0.cm
  \caption{Double-lens horn from the IHEP beam, taken from \cite{baratov1977d}. 1.-inner conductor, 2.-neck, 3.-flange, 4.-insulating ring, 5.-clamp, 6.-flange, 7.-half ring, 8.-transmission stripline, 9.-current distribution ring, 10.-outer conductor, 11.-insulation, 12.-air cooling slot.}
  \label{fig:baratov1977d-fig2}
\end{figure} 

\begin{figure}[p]
  \centering
  \includegraphics[width=7.5in,angle=90]{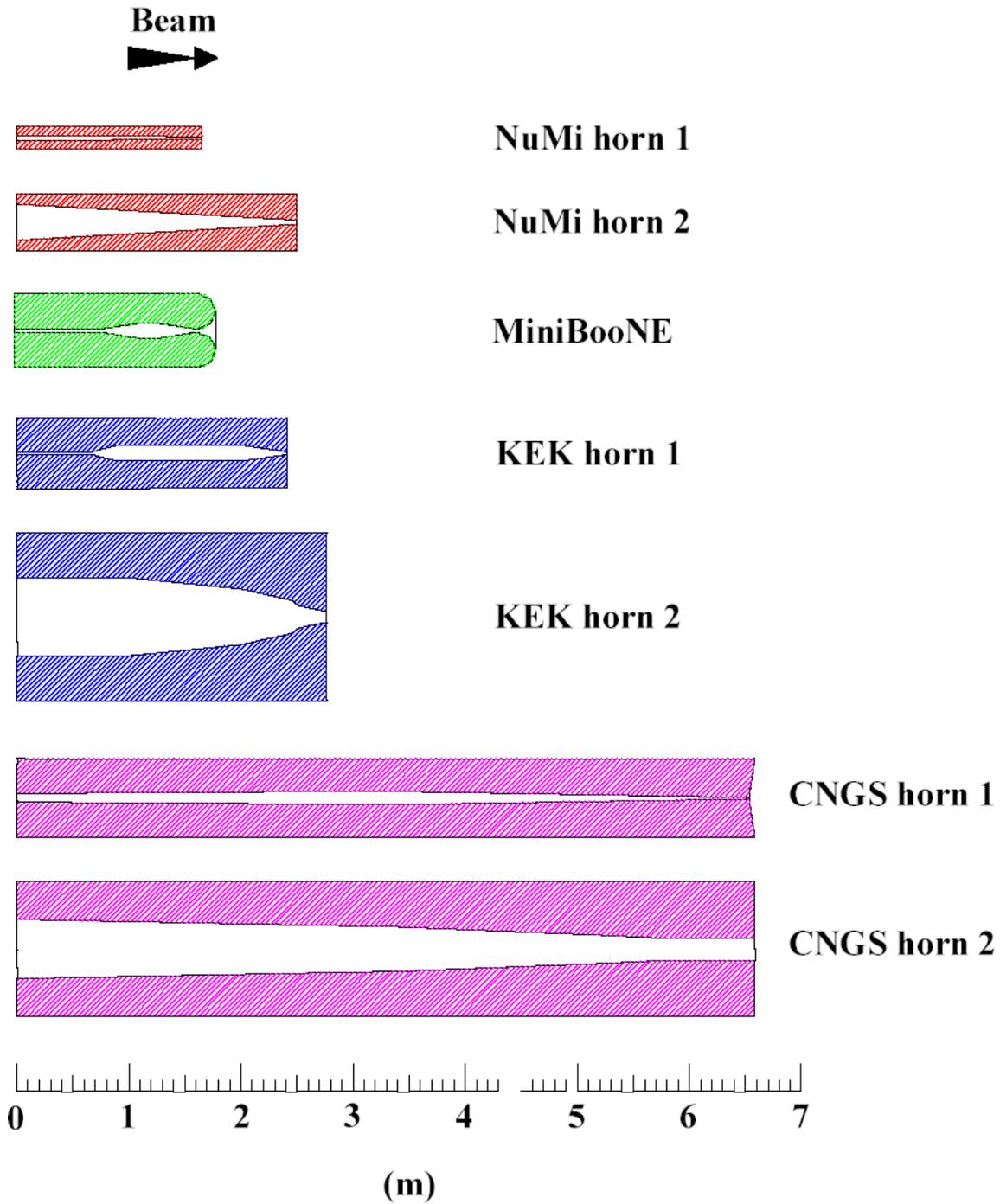}
\vskip -.5 cm
  \caption{Scale drawings of magnetic horns from the KEK, FNAL-NuMI, FNAL-MiniBooNE, and CERN-CNGS neutrino beams.  All but FNAL-MiniBooNE were multi-horn beam lines.  Adapted from \cite{rangod2002}.}
  \label{fig:rangod-horn-figure}
\end{figure}

Serpukhov adopted a three-horn beam \cite{baratov1977a,baratov1977d}, which had the distinction of a two-lens horn, shown in Figure~\ref{fig:baratov1977d-fig2}:  the first horn consisted of two tapered regions with two ``necks,'' giving the equivalent of a pair of lenses.  In this sense the IHEP beam was actually a four-lens system (see Figure~\ref{fig:baratov1977d-fig1}).\footnote{A double-neck conical horn was attempted at Fermilab,\cite{nezrick1975} but this was replaced in favor of a single horn \cite{grimson1978}.}  Horns for the more recent beam lines are shown in Figure~\ref{fig:rangod-horn-figure}.

\begin{figure}[t]
\vskip 0.cm
  \centering
  \includegraphics[width=4.8in]{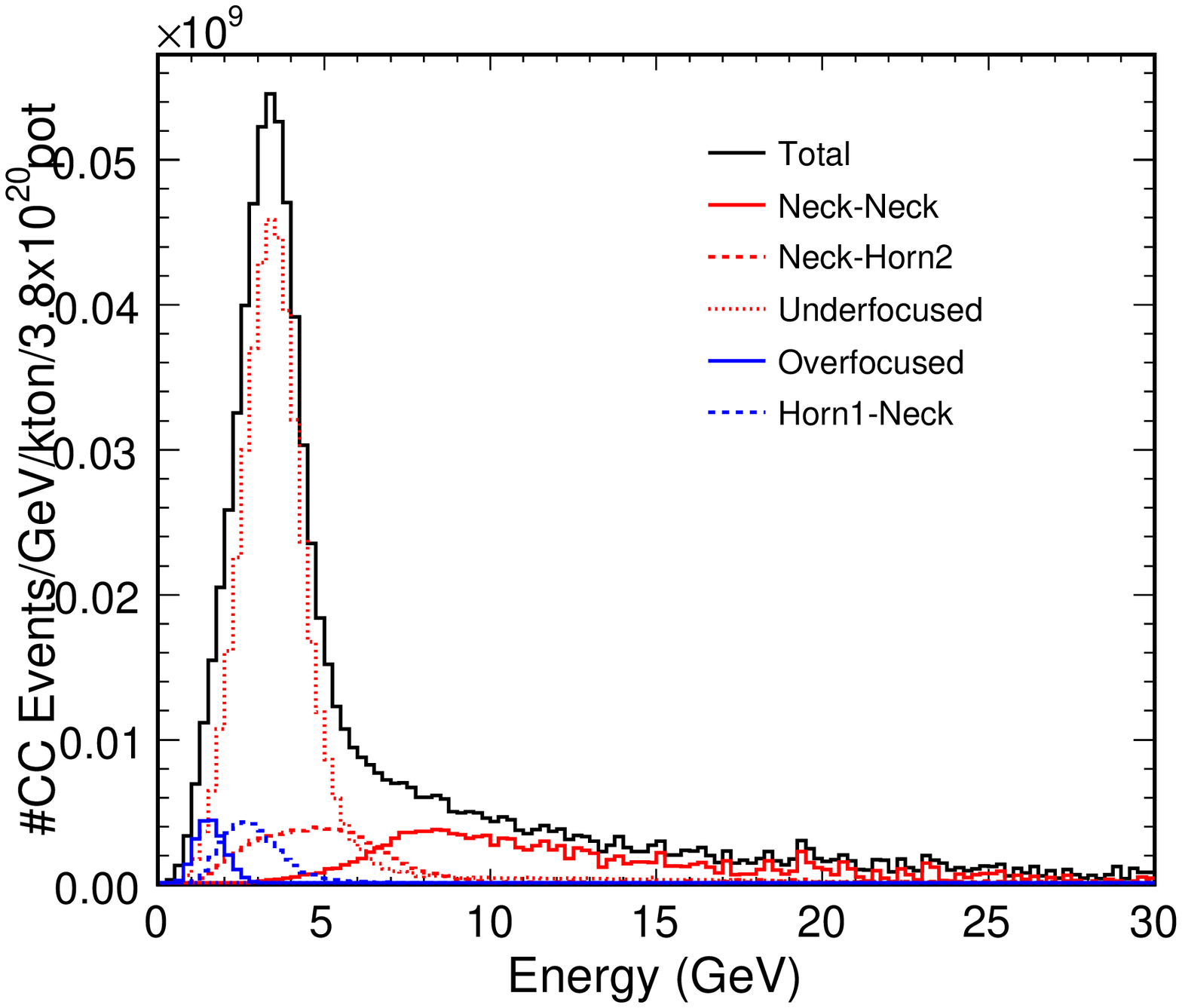}
\vskip -.3 cm
  \caption{Neutrino spectrum from the two-horn beam at the NuMI facility at FNAL.  The components of the spectrum correspond to the different possible pion trajectories of Figure~\ref{fig:two-horn-focusing}.  Taken from \cite{pavlovic2007}.}
  \label{fig:5-beams}
\end{figure}

\begin{figure}[b]
\vskip -.5cm
  \centering
  \includegraphics[width=2.in,height=3in]{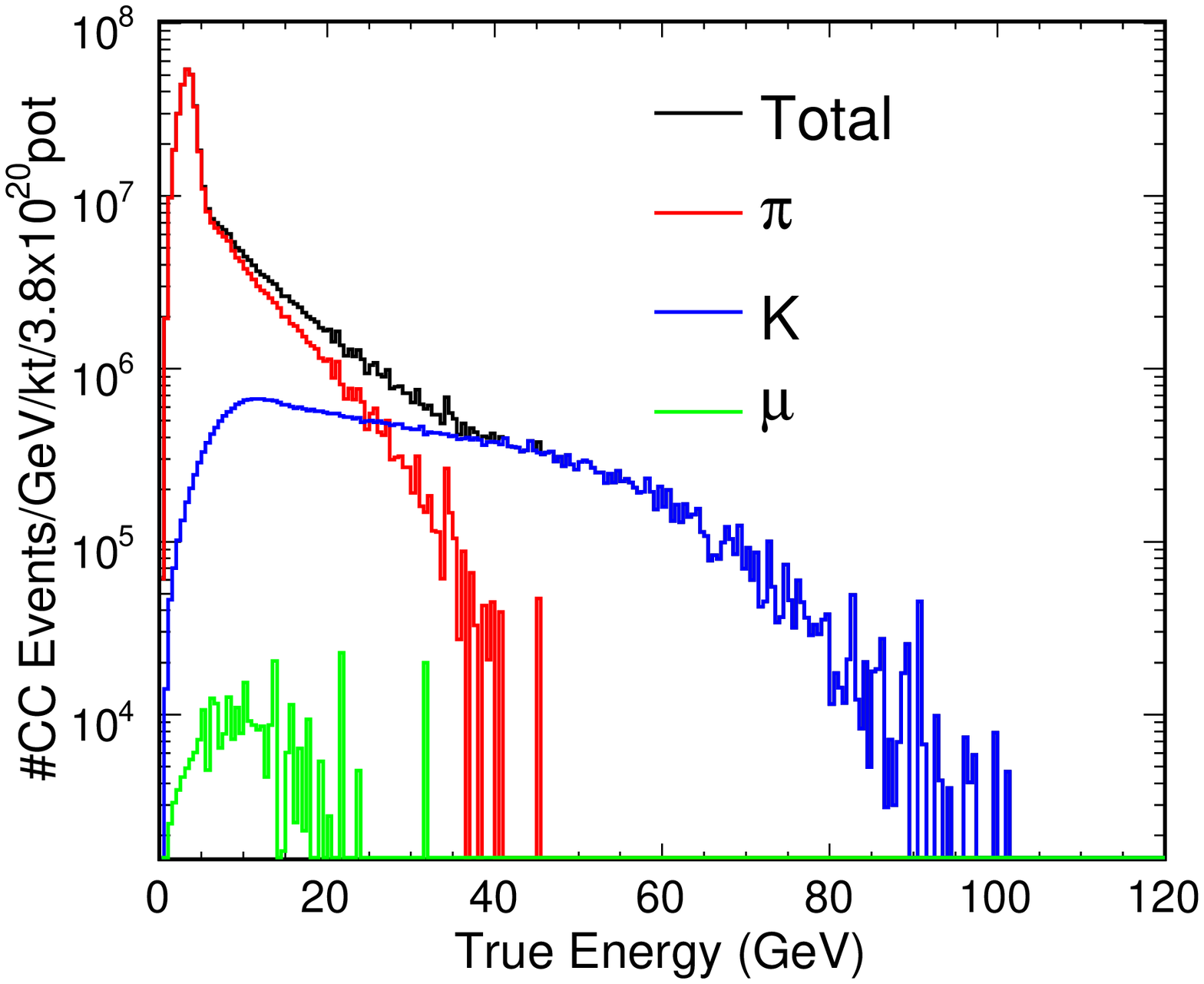}
  \includegraphics[width=2.in,height=3in]{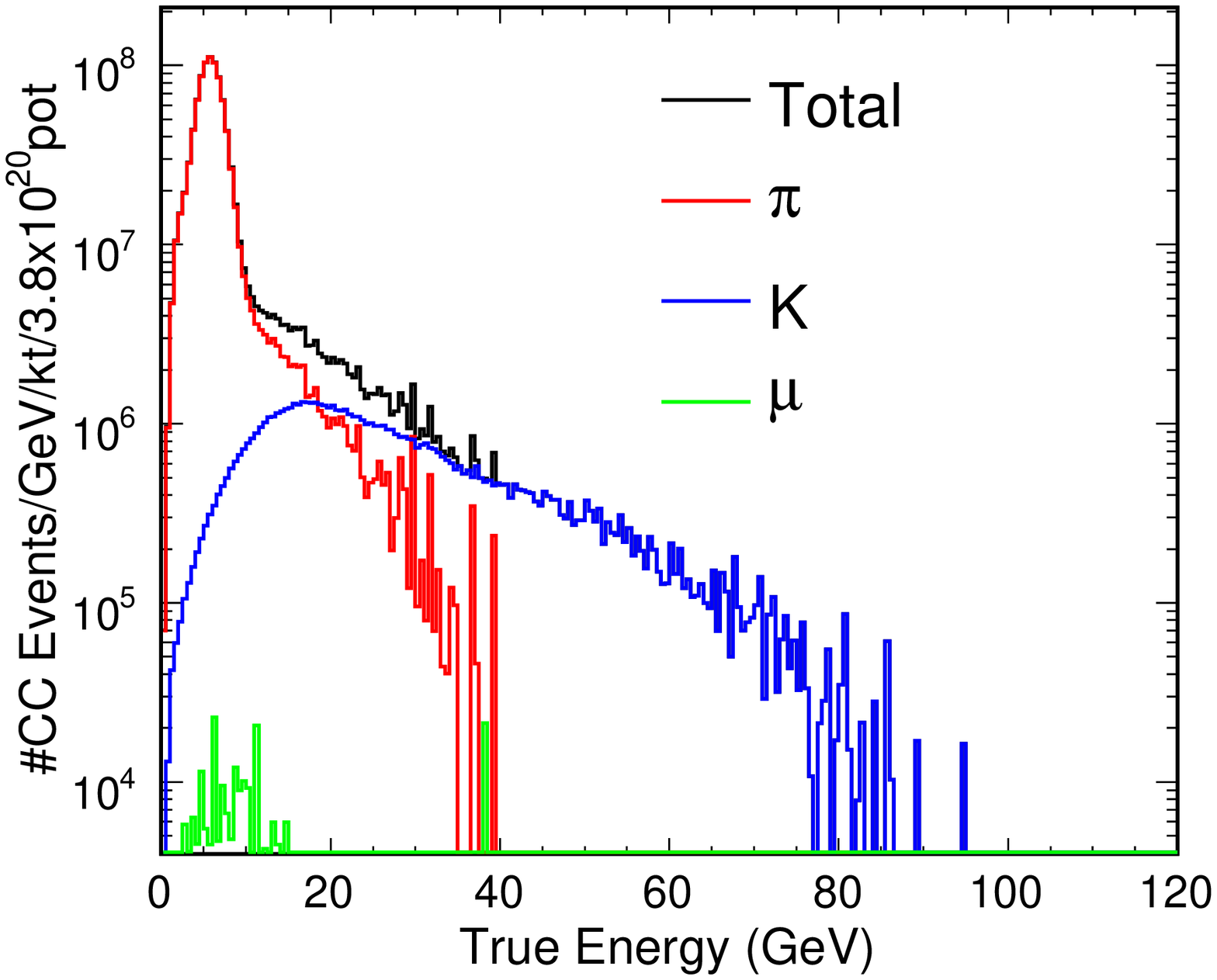}
  \includegraphics[width=2.in,height=3in]{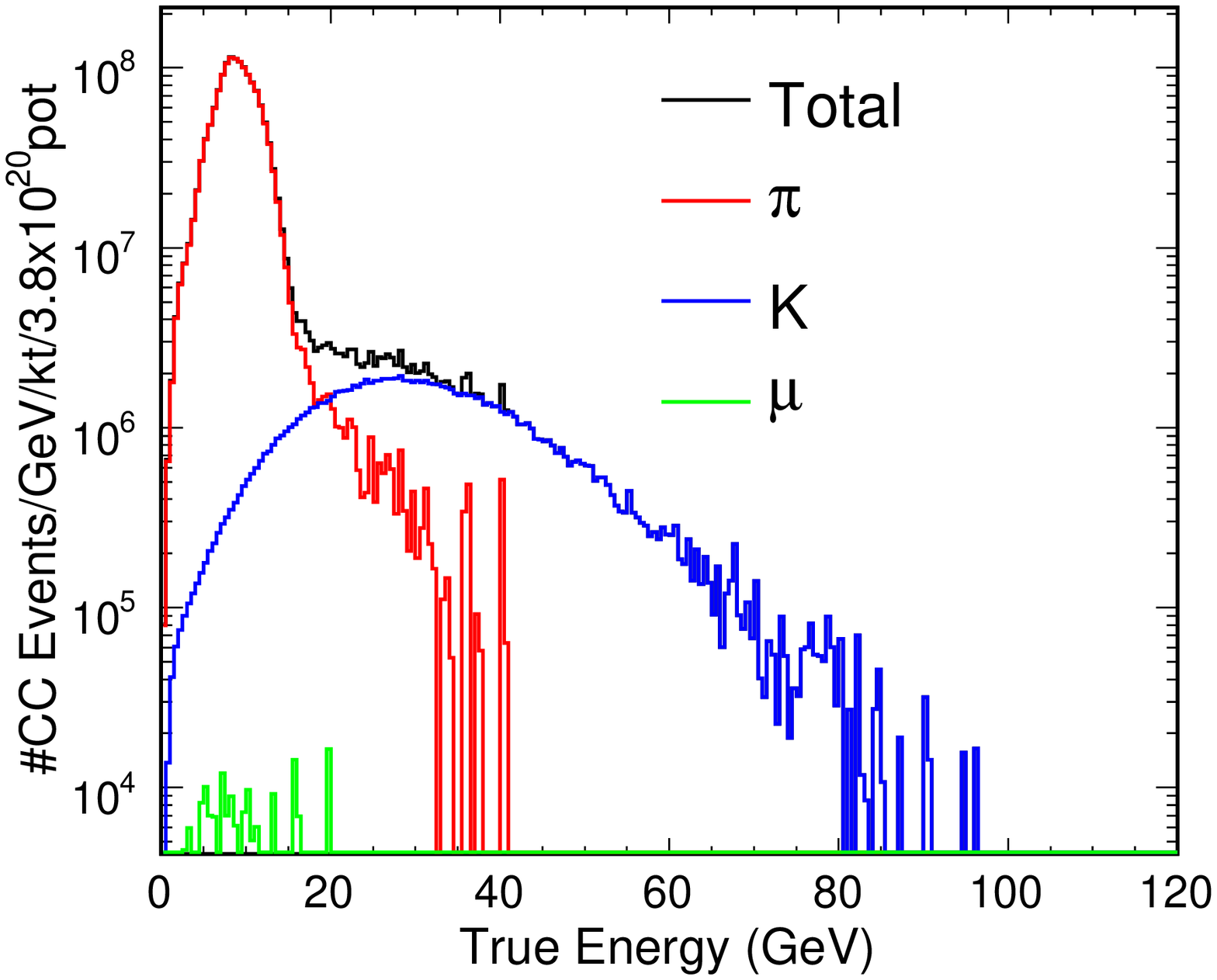}
\vskip -.3 cm
  \caption{Neutrino spectrum from the two-horn beam at the NuMI facility at FNAL, showing separate contributions from $\pi^+$, $K^+$, and $\mu^-$ decays.  The three graphs are from three settings of the NuMI beam line designed to give a low (left), medium (middle), or high (right) energy tune (see text and Figure~\ref{fig:zarko-linear-dependence}).  Taken from \cite{pavlovic2007}.}
  \label{fig:compare-parents}
\end{figure}

Figure~\ref{fig:5-beams} shows the predicted neutrino spectrum from the two-horn system of NuMI at FNAL.  
Also shown are the components of this spectrum corresponding to the different pion trajectories of Figure~\ref{fig:two-horn-focusing}.  As the angle of the neutrino parent decreases, one expects its momentum $p\approx\langle p_T\rangle/\theta$ to increase.  The pions focused by only horn 1 give softer neutrinos than those focused only by horn 2.  It is of note that the peak of the neutrino energy spectrum comes from particles which pass through the focusing system, while the ``high energy tail'' comes from particles which pass through the field-free apertures of the horns.  Figure~\ref{fig:compare-parents} shows the two components from $\pi$ and $K$ decays common to horn-focused beams.

\begin{figure}[t]
\vskip -1.cm
  \centering
  \includegraphics[width=3in]{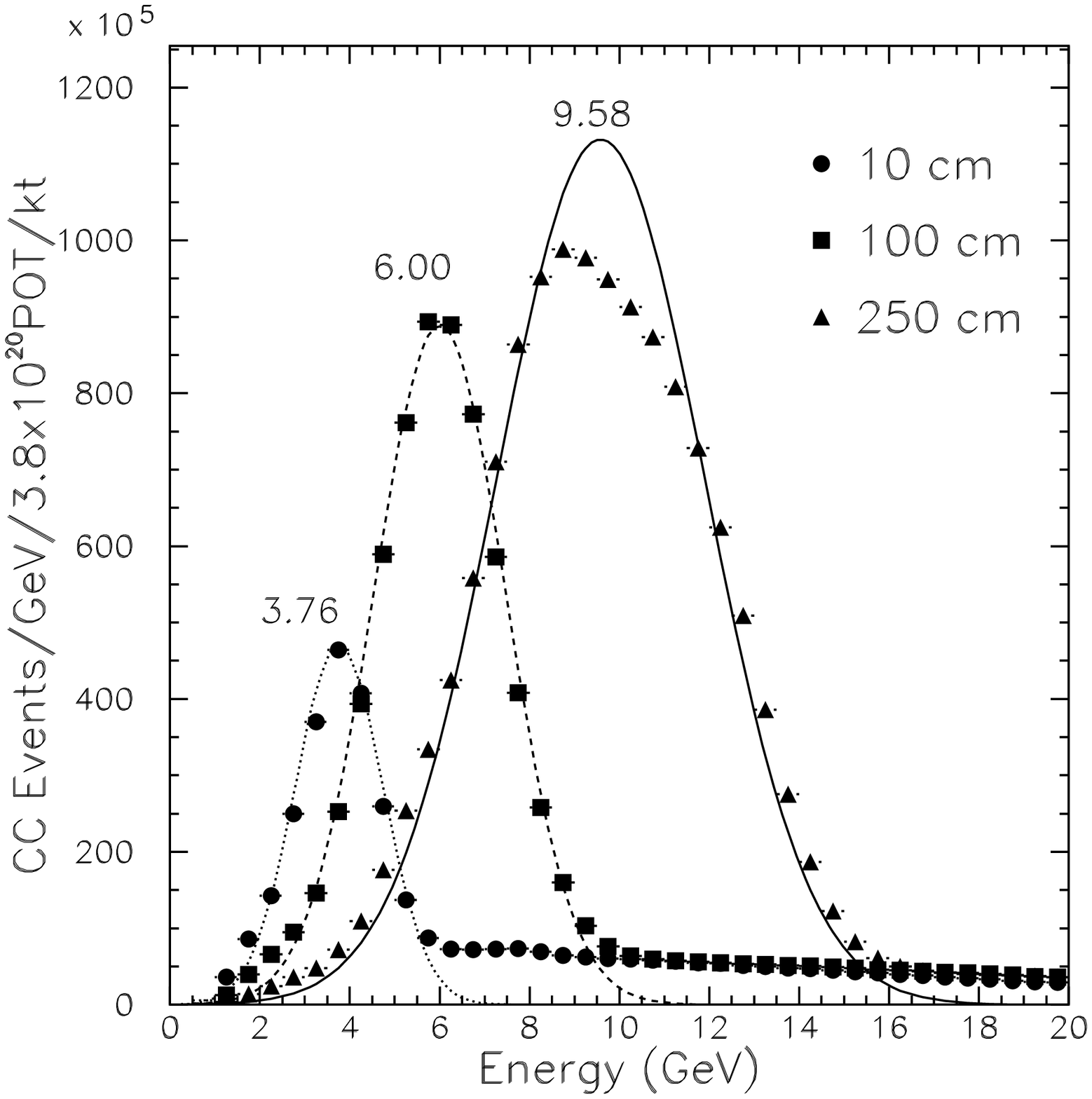}
  \includegraphics[width=3in]{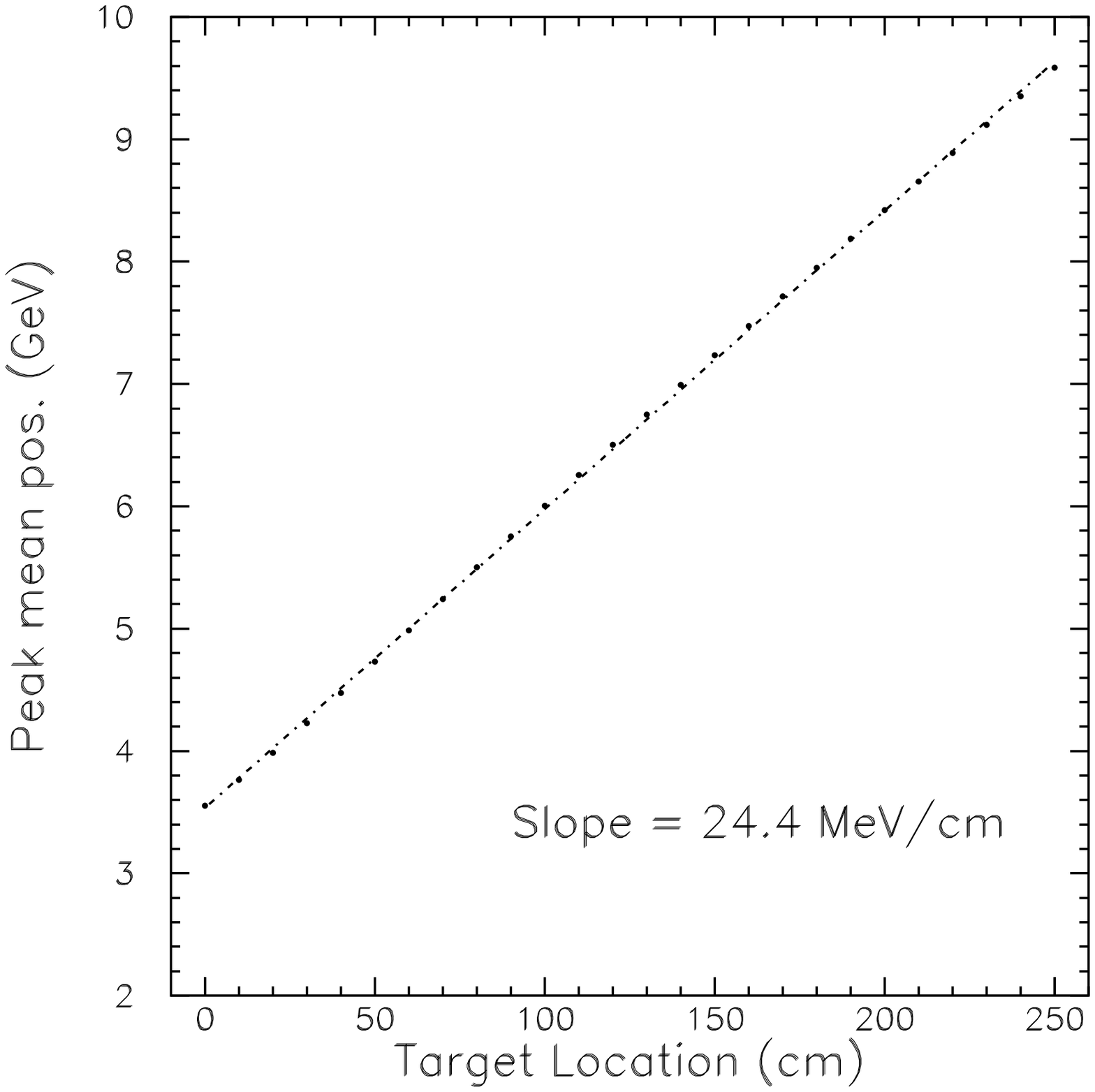}
\vskip -0.3 cm
  \caption{(left) Neutrino energy spectrum at the detector hall 1040~m from the NuMI target for several positions of the target upstream of the first horn.  Each spectrum consists of a focusing peak (fitted with a gaussian curve) and a high-energy tail from unfocused parents.  (right) Peak neutrino energy (from the fitted gaussian) as a function of the target location.  Taken from \cite{pavlovic2007}.}
  \label{fig:zarko-linear-dependence}
\end{figure}

The NuMI beam at Fermilab implemented a ``continuously variable'' neutrino energy capability by mounting the target on a rail drive system that permits up to 2.5~m travel along the beam direction \cite{kostin2002}.  The target's remote control permits change of the neutrino energy without unstacking of the shielding elements.  The utility of such a system is that it can assist in understanding detailed systematics of the neutrino energy spectrum observed in the detectors \cite{michael2006}.  The principle of the variable energy beam relies upon Equation~\ref{eq:para-horn-f}:  since $f\propto p$, the momentum at which point-to-parallel focusing is achieved will increase as the source distance is increased.  Thus in the thin-lens approximation one expects linear dependence of the peak focused neutrino energy upon the target position $\ell$.  Such is borne out by simple Monte Carlo calculation (see Figure~\ref{fig:zarko-linear-dependence}), and by observation in the NuMI/MINOS neutrino data \cite{michael2006}.\footnote{The authors of Ref. \cite{baratov1977a} and \cite{barish1977} note that variations of the horn current and the target positions can be used to vary the neutrino energy.  However, these groups appear to have employed only variations in current, and with a goal of increasing neutrino event rate at the detectors.  The authors of \cite{dohm1975,abramov2002} note that linear lenses permit different target-horn placements to obtain low-, medium-, or high-pass beams.}  Further discussion is found in Section~\ref{two-det}.




\clearpage

\clearpage
\newpage

\subsection{Quadrupole-Focused Beams}

Quadrupole-focused beams are generally less efficient than horn focusing, but they are relatively inexpensive and simpler to design, relying on magnets for conventional accelerator rings and they need not be pulsed, permitting use in slow-spill beams.  

\begin{figure}[t]
\vskip 0.cm
  \centering
  \includegraphics[width=3.8in]{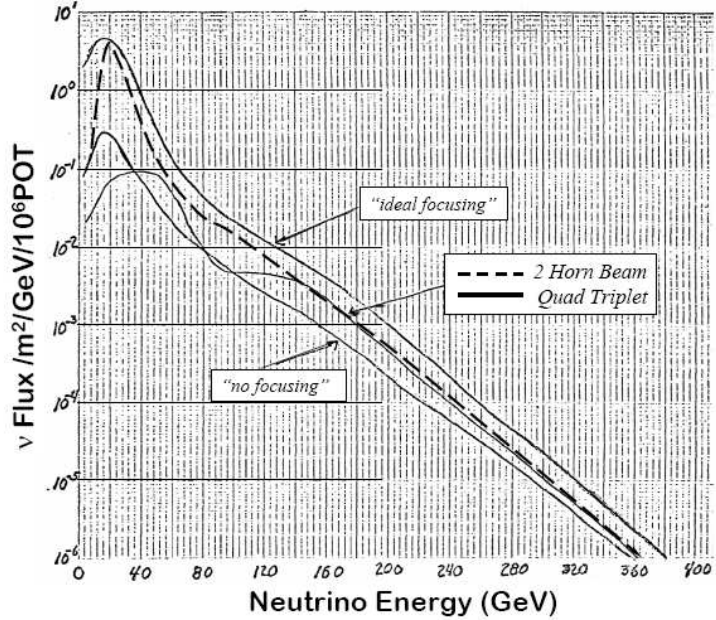}
\vskip 0.cm
  \caption{Comparison of the quadrupole triplet focused beam with the two-horn focused beam proposed for 500~GeV proton beams at FNAL.  The fluxes for ``ideal focusing'' and no focusing are also shown.  The quadrupole beam becomes equally efficient as the horn beam at high energy, effectively ``hardening'' the neutrino spectrum.  Adapted from \cite{carey1970}.  }
  \label{fig:tm472-fig4-6}
\end{figure} 

\subsubsection{Quadrupole Triplet}
While a single quadrupole magnet acts like a focusing lens in one plane and a defocusing lens in the other, pairs of quadrupoles act like a net focusing lens in both planes.  Quadrupole triplets, furthermore, help make the containment more similar in both planes \cite{humphries1986,roberts1968,carey1971}.    The aperture of a quadrupole is typically much smaller than for a horn, but for high energy neutrino beams such is not a limitation:  recalling that secondaries off the target emerge ({\it c.f.} Equation~\ref{eq:pi-divergence}) with angular spread $\theta=(0.300~\mbox{GeV}/c)/p$, a quadrupole's acceptance is well-matched to high-momentum secondaries.  Figure~\ref{fig:tm472-fig4-6} compares the neutrino flux from a horn-focused and quad triplet beam at a 500~GeV/$c$ accelerator, for example.  In principle, a quadrupole system provides an exact focus for a particular momentum $\langle p\rangle$ of the secondary beam, thus the double-peak structure in Figure~\ref{fig:tm472-fig4-6} results from the decays of $\pi$'s and $K$'s travelling in the secondary beam at the focused momentum $\langle p\rangle$.    

%
%
\begin{figure}[t]
\vskip 0.cm
  \centering
  \includegraphics[width=4.5in]{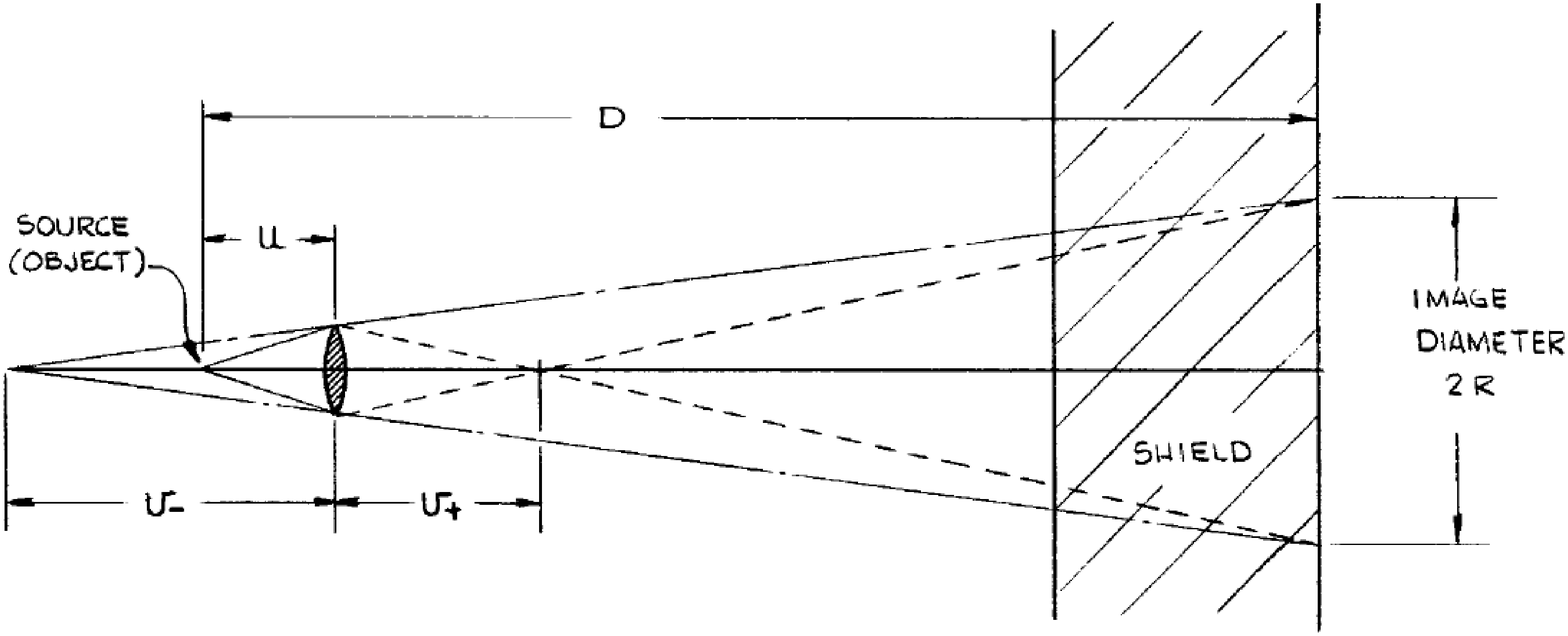}
\vskip 0.cm
  \caption{Illustration of limits of acceptance of a quadrupole lens pair for focusing neutrino parents toward a detector at a distance $D$ from the source (target).  The target momentum $\langle p\rangle$ is focused toward the detector, while the momentum limits $p_\pm$ are defined by the points at which the real $v_+$ and virtual $v_-$ images fill the detector of radius $R$.  Taken from \cite{roberts1968}.}
  \label{fig:fn124-fig1}
\end{figure} 

%
%

Despite providing an exact focus for particles at the design momentum $\langle p\rangle$, a quadrupole system is actually wide-band for detectors not too far away from the source \cite{roberts1968}.  As shown in Figure~\ref{fig:fn124-fig1}, particles over-focused or under-focused illuminate a detector of radius $R$ at a distance $D$ from the source.  The momentum limits of the quadrupole system are defined by the ``cone of confusion,'' those rays  coming from either the real or virtual image.  

\begin{table}[b]
\begin{center}
\begin{tabular}{ccc}
$\theta_{\mbox{in}}$ (mrad) & $p_+$ (GeV/$c$) & $p_-$ (GeV/$c$) \\\hline\hline
2		&  183		& 318  \\
3		&  195		& 276  \\
4		&  201		& 260  \\
5		&  205		& 252  \\
6		&  208		& 246  \\
10		&  215		& 237  \\\hline
\end{tabular}
\caption{Momenta $p_\pm$ of particles over- and under-focused by a quadrupole triplet channel.  The quads are set to provide point-to-parallel focusing for $\langle p\rangle=225$~GeV/$c$.  The momentum limits are shown for a detector of radius $R=1$~m at a distance $D=1000$~m for several possible angular apertures $\theta_{\mbox{in}}=a/u$ of the quadrupole channel (see Figure~\ref{fig:fn124-fig1}).
\label{tab:mom-bite}}
\end{center}
\end{table}

\begin{figure}[t]
\vskip 0.cm
  \centering
  \includegraphics[width=4.9in,height=2.5in]{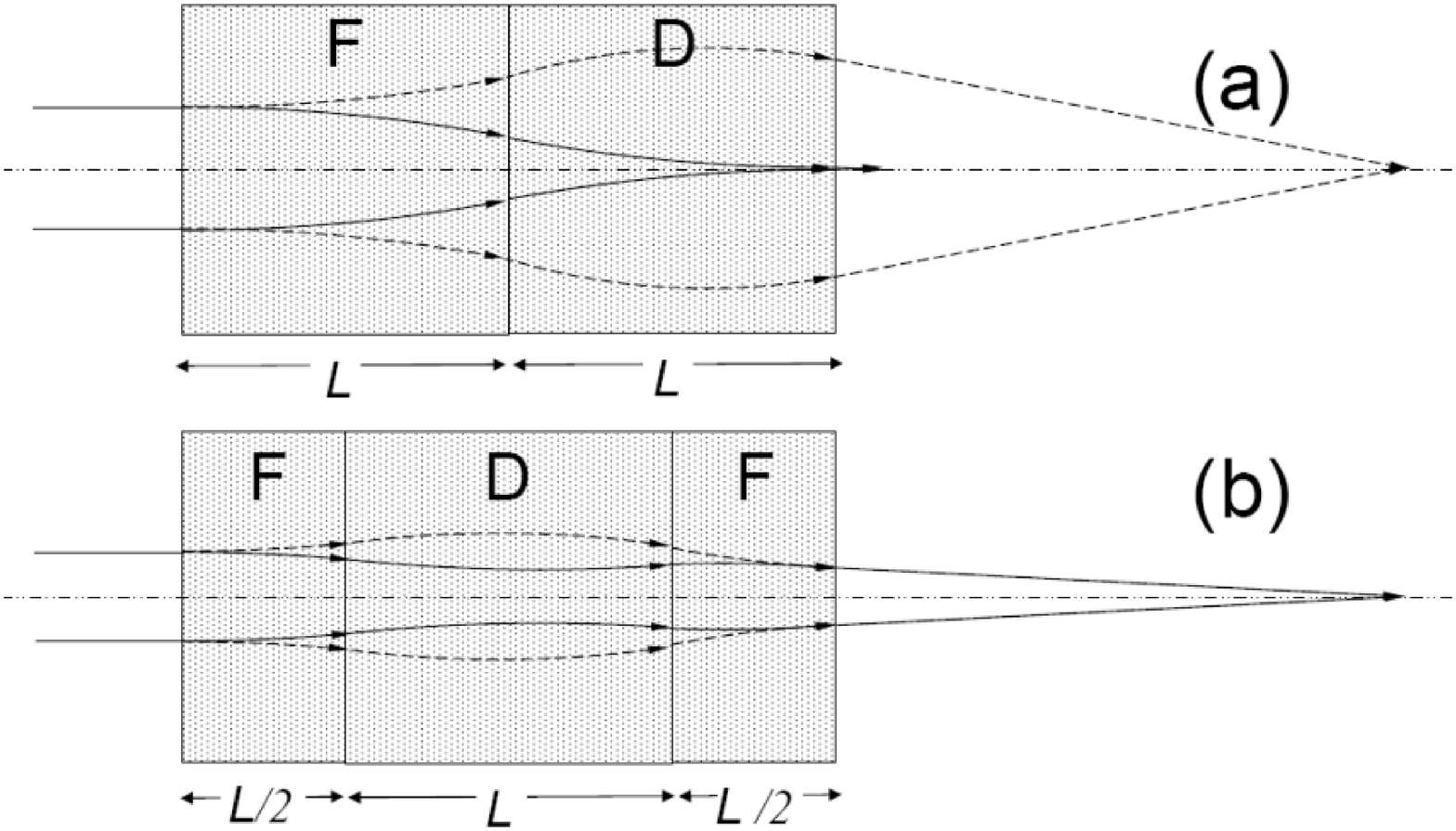}
\vskip 0.cm
  \caption{Trajectories through a (a) quadrupole doublet and (b) quadrupole triplet.  The ``F'' and ``D'' designate quads with focusing and defocusing, respectively, for the particle trajectories in the plane indicated by the solid rays.  The dashed rays indicate the particle trajectories in the opposite plane rotated 90$^\circ$ to that shown, for which the same quads are ``D'' and ``F'', respectively.}
  \label{fig:quad-focusing}
\end{figure}

The span of over- and under-focused particles by a quadrupole system is responsible for the wide-band focusing.  An optical source located a distance $u$ upstream of a lens of diameter $2a$ fills the detector with those rays emanating from the real and virtual points of focus at $v_\pm$\footnote{In contrast to geometric optics calculations, here $v_\pm>0$ is defined.}:
           \begin{equation} 
              v_\pm=\frac{(D-u)a}{R\pm a}
              \label{eq:foci}
           \end{equation}
for which the focal lengths $f_\pm$ are
\begin{equation} 
    f_\pm=1/(1/u\pm1/v_\pm) 
    \label{eq:fplusminus}
\end{equation}  
The focal length of the quad triplet shown in Figure~\ref{fig:quad-focusing} is: 
\begin{equation} 
    f^*=6/k^2L^3= 6 \frac{p^2a^2}{0.09B_0^2L^3} 
    \label{eq:foc-doublet} 
\end{equation}
where $k=0.3B_0/pa$, and $p$ is the particle momentum (in GeV/$c$), $a$ the quadrupole aperture, $B_0$ the maximum field at the pole tip (in Tesla).  Equations~\ref{eq:foc-doublet} and \ref{eq:foci} can be inserted into Equation~\ref{eq:fplusminus} to determine the limits $p_\pm$ of the focusing.  With the quads set to focus a particular momentum $\langle p\rangle$, then $u$ is defined by $u=\langle p\rangle^2a^2/6(0.09B_0^2L^3)$, and the momentum limits are given by 
\begin{equation}
   \left( \frac{p_\pm}{\langle p\rangle} \right)^2\approx \frac{\theta_{\mbox{in}}}{\theta_{\mbox{in}} \pm \theta_{\mbox{out}}}
   \label{eq:mom-bite}
\end{equation}
where $\theta_{\mbox{in}}=a/u$ is the angular aperture of the quadrupole for incoming particles and $\theta_{\mbox{out}}=R/D$ is the desired angular illumination of the beam.  As seen in Table~\ref{tab:mom-bite}, increasing the angular aperture of the triplet decreases the ``depth of focus'' (the momentum bite admitted by the quadrupoles), in analogy with geometric optics \cite{roberts1968}.  As $D\rightarrow\infty$, the momentum bite also goes to zero, so quad focusing is appropriate only for ``short baseline'' experiments.

The above discussion assumes neutrinos follow exactly the secondaries' direction.  At $\langle p\rangle=225$~GeV/$c$, the neutrino angle with respect to the pion is $\theta_\nu\sim1/\gamma_\pi=0.6$~mrad, to be compared with $\theta_{\mbox{in}}=2-10$~mrad and $R/D=1$~mrad considered in Table~\ref{tab:mom-bite}.

Quadrupole triplets are used in neutrino beams because of their near-identical containment conditions in the horizontal and vertical views of the particles' trajectories.  A doublet of two quads of length $L$ and focal length $f=(kL)^{-1}$ will not have equal focal planes in both views, as indicated in Figure~\ref{fig:quad-focusing}(a), {\it i.e.} incident parallel rays will converge to a focal plane that is different in each view.  The transfer matrix for a doublet is \cite{humphries1986}
\begin{equation}
A_{doublet} = \left(\begin{array}{cc} (1\pm kL^2) & 2L \\ -2k^2L^3/3 & (1\mp kL^2) \end{array}\right) + \mathcal{O}(\sqrt{k}L)^4
\label{eq:doublet-matrix}
\end{equation}
where the upper (lower) signs in the $\pm$ terms indicate the FD (DF) planes.  The location of the focal plane is given by $F=-a_{11}/a_{21}$, so the differing $a_{11}=(1\pm kL^2)$ terms for the FD (DF) planes create an astigmatism.  The equal focal lengths $f^*=-a_{21}=3/2k^2L^3$ in each view guarantee only equal angles exiting the doublet for incident parallel rays (or for the case of a particle source emanating from a neutrino target, we would view the drawing in reverse:  the equal focal lengths guarantee only point-to-parallel focusing for equal emission angles off the target).  For a neutrino beam, the secondary beam emerges from the quad doublet larger in the DF plane than the FD plane, which poses an aperture restriction in one view because quads are typically symmetric about the beam axis.  

The quad triplet shown in Figure~\ref{fig:quad-focusing}(b), with ``F'' cells of length $L/2$ and a ``D'' quad of length $L$ has a transfer matrix equal in both the ``FDF'' and ``DFD'' views \cite{humphries1986}:
\begin{equation}
A_{triplet} = \left(\begin{array}{cc} 1 & 2L \\ -k^2L^3/6 & 1) \end{array}\right) + \mathcal{O}(\sqrt{k}L)^4
\label{eq:triplet-matrix}
\end{equation}
The fact that the term $a_{11}=1$ in $A_{triplet}$ is responsible for the near identical focusing in both planes.  As noted in \cite{carey1971}, subsequent quadrupole cells taking on adiabatically larger apertures and smaller focusing field strengths, serves to extend the momentum range of containment.  


\begin{figure}[t]
\vskip 0.cm
  \centering
  \includegraphics[width=3.2in,height=3.3in]{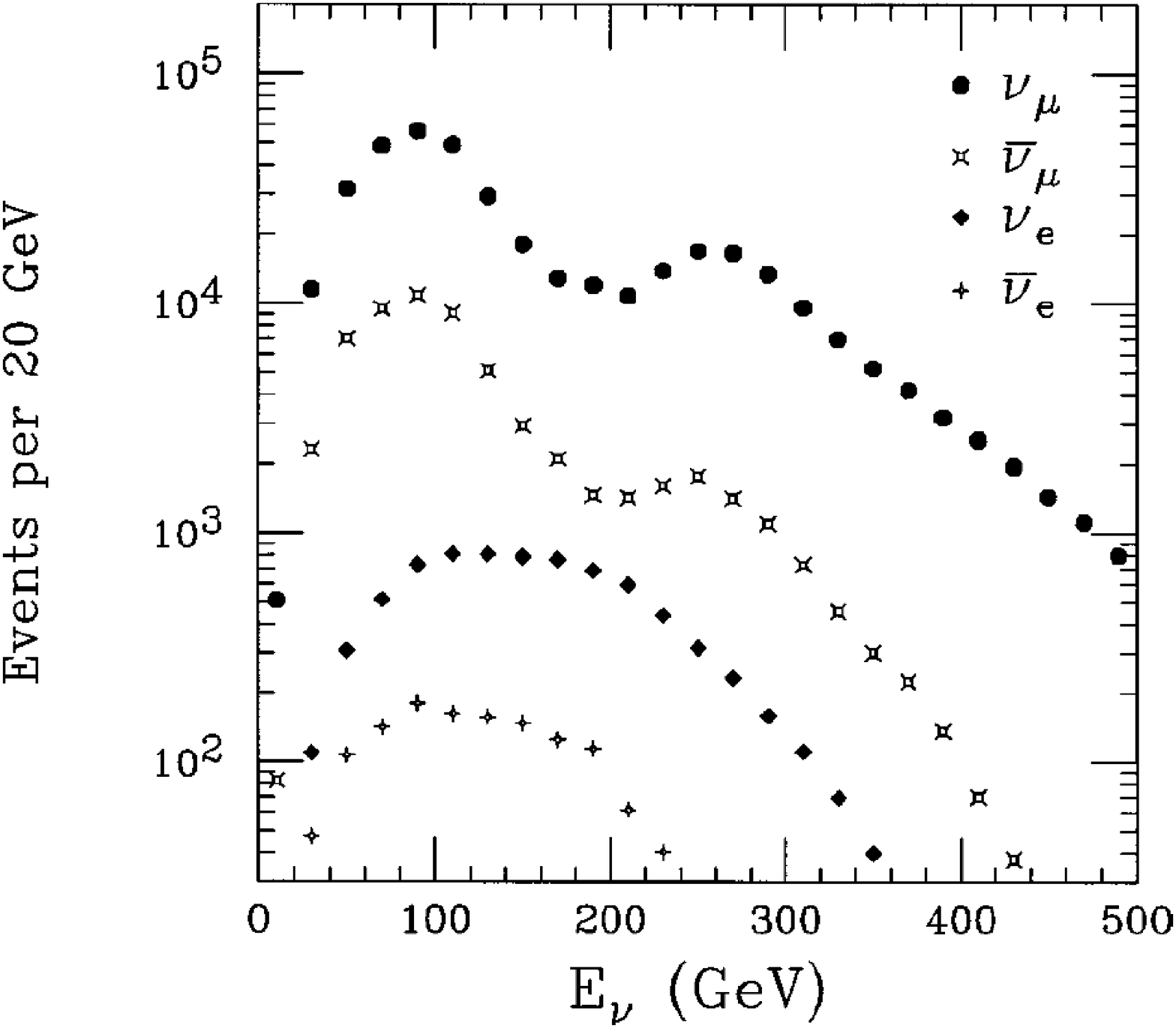}
  \includegraphics[width=3.2in,height=3.3in]{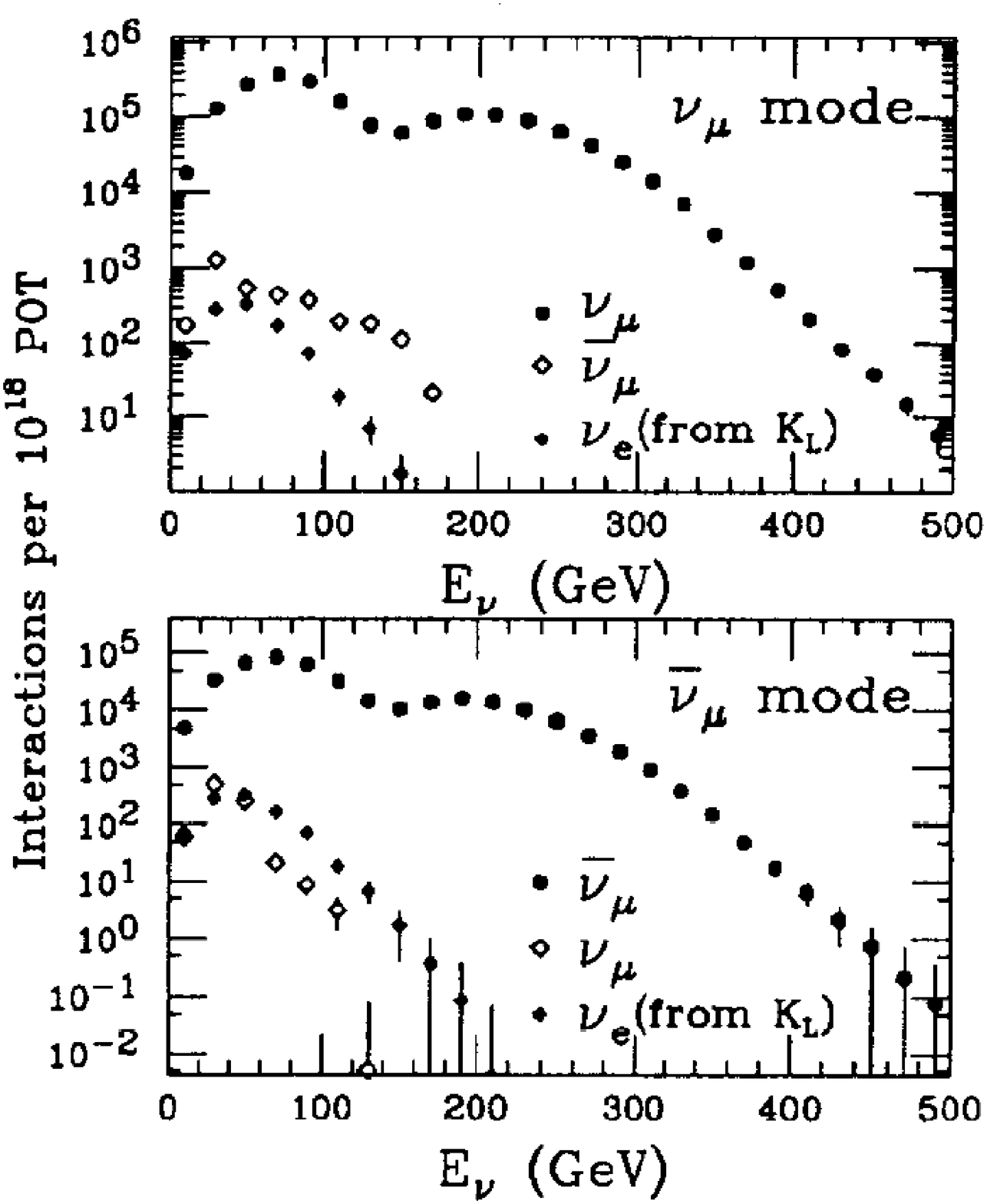}
\vskip 0.cm
  \caption{Comparison of the quadrupole triplet beam (left, taken from \cite{conrad1998}) used at the 800~GeV proton beam from the Fermilab Tevatron with the ``sign-selected quadrupole triplet'' (right, taken from \cite{bernstein1994}).}
  \label{fig:tm1884-figs}
\end{figure}

\subsubsection{Sign-selected Quadrupole Triplet}

A quadrupole system by itself focuses both signs of secondaries, thus in principle equal fluxes of $\nu_\mu$ and $\overline{\nu}_\mu$ are obtained.  In experiments in which pure $\nu_\mu$ or $\overline{\nu}_\mu$ beams are desired, sign-selection of the secondaries must be done with a dipole to sweep out the wrong sign.  The NuTeV experiment at Fermilab employed such a ``sign-selected quadrupole triplet'' \cite{bernstein1994}.  In practice, the aperture limit of the dipole, plus the lack of focusing along the dipole's length, limits the wide-band acceptance of such a system by a small amount (NuTeV tuned to 225 GeV momentum selection, FWHM about 150 GeV).  As can be seen in Figure~\ref{fig:tm1884-figs}, the sign-selection significantly reduces the wrong-sign contamination.  Wrong-sign elimination is especially important if running in $\overline{\nu}$ mode because of the lower anti-neutrino cross sections.  Another important development in the NuTeV SSQT was the ability to target the proton beam at an off-angle with respect to the neutrino line, thus reducing $\nu_e$ contamination in the beam from $K_L$ decays.  The wrong-sign and $\nu_e$ contaminations are significantly less than in a horn focused beam (for which $\overline{\nu}_\mu\sim10\%$ because of unfocused particles throught the necks and $\nu_e\sim1\%$ from muon and $K$ decays.).

\subsection{Other focusing systems}
\label{crazylens}

\subsubsection{Plasma Lens}

The BNL-Columbia group \cite{forsythe1965} proposed an alternative to the horn called the ``plasma lens.''  Based on an idea from Panofsky \cite{panofsky1950}, the idea is to place a cylindrical insulating vessel around the beam axis downstream of the target.  The vessel has electrodes pulsed at $\sim10$~kV  at its end and partial atmosphere N$_2$ or Ar gas inside.  A plasma discharge with current densities of $\sim10^5$~Amp/cm$^2$ is initiated at the outer wall and spreads throughout the tube.  The axial current thus produces a toroidal magnetic field, much like the horn\footnote{In fact, van der Meer seems to have known about Panofsky's idea of an axial current \cite{vandermeer1959}.}.  Particles of one sign only are focused.  

Some notable differences between a horn and a plasma lens:
\begin{itemize} 
\item The plasma lens in principle has no hole in its center (unlike the horn).  In practice, neutrino beams supplied by proton beams with 400-4000~kW power would probably find this infeasible.
\item One can control the radial distribution of current density in the plasma to ``tune'' the magnetic field.
\end{itemize}

Assuming a uniform current density $j$ (in Amp/cm$^2$) of radius $R$ along the beam axis, then $B_\theta=\frac{j}{5}\frac{r}{R^2}$ is present for $r\leq R$.  A particle passing through this region has motion
$$\frac{d^2r}{dz^2} + k r^2 = 0, ~~~~~k^2=\frac{60\pi j}{p}$$
where the particle momentum is in eV.  The solution to the particles motion is 
$$r=A\sin kz, ~~~~~ A=\frac{\Theta_0}{k}$$
where $\Theta_0$ is the maximum entrance angle contained by the lens.  Particles are focused parallel to the beam axis when $kz=\pi/2$, setting the desired length of the column to be $L=\pi/2k$.  The maximum radius is defined by the definition of $A$:
$$R=\frac{\Theta_0}{k}\sin kL = \frac{\Theta_0}{k} ~~~~~ \hbox{(for }kL=\pi/2)$$
so the current required to focus a beam of particles emitted into the lens at $\theta<\Theta_0$ is 
$$I=\frac{p\Theta_0^2}{60}.$$
For $p=3$~GeV/$c$ and $\Theta_0=6^\circ$, \footnote{This is a somewhat realistic example given that $\langle p_T \rangle \approx200$~MeV/$c$ for all pions.} we get $I=5\times10^5$ Amperes. 

The authors report a $\times3$ increase in neutrino flux.  The operational experience gained in the beam is not clear;  others report initial technical difficulties \cite{dusseux1972}.

\begin{figure}[t]
\vskip 0.cm
  \centering
  \includegraphics[width=4.7in]{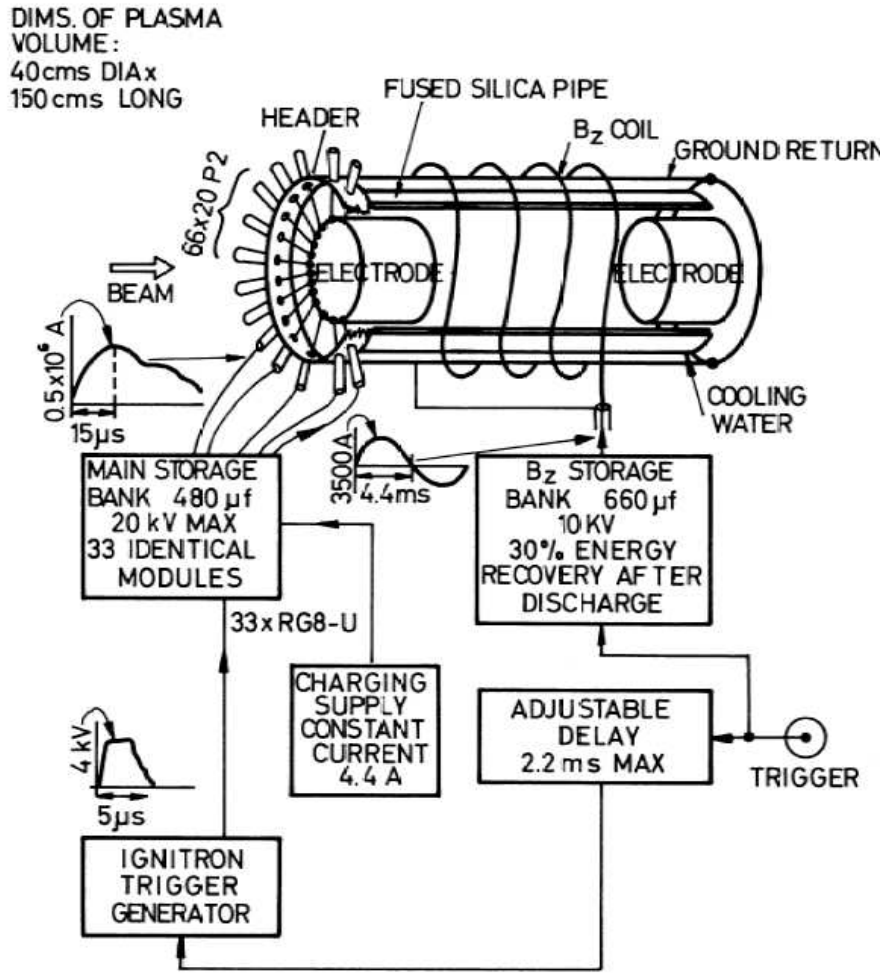}
\vskip 0.cm
  \caption{Schematic diagram of the plasma lens by the BNL-Columbia group \cite{forsythe1965}.}
  \label{fig:forsythe1965-fig1}
\end{figure} 

\subsubsection{DC-Operated Lenses}

The horns of various neutrino beams have been operated at pulse-to-pulse cycle times of 0.2~s (FNAL-MiniBooNE), 2~sec (CERN-PS, BNL-AGS, FNAL-NuMI), to 20~sec (FNAL-Tevatron), designed to operate in conjunction with the cycle time of the synchrotron source.  Pulsed devices are not practical at a linear proton accelerator like those at Los Alamos or the SNS, or a rapid-cycle machine like an FFAG, whose Megawatts of beam power could prove advantageous for neutrino production\cite{lanl1979}-\cite{lanl1982}, nor are they practical for ``slow-spill'' beam experiments.  Thus, DC-operated lenses are of advantage.   Only brief mention shall be made here.
\clearpage
\noindent\begin{large}{\it Magnetic Spokes}\end{large}

\begin{figure}[t]
\vskip 0.cm
  \centering
  \includegraphics[height=1.99in]{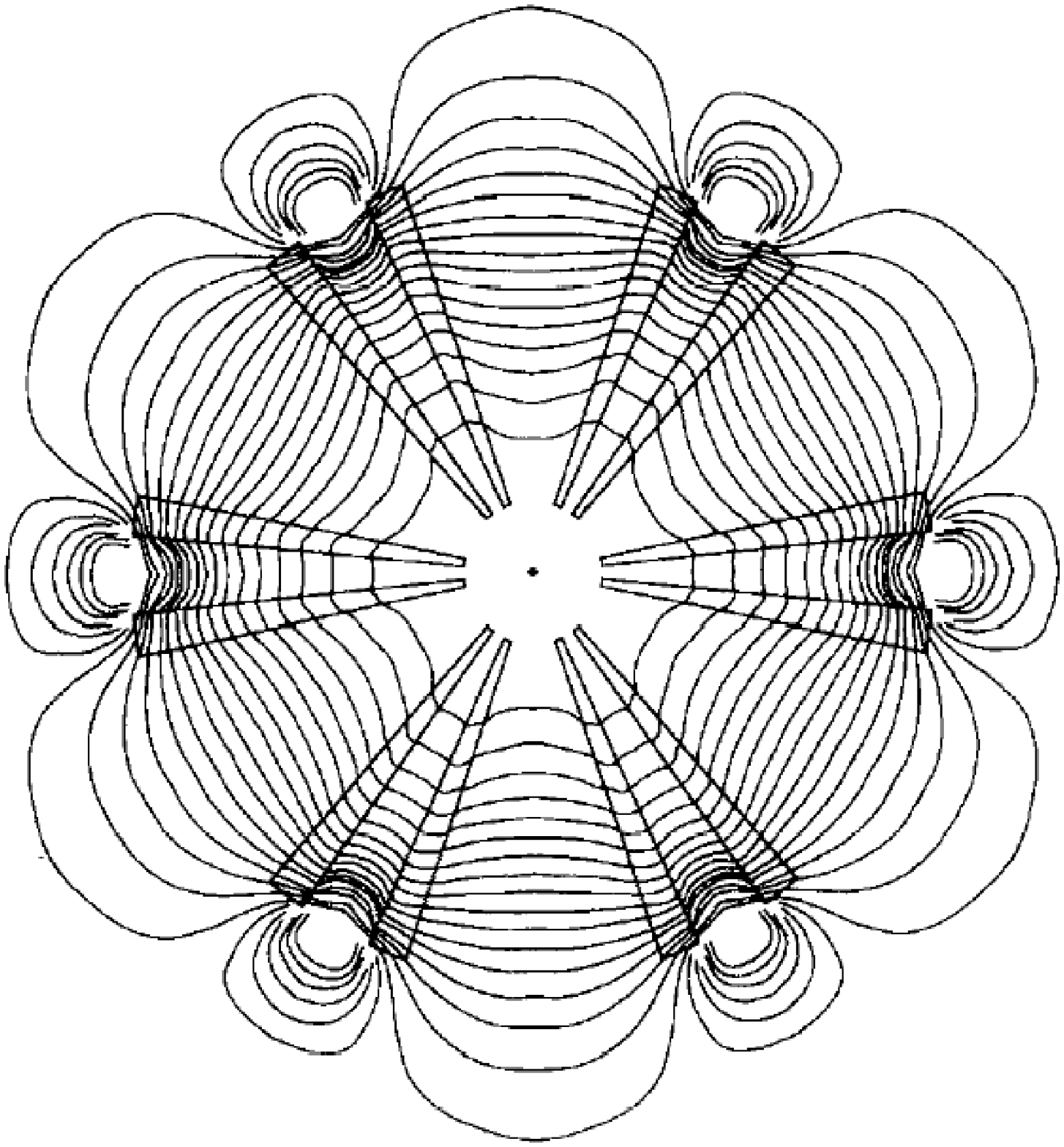}
  \includegraphics[height=1.99in]{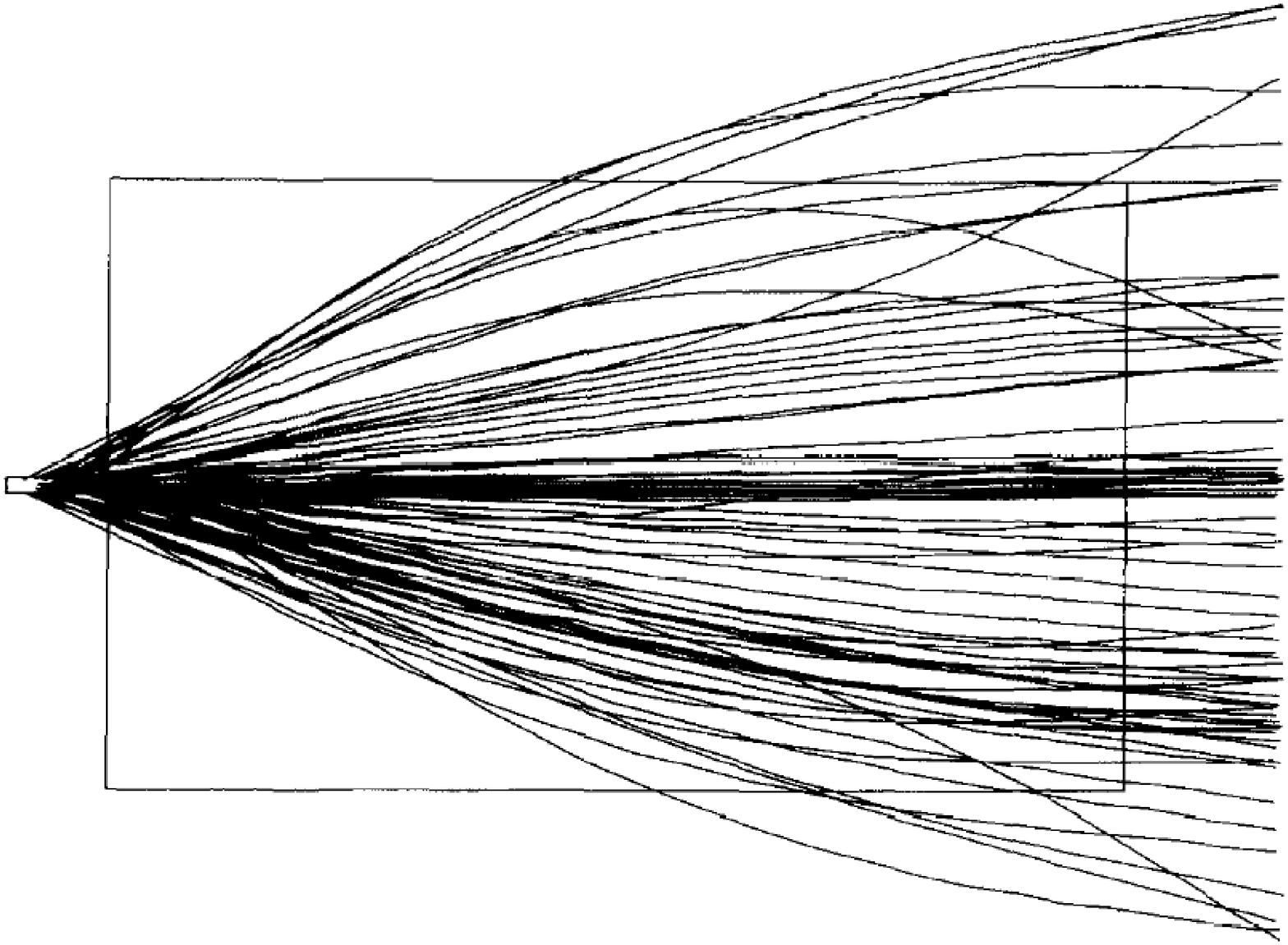}
\vskip 0.cm
  \caption{Schematic diagram of the DC-operated ``magnetic spokes'' lens of Ref. \cite{koetke1996}.}
  \label{fig:koetke1996-figs}
\end{figure}

The authors of \cite{koetke1996} note that the required $\int B\cdot d\ell$ to focus pions grows $\propto r$ as a function of the pion production angle off the target.  For a cylindrical lens, whose focusing length $\ell$ doesn't vary with $r$, this criterion requires $B\propto r$.  Given that $B\mu_0(NI)/(2\pi r)$, the authors of \cite{koetke1996} chose a current distribution $NI=(0.5)njr^2\theta,$ where the current is achieved by mounting conductors on $n$ wedged-shaped ``fins'' (see Figure~\ref{fig:koetke1996-figs}), each with opening angle $\theta$, and carrying a uniform current density $j$ down each side of the fins.  With $j$ uniform, then $NI\propto r^2$ is achieved by having the number of conductors increase as $r^2$.\footnote{The authors mistakenly state that the kick from a magnetic horn varies as $\propto 1/r$.  While it is true that the horn $B\propto1/r$, the pathlength of the particle through a parabolic horn grows as $r^2$, giving $\int B\cdot d\ell \propto r$, as required.}

To reduce pion absorbtion, the fins number only 8, each 8$^\circ$.  The return winding is achieved by the cables returning at the outer radial edge of the fins.  The authors calculations show a net increase over a bare target beam of a factor of $4$ with a 2.5~m long magnet carrying 20~A.  Results of the calculated fields and several pion trajectories in this field are shown in Figure~\ref{fig:koetke1996-figs}.

\vskip .5 cm
\noindent\begin{large}{\it Solenoid Lens}\end{large}

As has been noted by many authors ({\it e.g.} \cite{batalov1986}), a solenoid with axis of symmetry along the proton beam and target direction has the effect of transforming radial components of momemntum into azimuthal (angular) momentum.  So, while it prevents the secondary beam from becoming larger, it does not by itself focus the secondaries toward a detector.  The focusing comes from producing a gradient in the solenoid field.  Ref. \cite{diwan1999} shows results of a tapered solenoid which produces a field $B(z)=B_0/(1+az)$, for example.  As emphasized in \cite{mcdonald2003}, the gradient provides the focusing through conservation of canonical momentum $(\frac{d{\bf P}}{dt})_{\phi}=\frac{d}{dt}[r(P_{\phi}+\frac{e}{c}A_{\phi})]=0$.  An advantage of this lens is that it is further from the direct path of the beam, while a disadvantage is that it focuses both signs of secondaries.  The solenoid focuses certain pion (hence neutrino) momenta, which can be an advantage over a broad-band beam \cite{mcdonald2003}.

\clearpage
\newpage
\section{Focusing of Narrow-Band Beams}
\label{nbb}

In many experiments it is desirable to produce fewer neutrinos with more carefully-selected properties:  for example, wide-band horn beams have large ``wrong-sign'' content ($\overline{\nu}_\mu$'s in a $\nu_\mu$ beam).  Or, it might be desirable to select neutrinos of a given energy for study of energy-dependence of cross sections or neutrino oscillation phenomena at a particular neutrino energy.

\subsection{Dichromatic Beam}

\begin{figure}[t]
\vskip 0.cm
  \centering
  \includegraphics[width=6.in]{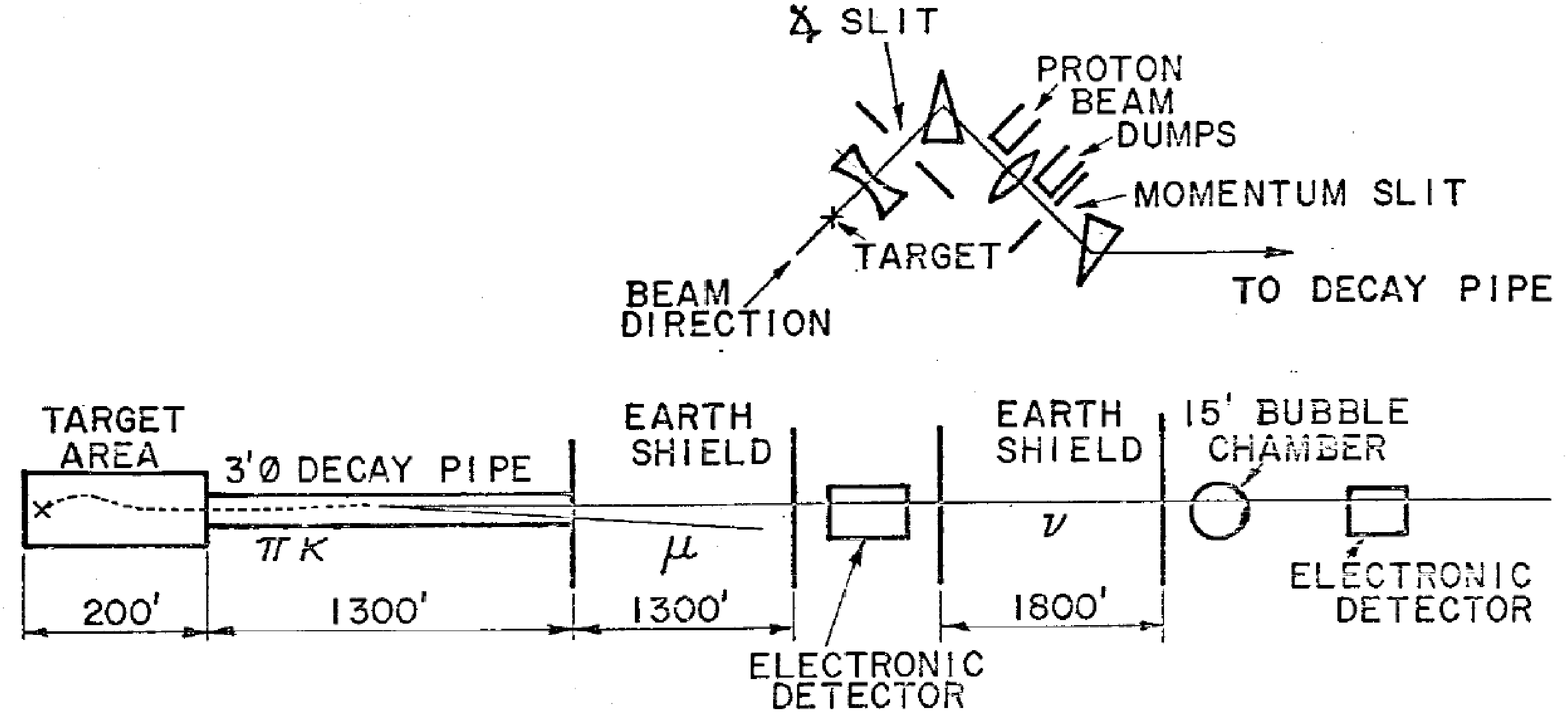}
\vskip 0.cm
  \caption{Plan view of the first di-chromatic neutrino beam at Fermilab.  Dipoles momentum- and sign-select the secondaries, while quadrupoles provide point-to-parallel focusing as they head into the decay tunnel. Taken from \cite{barish1981}.}
  \label{fig:limon1974-fig1}
\end{figure}

Fermilab was the first to pioneer the so-called ``di-chromatic neutrino beam'' \cite{limon1974}, and the high event rates possible yielded rapid physics results \cite{barish1973,barish1975,benvenuti1975}.  Such a beam, shown schematically in Figure~\ref{fig:limon1974-fig1}, uses dipole magnets downstream of the target to sweep out quickly wrong-sign secondaries from the neutrino channel.  In the first di-chromatic beam, two quadrupole magnets were used to provide point-to-parallel focusing of those secondaries of the momentum selected by the dipoles.  The monochromatic secondary beam of pions and kaons is sent into the decay tunnel, where they decay.  Following the construction of the SPS at CERN, a similar dichromatic beam was built there \cite{heijne1983}, with physics results coming from the CDHS\cite{berge1987}, and CHARM \cite{allaby1988} detectors.

\begin{figure}[t]
\vskip 0.cm
  \centering
  \includegraphics[height=4.3in]{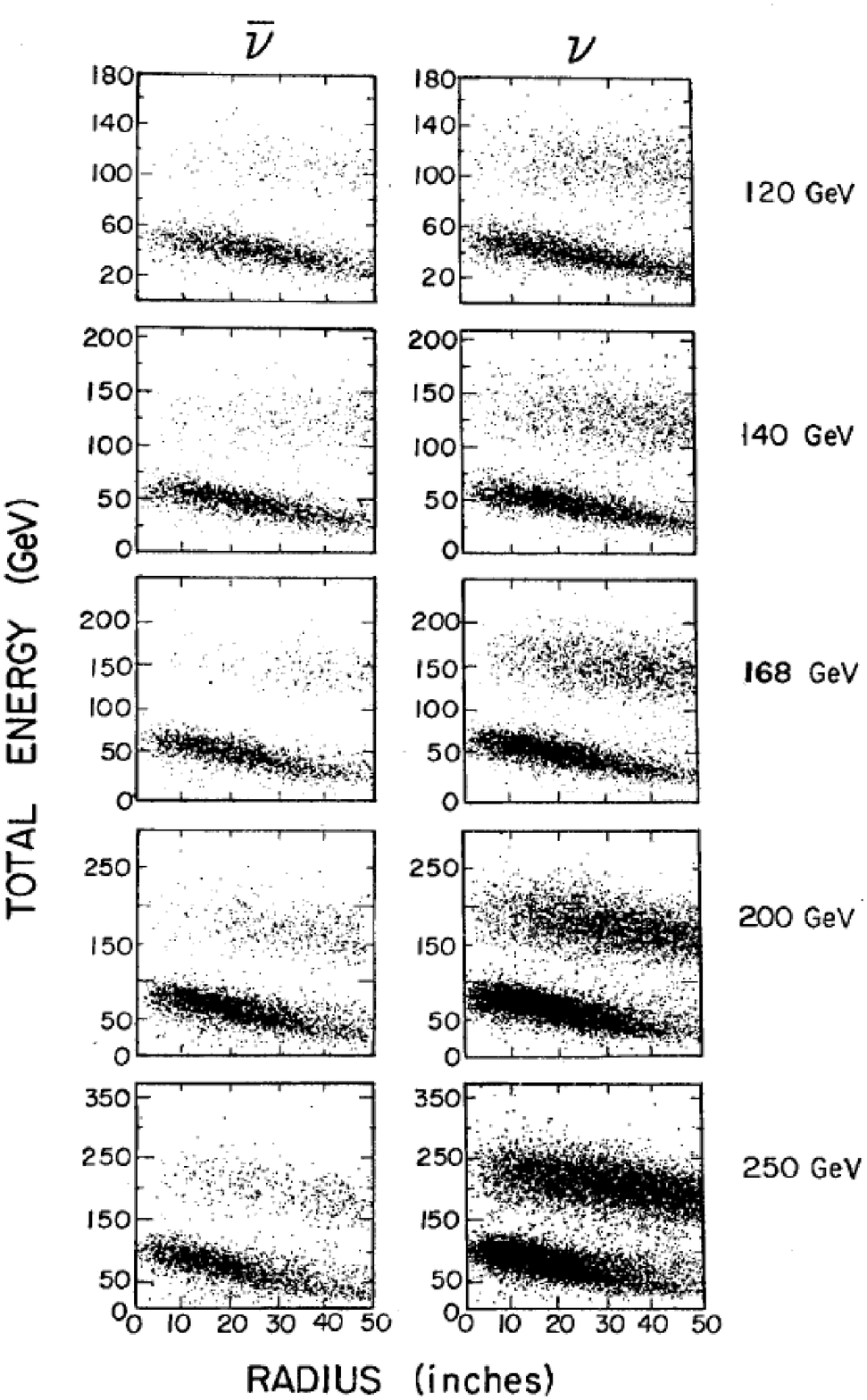}
  \includegraphics[height=4.2in]{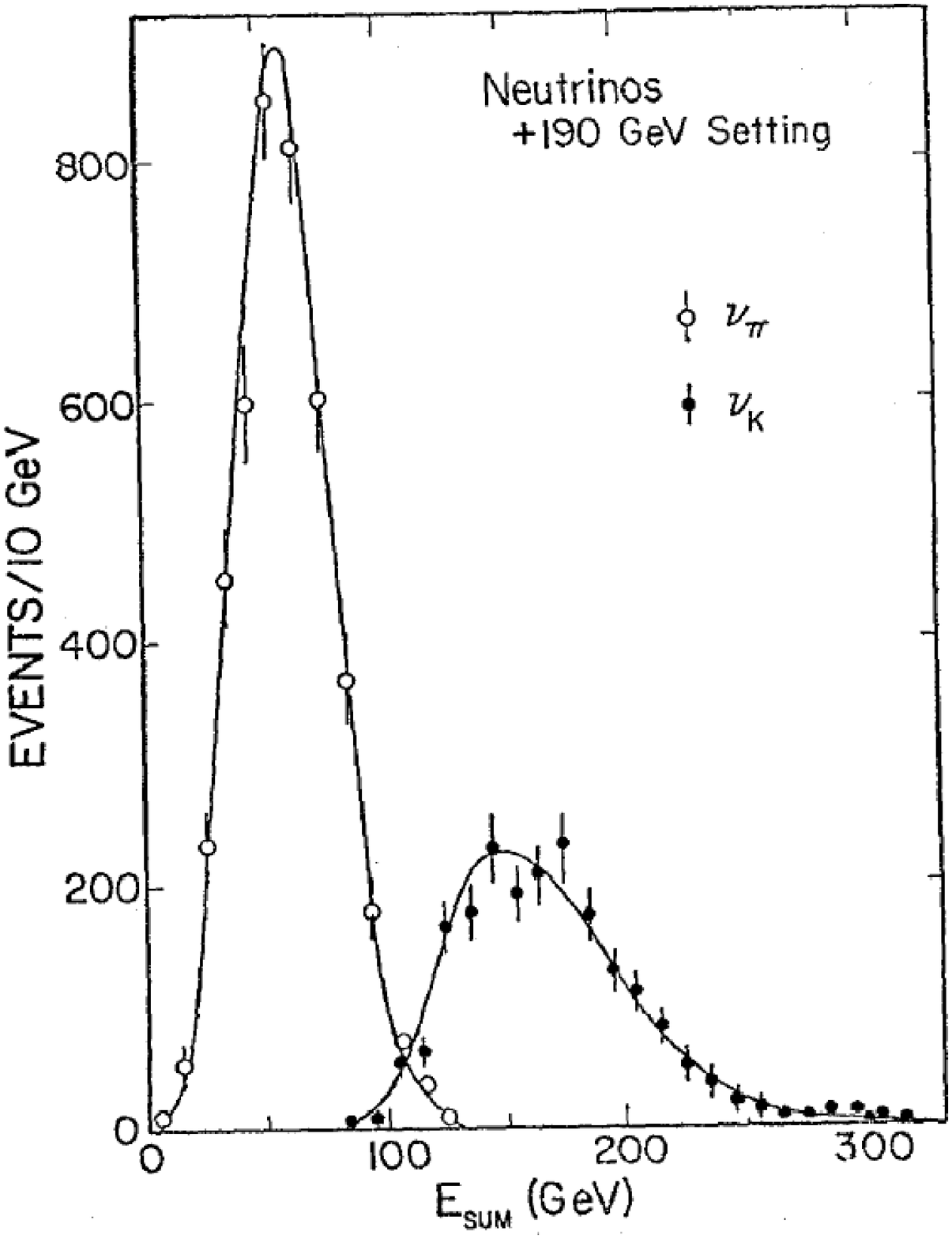}
\vskip 0.cm
  \caption{(left) Graph of $E_\nu\approx p_\mu+E_{\mbox{shwr}}$ in the CITFR calorimeter as a function of transverse radius of the neutrino interaction with repect to the beam axis \cite{barish1981}. (right) Graph of the visible energy in the CITFR for the $+190$~GeV/$c$ beam setting, with the $\pi$ and $K$ radial bands shown separately \cite{barish1978}.}
  \label{fig:barish1981-fig15}
\end{figure}

The term ``di-chromatic'' comes from the two distinct neutrino energies produced in such a decay channel.  The decay of a pion or kaon secondary results in a neutrino of energy
\begin{equation}
E_{\nu}=\frac{(1-(m_\mu/m_{(\pi,K)})^2)E_{(\pi,K)}}{(1+\gamma^2\theta^2)},
\label{eq:enu-vs-epi}
\end{equation}
where $\theta$ is the angle between the neutrino and meson direction, and $\gamma=E_{(\pi,K)}/m_{(\pi,K)}$.  The momentum of the secondary beam is fixed, but the presence of both pions and kaons lead to two possible values for the neutrino energy.  The possibility for off-angle decays of the $(\pi,K)$ beam can change $E_\nu$.  Figure~\ref{fig:barish1981-fig15} shows this kinematical relationship in the Caltech-Fermilab neutrino detector located 1300~feet from the end of the decay pipe:  neutrino interactions reconstructed in their detector at large transverse distances ({\it i.e.}: large $(\pi,K)$ decay angles) from the beam central axis show a smaller total energy deposition in the detector, though two distinct bands are observed, arising from pion and kaon decays.

The channel downstream of the target starts producing neutrinos as soon as secondaries decay.  Decays before the momentum- and sign-selection are achieved result in a ``wide-band background'' under the two energy peaks in Figure~\ref{fig:barish1981-fig15}.  For this reason, the proton beam is brought onto the target at an angle off the axis of the decay tunnel, resulting in such ``wide-band'' secondaries decaying preferentially away from the neutrino beam's axis.  Further, momentum-defining collimators are placed along the neutrino channel to better eliminate off-momentum secondaries from the beam.  In fact, these considerations, plus the upgraded capabilities of running the Fermilab Main Ring at 400~GeV/$c$ primary momentum, led to an upgrade of this dichromatic beam \cite{edwards1976a,edwards1976b} with larger primary targeting angle to reduce the wide-band backgrounds and better momentum selection to reduce wrong-sign contamination.


\subsection{Horn Beam with Plug}

\begin{figure}[t]
\vskip -1.cm
  \centering
  \includegraphics[width=6.3in]{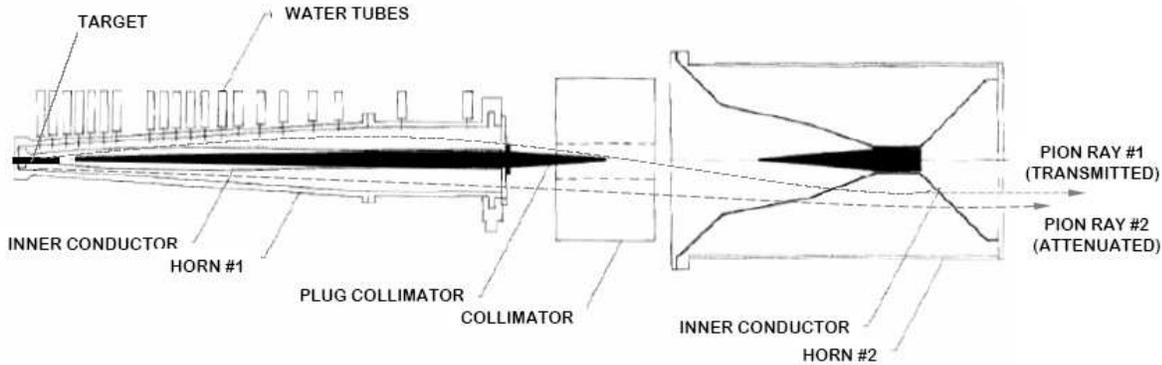}
\vskip -0cm
  \caption{Two-horn system for the BNL narrow-band beam, with collimators and beam plugs to stop unwanted pion trajectories.  Pion ray \#1 is transmitted, while pion ray \#2 is stopped by the collimator between the two horns. Likewise, rays passing through the two beam plugs would be expected to be attenuated. Figure adapted from \cite{sims1987}.}
  \label{fig:sims1987-fig1}
\end{figure} 

The wide-band horn-focused beam, referring to Figures~\ref{fig:5-beams} and \ref{fig:two-horn-focusing}, produces a span of neutrino energies corresponding to a variety of particle trajectories through the focusing system. To cut off the largest range of neutrino energies, it is desirable to eliminate those particles which travel through the field-free ``necks'' of the horns.  Such was attempted at CERN \cite{pattison1969,bloess1971} by placing a Tungsten block (beam ``plug'') at the end of the usual target to help attenuate those high energy pions which tend to leave the target at small angles ($\theta_\pi\approx2/\gamma_\pi$).

The collimation for a narrow-band beam was refined in a series of experiments at BNL \cite{carroll1985,carroll1987}, in which two beam plugs and a collimator located in between the horns were used to attenuate all but the desired trajectories, as shown in Figure~\ref{fig:sims1987-fig1}.  Referring to Figure~\ref{fig:two-horn-focusing}, further eliminating those particles which do not cross the beam center line between the two horns has the effect of cutting all but the smallest momenta, as is achieved with the collimator between the two horns in Figure~\ref{fig:sims1987-fig1}.  A similar proposal was made at Fermilab \cite{nezrick1975}.


\subsection{Horn Beam with Dipole}
\begin{figure}[t]
\vskip 0.cm
  \centering
  \includegraphics[width=4.3in]{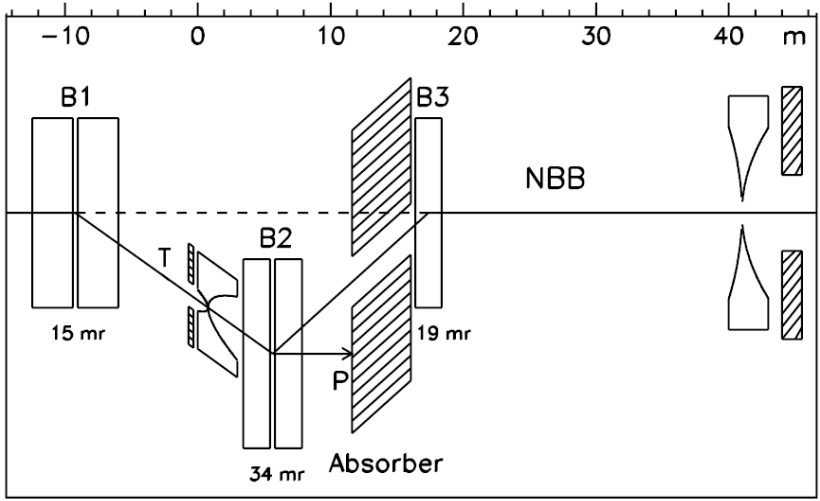}
\vskip 0.cm
  \caption{NBB achieved by a momentum-selecting dipole placed in between the two horns of a WBB.  Taken from \cite{abramov2002}.}
  \label{fig:abramov2002-fig2}
\end{figure} 

As noted in \cite{abramov2002}, a dipole magnet placed in between the two horns of a wide-band beam has the effect of achieving better momentum and sign selection.  As shown in Figure~\ref{fig:abramov2002-fig2}, a dump for the primary beam must in this case be placed in the target hall, just like in the dichromatic beam, which is somewhat of a challenge for high-intensity neutrino beams.  In practice, the aperture restriction of the dipole does attenuate some of the pion flux.  

\subsection{Off-Axis Neutrino Beam}
\label{off-axis}

The idea for an off-axis neutrino beam was first proposed by BNL experiment E889 \cite{beavis1995}.  Many of the kinematic features of off-axis pion decay were worked out in Ref. \cite{ramm1963}.  The estimates of the on-axis WBB flux in Section~\ref{wbb} made implicit use of the fact that the energies of neutrinos emitted along the axis of travel of the secondary pion or kaon is linearly related to the meson energy.  The problem of achieving a particular energy NBB thus reduces to focusing a particular energy meson beam.  

\begin{figure}[t]
\vskip -0.cm
  \centering
  \includegraphics[height=2.4in]{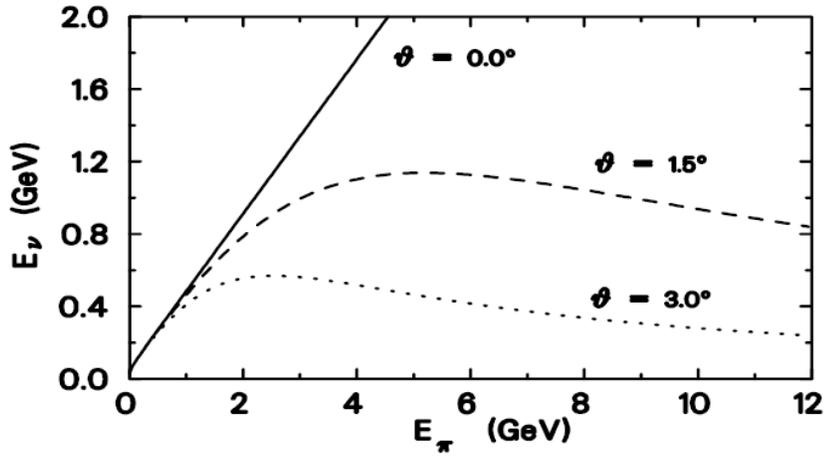}
\vskip -0cm
  \caption{Neutrino energy from pion decay as a function of pion energy, for several choices of decay angle between the neutrino and pion direction.  Taken from \cite{beavis1995}.}
  \label{fig:beavis1995-fig4}
\end{figure}

In the limit that mesons are focused and travel parallel to the decay pipe axis, the BNL E889 team noted that under some circumstances nearly all mesons of any energy could contribute to generating the same energy of neutrino.  While Equation~\ref{eq:enu-vs-epi} states that the neutrino and meson energy are in fact linearly related for on-axis decays ($\theta=0^\circ$), the relationship is more complex for neutrinos observed to emerge at some angle with respect to the beam due to the denominator.  Equation~\ref{eq:enu-vs-epi} is graphed for several particular decay angles in Figure~\ref{fig:beavis1995-fig4}.

Figure~\ref{fig:beavis1995-fig4} has an interesting interpretation:  for on-axis decays, the neutrino energy is related to the meson energy.  For off-axis decays, this relationship is weaker.  Thus, for large off-axis angles, nearly any pion energy makes about the same energy of neutrino.  A broad-band pion beam, therefore, can be used to generate a narrow-band neutrino spectrum.  

\newpage
\clearpage

\begin{figure}[b]
\vskip -0.cm
  \centering
  \includegraphics[height=4.5in]{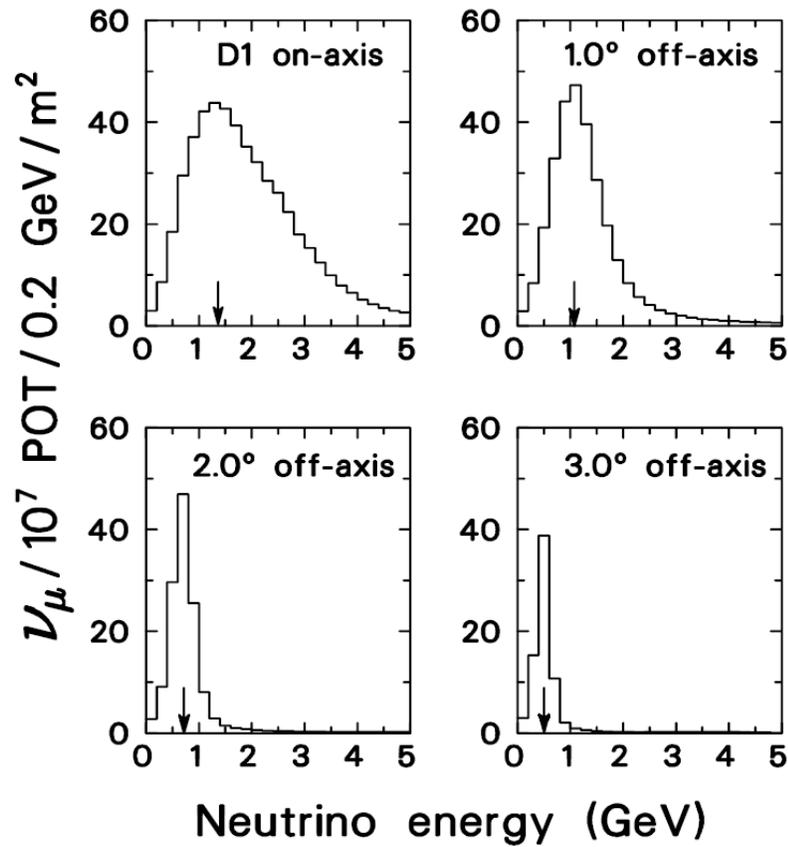}
\vskip -0cm
  \caption{Neutrino energy spectra at a distance of 1~km from the proposed BNL beam for several off-axis angles.  Taken from \cite{beavis1995}.}
  \label{fig:beavis1995-fig2}
\end{figure} 

\newpage
\clearpage

\begin{figure}[t]
\vskip -0.cm
  \centering
  \includegraphics[height=3.2in]{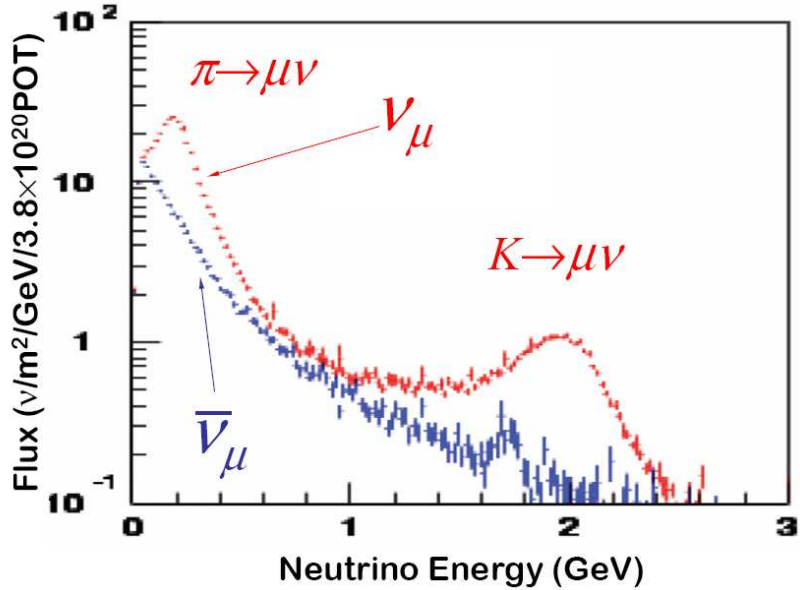}
\vskip -0cm
  \caption{Calculated flux from the NuMI beam at $z\approx750$~m and 110~mrad off-axis angle, corresponding to the location of the MiniBooNE detector.  Two peaks, from $\pi$ and $K$ decays, can be seen.  Taken from \cite{aguilar2005}.}
  \label{fig:numi-miniboone}
\end{figure} 

The BNL team proposed such a NBB spectrum for a search for $\nu_\mu\rightarrow\nu_e$ oscillations, since NC $\nu_\mu$ interactions of any energy can leave small energy depositions in a detector which mimic $\nu_e$ interactions.  Thus, cutting down all $\nu_\mu$ energies which contribute to NC background is of value.  They proposed placing a detector a couple of degrees off the beam axis for their new beam line, thereby choosing the particular NBB energy to be achieved.  

Figure~\ref{fig:beavis1995-fig2} shows, for the beam configuration and detector distance in the BNL proposal, the neutrino energy spectrum from pion decays at several off-axis locations. In addition to the lower, narrower, energy spectrum at larger off-axis angles, it may be noted that, at certain energies, the flux at the peak actually exceeds the flux at that same energy in the on-axis case.  Thus, the fact that all pions contribute to approximately the same neutrino energy can, in part, compensate for the loss of flux at off-axis angles, from Equation~\ref{eq:flux-angle}.

The proposal, not approved, has since been adopted by teams at JPARC \cite{itow2001} and Fermilab \cite{nova2002}, which will employ the narrow-band off-axis beam to search for $\nu_\mu\rightarrow\nu_e$ oscillations.  The first detection of neutrinos from an off-axis beam is at Fermilab, where the MiniBooNE detector is situated 110~mrad off-axis of the NuMI beam line at a distance of $\sim750$~m from the NuMI target.  Neutrinos from NuMI have been observed in MiniBooNE \cite{aguilar2005}.  The off-axis angle is sufficiently large that both peaks from $\pi$ and $K$ decays can be seen (see Figure~\ref{fig:numi-miniboone}), permitting use of the MiniBooNE detector to derive the $\pi/K$ ratio of the NuMI beam.\footnote{The ability to resolve separate pion and kaon peaks at large off-axis angles, as well as the low systematic uncertainties in predicting the flux of an off-axis beam were studied in \cite{para2001}.}

\clearpage
\newpage
\section{Decay Volumes}
\label{decayvol}

\subsection{Decay Tube}

\begin{figure}[t]
\vskip 0.cm
  \centering
  \includegraphics[width=4.5in]{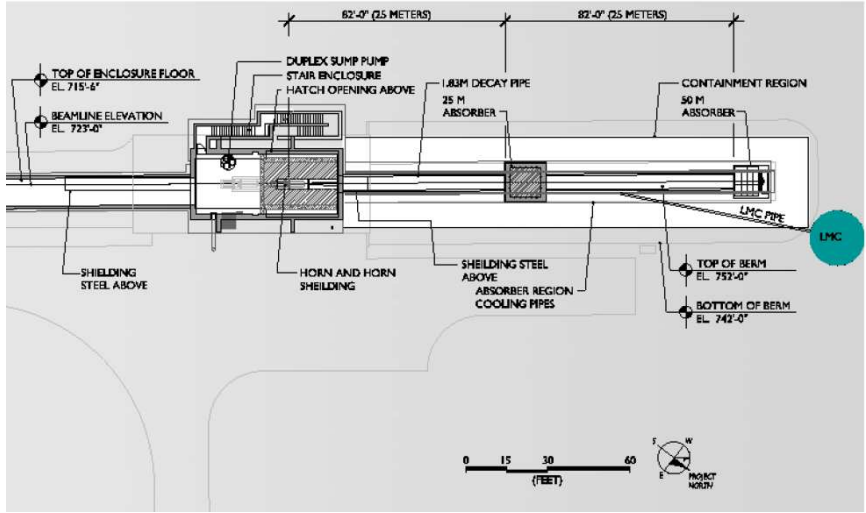}
\vskip 0.cm
  \caption{Schematic diagram of the MiniBooNE beam line at FNAL, taken from \cite{brice2004}.}
  \label{fig:brice2004-dump}
\end{figure}

Decay volumes are drift spaces to permit the pions to decay.  For a 5~GeV pion, $\gamma\approx35$, and $\gamma\beta c\tau_\pi\approx280$~m.  This sets the scale for how long the decay tube should be if just 63\% of the pions for a 2~GeV neutrino beam are to be allowed to decay.  As noted by \cite{vandermeer1963}, the decay pipe radius is also of importance, and has to be as wide as practical for efficient low neutrino energy beams: in general low-energy pions are not as well focused in a horn focused beam, and have a divergence which will send them into the decay volume walls before decaying.

Decay tubes are often evacuated.  The same 280~m mean flight path, in air, represents 0.9 radiation lengths ($X_0=304$~m for air at STP), and 0.26 nuclear interaction lengths ($\lambda_{int}^{air}=1080$~m).  Thus, a pion drifting in air at atmospheric pressure would have a $\approx26$\% chance of being absorbed by a collision, and those that are not lost will suffer multiple Coulomb scattering of a typical magnitude of 2.8~mrad.  Such scattering angles are already significant compared to the $\sim1/\gamma$ opening angle between the muon and neutrino in pion decay, which is 14~mrad for a 10~GeV/$c$ pion decaying to a 4~GeV neutrino.  Other decay tubes, such as KEK \cite{ahn2006}, are filled with He gas to reduce absorption and scattering.

\begin{figure}[t]
\vskip 0.cm
  \centering
  \includegraphics[width=5in]{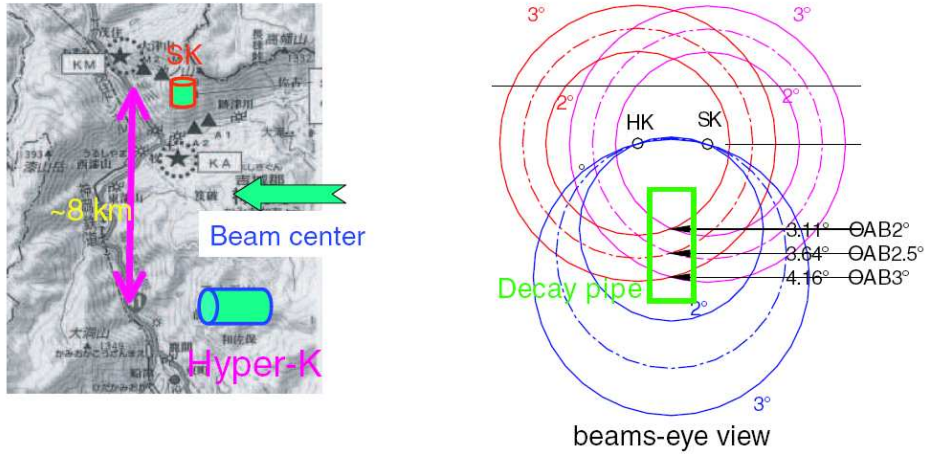}
\vskip 0.cm
  \caption{Schematic of the decay volume for the T2K beamline, taken from \cite{kearns2005}.}
  \label{fig:kearns2005-fig62}
\end{figure} 

Because scattering can do much to defocus the secondary beam already focused by the horns, particular care is given to the entrance windows to decay volumes.  The needs of mechanical strength for the large evacuated chamber must be balanced against placing significant scattering material in the beam.  NuMI's 2~m diameter decay pipe has a composite window with a 3~mm thick, 1~m diamter Aluminum center and thicker steel annulus at larger radius to reduce pion scattering and heating from the unreacted proton beam.  The CNGS beam has a thin Ti window \cite{elsener2005}.

T2K \cite{kearns2005} has a flared decay volume which enlarges at its downstream end, as shown in Figure~\ref{fig:kearns2005-fig3}.  This beamline is envisioned to support experiments at two remote locations, one at Super-Kamiokande and also a future ``Hyper-Kamiokande'' site.  It is envisaged to be an off-axis beam (see Section~\ref{off-axis}) of about 2-3$^\circ$ to both these sites.  The flared beam pipe permits tuning of the off-axis angle (hence $\langle E_\nu \rangle$) as the experiments require.

\subsection{Hadron Hose}
\label{hose}

\begin{figure}[t]
\vskip 0.cm
  \centering
  \includegraphics[width=5.5in]{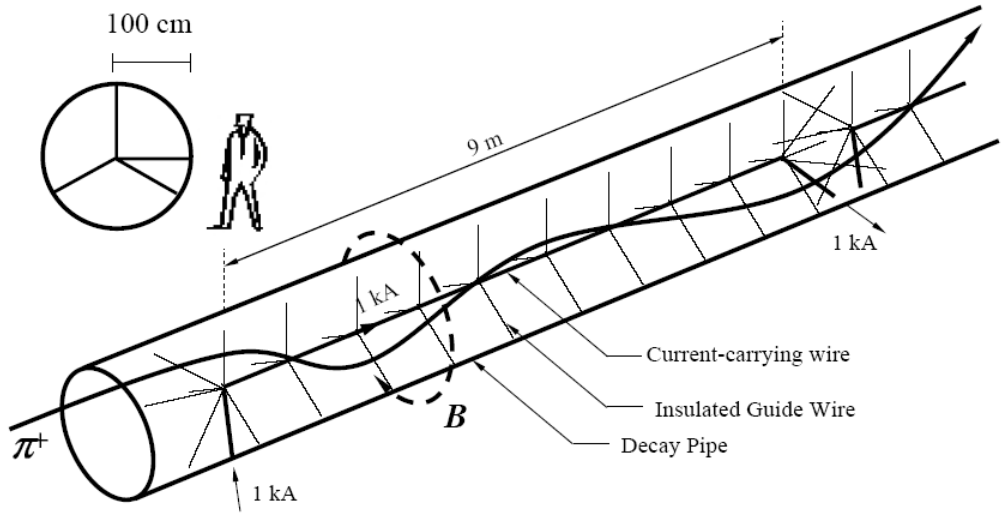}
\vskip 0.cm
  \caption{Schematic diagram of the hadron hose focusing device proposed by \cite{hylen2003}, based on the ``beam guide'' of van der Meer \cite{vandermeer1962}.  Secondary particles entering the decay volume spiral around the current-carrying wire until they decay to neutrinos.}
  \label{fig:hose-figure}
\end{figure}

Fermilab proposed building a focusing device along the length of the decay pipe which would enhance the neutrino flux and reduce systematic uncertainties in predicting the energy spectrum of neutrinos \cite{hylen2003}.  Based on the ``beam guide'' idea originally proposed by van der Meer \cite{vandermeer1962}, the device consists of a single or multiple wires travelling axially down the length of the decay volume which are pulsed with $\sim1$~kA of current, providing a weak toroidal field, but long focusing length (the full particle trajectory before decay).  As indicated in Figure~\ref{fig:hose-figure}, such focusing draws particles diverging toward the decay volume walls back toward the beam center, where they can decay without absorption on the walls.

The hadron hose can increase the neutrino event rate to experiments by 30-50\% because pions heading toward the decay pipe walls are drawn back toward the beam centerline.  Improved probability for pion decay can also be achieved simply by constructing a larger diameter decay volume, but such is quite expensive due to the extensive shielding which must surround the decay volume in high-power neutrino beams.  Thus, the hadron hose may be viewed as an active decay volume, a low-cost alternative to the large-diameter passive decay volume.

The hadron hose provides a second benefit which is less obvious:  the spiral orbits essentially randomizes the decay angles of the pions leading to neutrinos in the detector.  This is beneficial for two-detector neutrino experiments, because the ``near'' and ``far'' detectors often observe slightly different energy spectra just due to the solid angle difference between the two detectors.  Recalling that the neutrino energy is $E_\nu=(0.43E_\pi)/(1+\gamma^2\theta^2)$, high energy pions which decay just in front of the near detector can result in neutrinos hitting the near detector for a wide span of angles $\theta$, lowering the neutrino energy as compared to the neutrinos reaching the far detector at $\theta\approx0$.  The randomization of decay angles, caused by the spiraling orbits in the hose field, is discussed further in Section~\ref{two-det}.

The focusing might naively be expected to converge all particles into the wire, causing large absorptive losses of pions: pions emerge from the target in the radial direction, and the radial restoring force causes many pion trajectories to cross the wire.  However, multiple Coulomb scattering of the pions and kaons in the upstream horns and entrance window to the decay volume leads to some azimuthal component of pion momentum, causing the pions to enter the decay volume and execute spiral orbits around the hose wire \cite{milburn2000}, as indicated schematically in Figure~\ref{fig:hose-figure}.  Analytic expressions for particle orbits in the hadron hose field have been computed \cite{regenstreif1964,milburn2000}.

Placing a high-current wire in the evacuated decay volume poses some technical challenges, as discussed in \cite{hylen2003}.  Namely, the wire's heat induced by $I^2R$ as well as energy deposition from beam particles must be dissipated sufficiently by blackbody radiation, the wire's voltage must be shown not to break down in the heavily ionized residual gas of the decay volume, and the long-term tension applied to the wire segments must be ensured not to cause plastic flow (``creep'') of the wire material such that a failure occurs.

\subsection{Muon Filter}
\label{muonfilter}

The muon filter is the part of the beam line required to range out muons upstream of the neutrino detector.  Keeping in mind that $\frac{1}{\rho}\frac{dE}{dx}\sim2$~MeV/(g/cm$^2$), and recalling for steel (often used in shielding) that $\rho\sim8$~g/cm$^2$, $\frac{dE}{dx}\sim1.6$~GeV/m for those nuisance muons.  The first neutrino experiment in fact had to lower the AGS accelerator energy to 15~GeV to reduce the maximum muon energy and thereby reduce the muon ``punch-through'' \cite{danby1962}.  The origninal neutrino line at Fermilab, which had an earthen ``berm'' sufficient to stop muons up to 200~GeV/$c$, had to be reinforced with 20~m of lead and 140~m of steel shielding following upgrades of the accelerator complex to run at 800-900~GeV proton energy \cite{carey1995}.

\begin{figure}[t]
\vskip 0.cm
  \centering
  \includegraphics[width=5.9in]{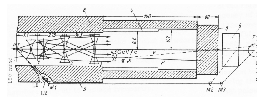}
\vskip 0.cm
  \caption{Plan view of the IHEP neutrino beam.  Taken from \cite{baratov1977a}.  In addition to the four-lens neutrino channel, a small side channel at $87^\circ$ could be activated for delivery of secondary mesons from the target to the neutrino experiments for calibration purposes.}
  \label{fig:baratov1977d-fig1}
\end{figure} 

The location of the upstream face of the muon filter defines the maximum pion or kaon drift time before decay to muon and neutrino.  It is expensive to construct a decay tube that allows most focused pions to decay.  For example, the CERN PS neutrino beam, with 80~m decay volume, would allow 25\% of pions and 90\% of kaons to decay, assuming that 5~GeV particles are being focused.  In the case of NuMI, with 725~m of drift space and $\sim$10~GeV/$c$ focusing, these numbers are 73\% and 100\%, respectively.  The length of the decay volume also impacts the level of $\nu_e$ content in the beam, since much of it arises from $\pi\rightarrow\mu\nu_\mu\rightarrow(e\nu_\mu\nu_e)\nu_\mu$ decays.

\begin{figure}[t]
\vskip 0.cm
  \centering
  \includegraphics[width=4.8in]{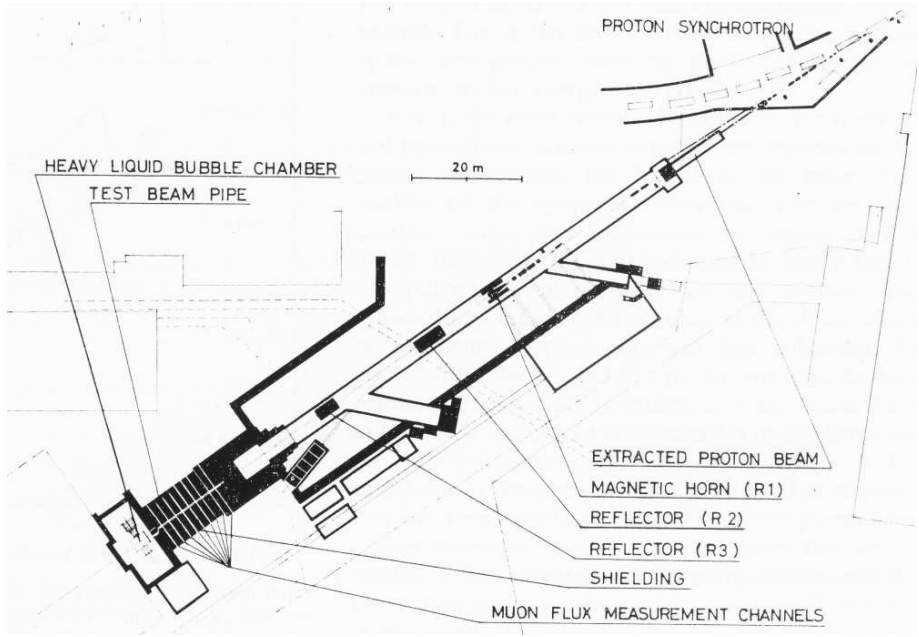}
\vskip 0.cm
  \caption{Plan view of the CERN PS neutrino beam from, taken from \cite{bloess1971}.  A small mercury-filled tube in the muon shield could be drained temporarily, exposing the downstream neutrino detectors to muons from the beam for calibration.}
  \label{fig:bloess1971-fig1}
\end{figure} 

The idea of a moving beam dump (see Figure~\ref{fig:brice2004-dump}) will be employed by MiniBooNE to demonstrate their $\nu_e$'s come from oscillated $\nu_\mu$'s from pion decay, not the ``instrinsic'' $\nu_e$ from the beam of $K\rightarrow\pi e\nu_e$ decays or $\mu$ decays.  The moving beam dump was first employed by \cite{danby1962} to show their neutrino candidates were from meson decay and not from interactions in the shielding.  Figure~\ref{fig:danby1962-fig1} shows a lead block that was placed close to the target to stop presumably all but a few $K$ decays, and indeed the neutrino rate decreased in proportion to expectations.  CERN's original WANF beam also apparently had a moveable mid-stream beam stop, but this was never employed \cite{heijne1983}.

Low-intensity beam lines combined the proton beam dump with the muon filter \cite{bloess1971,baratov1977f}.  However, for modern beam lines, a dedicated proton dump is required because of the intense beam power.  The NuMI beam, for example, is designed for a 400~kW proton beam, of which 70~kW heads for the beam dump, requiring water cooling, a special Aluminum core, etc.  Accident conditions are even more problematic:  the beam stop must allow for errant proton beam missing the target and striking the dump directly.  This is an even greater concern for upgrades to NuMI, CNGS, and JPARC, the latter two of which will use a graphite core.

A common problem in muon shielding is leakage, not attenuation \cite{plass1965b}.  Staggered assembly with no ``line of sight'' cracks is crucial to good shielding design.  This uniformity impacts the ability of downstream muon instrumentation to make meaningful measurements of muon intensity and lateral profile.  Such measurements, which can provide information on the neutrino flux and even the energy spectrum, are distorted by cracks which let lower-energy muons through, as has been observed at NuMI.

There have been a couple clever trickes to temporarily ``let down'' the muon shield of an experiment for the purposes of calibrating the neutrino detectors with particles (muons, pions) of known momentum.  The Serpukov beam could calibrate its spark chamber and bubble chamber experiments \cite{baratov1977a} using a small channel in their shielding at an angle $87^\circ$ to the primary beam axis.  Shown in Figure~\ref{fig:baratov1977d-fig1}, this channel permits secondaries from the target to be focused in a quadrupole-dipole system and be delivered directly to the experiments.  Such a calibration test beam was also utilized by the NuTeV experiment at Fermilab \cite{harris2000}.  Another trick employed at the CERN PS neutrino beam in 1967 was to install a mecury-filled tube which penetrated the entire 20~m muon shield \cite{pattison1969}, shown in Figure~\ref{fig:bloess1971-fig1}.  The mercury from this tube could temporarily be drained, exposing the heavy-liquid bubble chamber (HLBC) from Ecole Polytechnique to muons at the end of the decay volume.

\newpage
\clearpage

\section{Flux Monitoring}
\label{fluxmonitor}

\subsection{Primary Beam Monitoring}

The monitoring of the primary proton beam, as far as it impacts the physics of a neutrino experiment, is limited to requiring knowledge of the proton beam just upstream of the target.  Specifically, paramters such as the total intensity of the beam striking the target (both integrated over the lifetime of the experiment and on a per-pulse basis, since many experiments suffer rate-dependent effects), the position, angle, divergence and spot size of the beam as it is about to strike the target.  

The proton flux delivered to the neutrino target can be measured in a variety of ways.  Fast-extracted beams can use current toroids, and NuMI has recently demonstrated calibration of such a device to $\sim(1-2)\%$ over the first year of operation using precision test currents.  In the past, many experiments would often take their normalization from foil activation techniques, which measured the residual activity of gold \cite{barish1977}, A$\ell$ \cite{pattison1969}, or polyethylene \cite{burns1965b} foils placed in the proton beam upstream of the target.  Such techniques are typically precise to $(5-10)$\%, due to imprecise knowledge of production cross-sections for these radionuclides.  One motivation for using such foil techniques was to better match what the hadron production experiments did for proton normalization \cite{barish1977}, but this is becoming less imporatant as experiments are relying on more than one hadron production experiment.  

The proton beam profile has in various lines been measured by segmented ionization chambers \cite{baratov1977f,ahn2006,carroll1987,beavis1995}, Aluminum SEMs \cite{ferioli1997}, W wire SEMs \cite{tassotto2000}, ZnS screens\cite{pattison1969,barish1977,camas1995}.  Many of these techniques no longer work in high-power neutrino lines:  the large proton fluences motivate the need to reduce beam scattering and loss along transport line, as these cause irradiation and damage to transport line magnets.  Further, the proton beam's power can significantly degrade the performance of an interceptive device in the beam.  At NuMI, the profile is measured at the target with a segmented foil Secondary Emission Monitor (SEM) \cite{kopp2006a}.

\subsection{Secondary Beam Monitors}

Instrumentation placed directly in the secondary $(\pi,K)$ beam of a wide-band beam is relatively rare, since it must cope with quite high rates and can substantially affect the neutrino flux.  A few notable examples exist.  CERN proposed placing a spectrometer and Cherenkov counter system downstream of their horns to measure $\pi/K$ fluxes after the horn focusing.\cite{plass1965}.  Such was a ``destructive measurement,'' from the point of view of neutrino running, but would have yielded an {\it in situ} analysis of hadron production and focusing.  In Figure~\ref{fig:bloess1971-fig1}, this spectrometer is indicated by the Cherenkov counter just below (beam left) of the muon filter tilting at an angle which points back to a thin specrometer magnet (curved, in front of the horn R2).  A test of this system was conducted \cite{plass1967}, though backgrounds from $\delta$-rays in the beam and produced in the spectrometer appear to have been difficult.

\begin{figure}[t]
\vskip 0.cm
  \centering
  \includegraphics[width=4in]{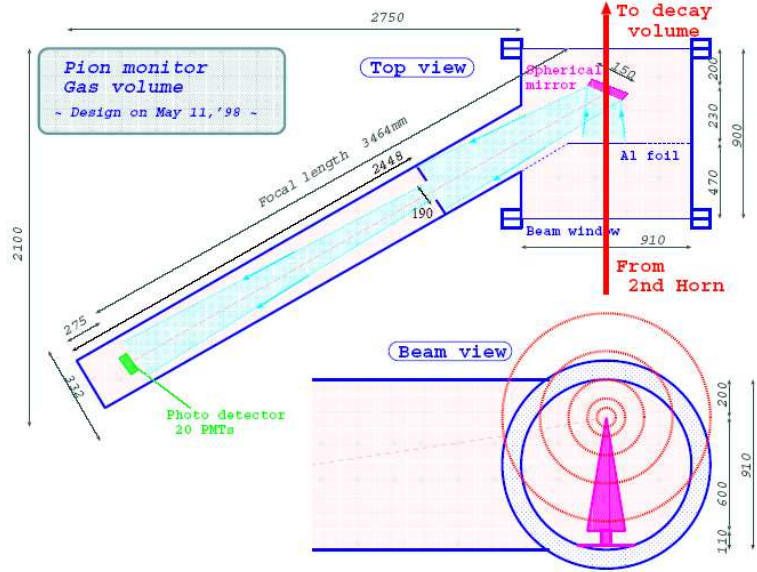}
\vskip 0.cm
  \caption{The K2K pion monitor.  Taken from \cite{ahn2006}.}
  \label{fig:ahn2006-fig12}
\end{figure} 

KEK placed a Cherenkov counter, shown in Figure~\ref{fig:ahn2006-fig12}, in their secondary beam for two brief periods during their run \cite{ahn2006,maruyama2000}.  This system placed a wedge-shaped sherical mirror at 30$^\circ$ to the beamline to direct Cherenkov light out to a PMT array several meters away from the beam axis.  Assuming that all particles in the beam are pions, the Cherenkov ring sizes provide the pions' momenta while their location on the PMT array provide the pions' direction off the beam axis.  Substantial ($\sim30\%$) substractions were made for electromagnetic particles in the beam.  With the $(p_\pi,\theta_\pi)$ information, a modified fit to the Sanford-Wang parameterization \cite{sanfordwang1967} is possible.  To avoid detections of protons in the Cherenkov counter, it could measure the neutrino spectrum above 1~GeV (pions above 2.3~GeV/$c$), which is approximately the location of the maximal flux (see Figure~\ref{fig:ahn2006-fig6}).

BNL\cite{chi1989} and the Fermilab NuMI beam \cite{kopp2006b} placed segmented ion chamber arrays directly in the secondary beam as beam quality monitors.  The NuMI chambers must contend with $\sim2\times10^9$~particles/cm$^2$/spill and are exposed to $\sim2$~GRad/yr dose, necessitating moving away from circuit board technology as in \cite{chi1989} to all-ceramic/metal design \cite{kopp2006b}.  Because of the large fluxes of photons, electrons, positrons, and neutrons in the secondary beam, neither chamber was used in a flux measurement.  The BNL chambers were placed midway down the decay volume, while the NuMI chambers were located right upstream of the beam absorber.  In the case of the NuMI beam, the flux at the hadron monitor is dominated by unreacted protons passing through the target, so the device serves as a useful monitor of the proton beam targeting, as well as a check of the integrity of the target.  The CERN WANF beam \cite{astier2003,heijne1983} had split-foil SEMs downstream of the target but upstream of the horns to ensure beam was on target, and the CNGS beam will do likewise \cite{elsener2005}.

\begin{figure}[t]
\vskip 0.cm
  \centering
  \includegraphics[width=6in]{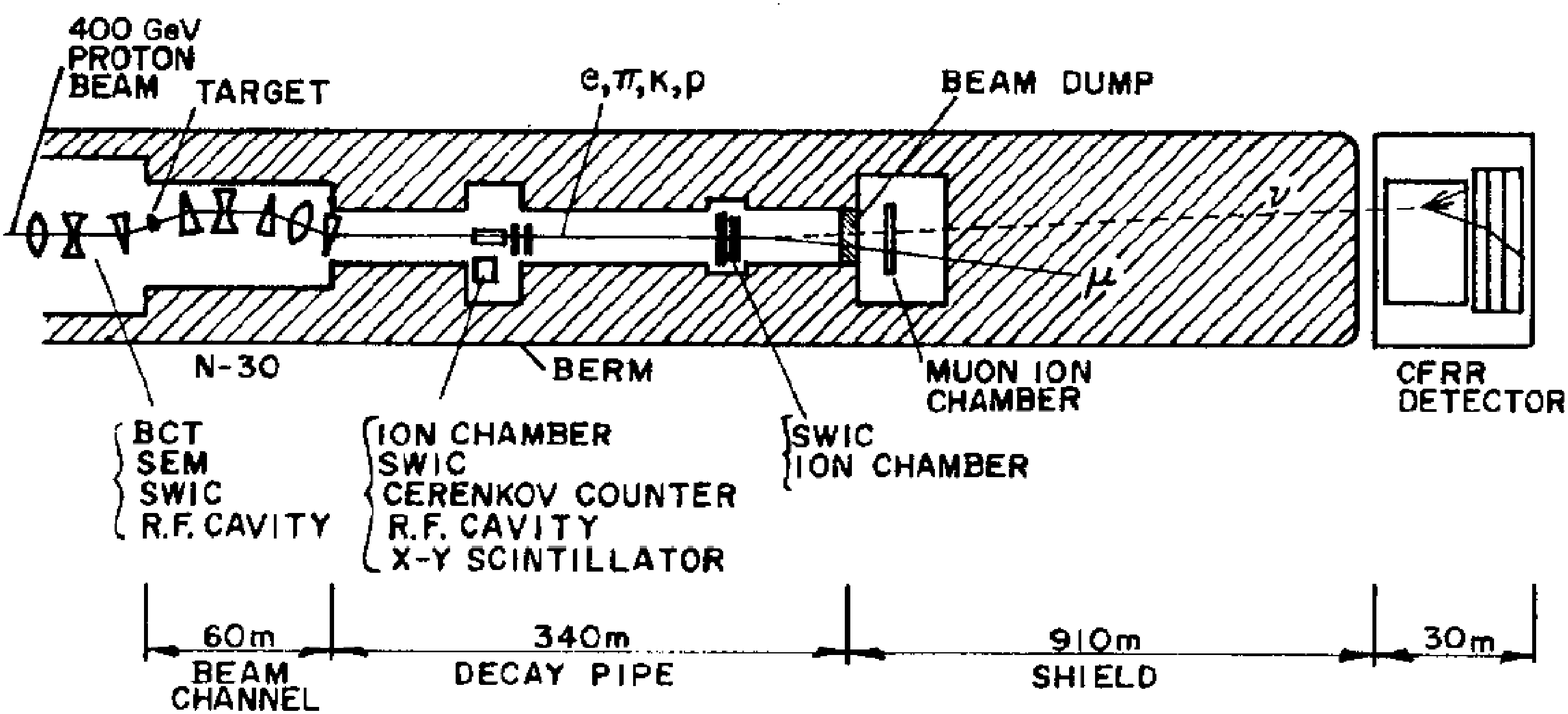}
\vskip 0.cm
  \caption{The FNAL dichromatic neutrino beam, with secondary beam instrumentation indicated.  An ion-chamber array and an RF cavity measure absolute particle flux in the secondary beam, and a Cherenkov counter measures relative abundances of $e,\pi,K,p$ in the secondary beam.  Taken from \cite{barish1981}.}
  \label{fig:barish1981-fig1}
\end{figure} 

The dichromatic beam at Fermilab had an elaborate secondary beam system which was crucial for making flux measurements and which enabled absolute neutrino cross sections to be measured.  The narrow, momentum-selected secondary beam permits reasonably small-diameter instruments to be inserted or removed from the secondary beam.  These detectors included two ion chambers which measured total particle flux, an RF cavity which was used to corroborate the ion chamber measurement, and a Cherenkov counter which could be scanned in pressure to measure the relative abundance of $e, \pi, K, p$ in the beam (subsequently normalized to the total flux determined by the ion chambers).  With this system in place, it was not, in principle, important to know the number of protons delivered to the target in order to estimate the neutrino flux.

The ion chambers were carefully calibrated.  Linearity with particle flux was demonstrated by comparison to the proton fluence on target measured by a beam current toroid.  Stability in time was shown by comparions to the two ion chambers' relative signals.  Studies were done to show that material upstream of the ion chambers contributed negligible signal in the form of $\delta$-rays (the ion chambers were placed well-downstream of any shielding), and the signal response was carefully studied as a function of relative particle abundances, since heavy protons cause a rise in signal due to nuclear interactions in the ion chamber materials.  

\begin{figure}[t]
\vskip 0.cm
  \centering
  \includegraphics[height=3.3in]{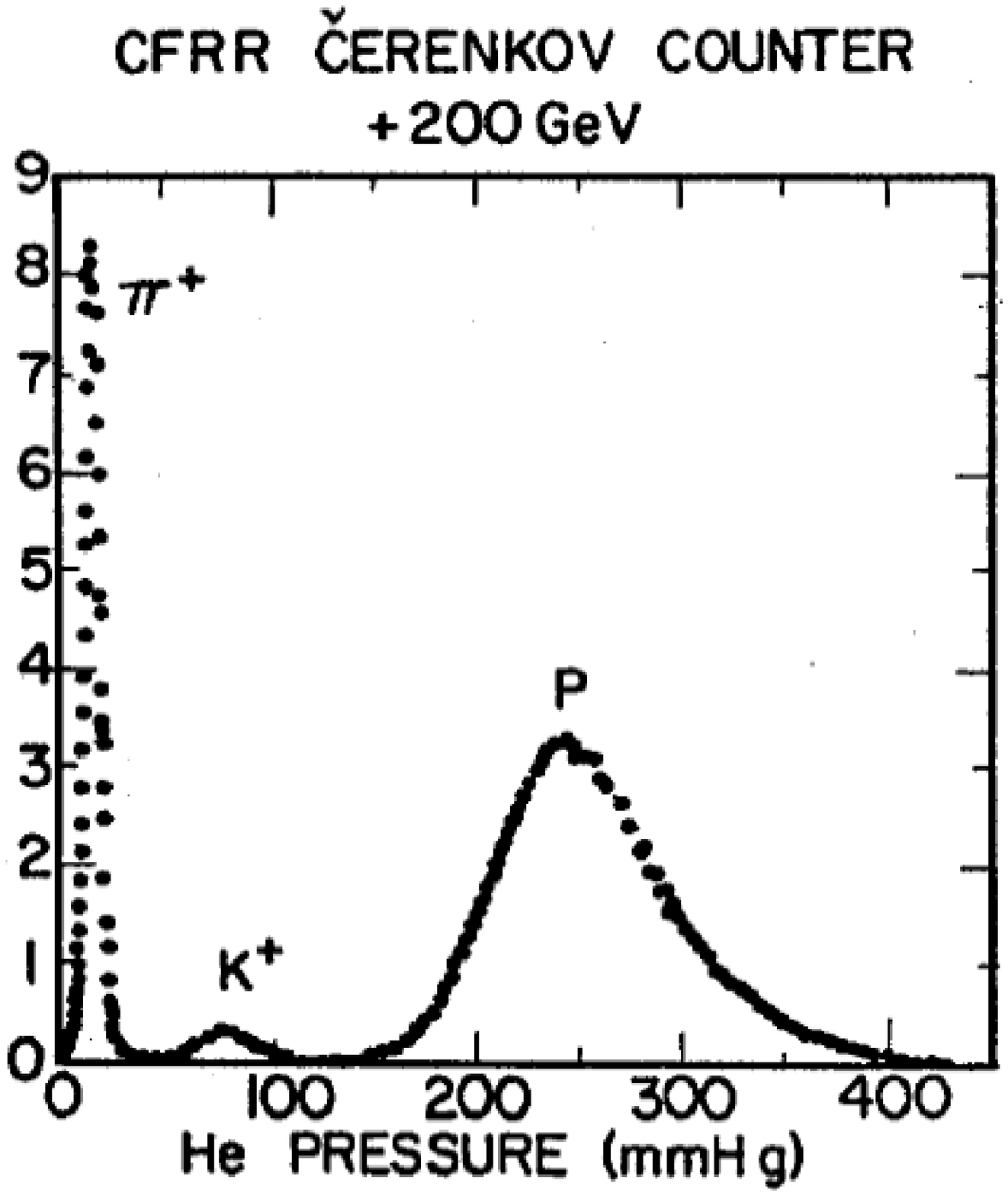}
  \includegraphics[height=3.3in]{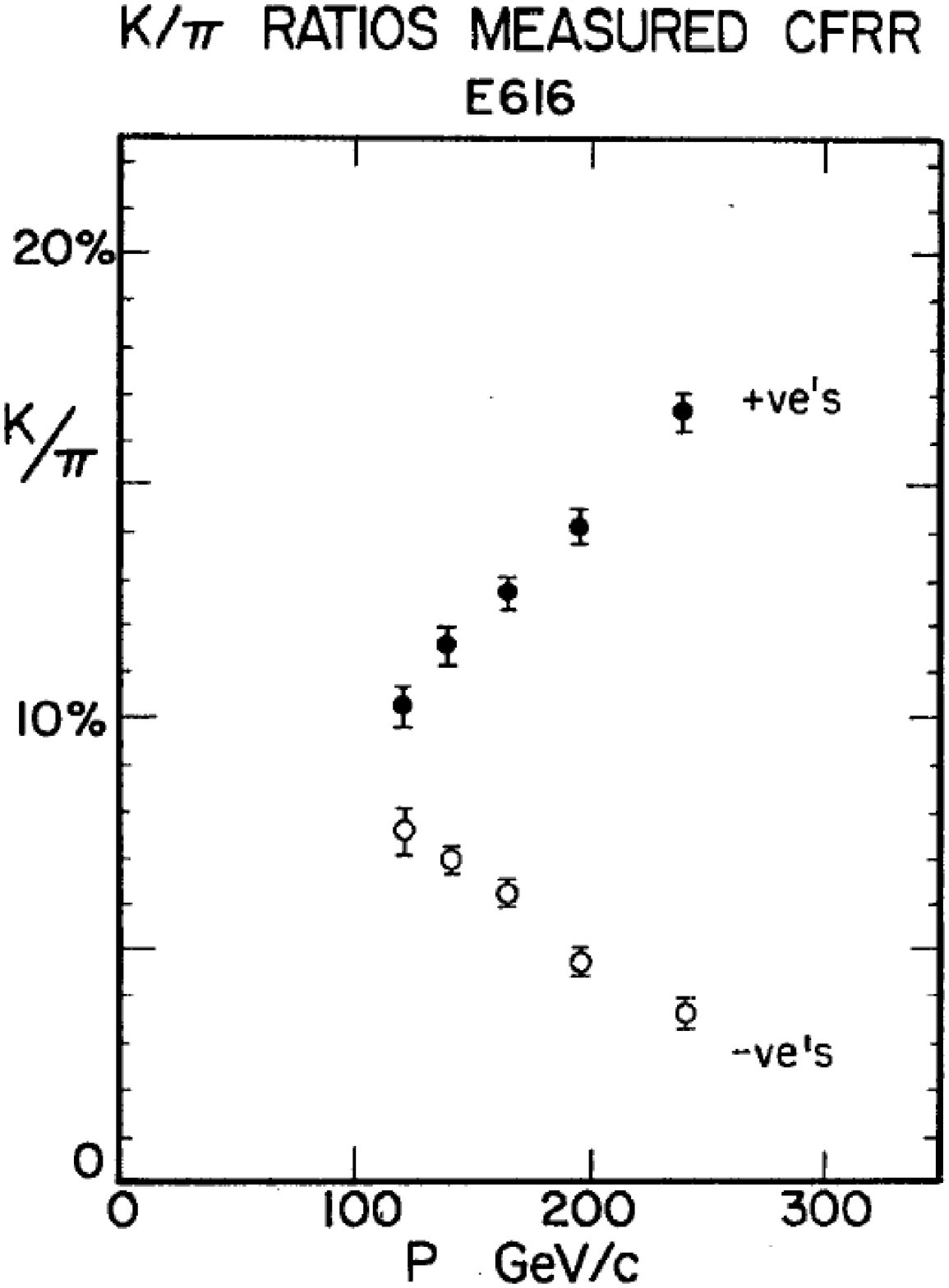}
\vskip 0.cm
  \caption{(left) Pressure scan of the Cherenkov counter in the CITFR dichromatic secondary beam, showing clear peaks in proportion to the $\pi^+$, $K^+$, and $p$ fluxes in the secondary beam.  The plot area is normalized to the total particle flux measured by the ion chambers.  (right) Ratio of $K/\pi$ fluxes {\it vs} secondary beam momentum for positive and negative beams.  Taken from \cite{barish1981}.}
  \label{fig:barish1981-fig7}
\end{figure} 


The Cherenkov counter was essential to this measurement because the $\pi\rightarrow\mu\nu$ and $K\rightarrow\mu\nu$ decays contribute to different energy neutrinos in the dichromatic beam, as shown in Figure~\ref{fig:barish1981-fig15}.  A plot of the relative abundance during $+200$~GeV/$c$ secondary beam running is shown in Figure~\ref{fig:barish1981-fig7}.  Measurement of these two individual fluxes absolutely, along with the known momentum bite of the dichromatic channel, allows absolute flux predictions which can then be compared with the event rates in Figure~\ref{fig:barish1981-fig15} to derive cross sections, independent of knowledge of protons on target.

\subsection{Muon Beam Monitoring}

There are two kinds of muon systems that have been built:  flux measuring systems and diagnostic systems.  Flux monitors attempt to use the tertiary muons to yield a measurement of the neutrino flux.  This is a plausible idea, since muons come from same decays as the neutrinos.  Not all beamline geometries are conducive to such flux measurements, however, because of either decay kinematics or because the shielding imposes limitations on the fractions of the muon flux visible to the muon detectors.  First, the muon detectors must be placed downstream in the muon filter, imposing a lower threshold of muon momentum and thereby cutting off direct measurement of the lowest part of the neutrino energy spectrum.  Second, the solid angle acceptance of such muon detectors (even in absence of intervening shielding), can be prohibitively small, especially at low energy where pion decays are wider angle and may not intercept the muon detectors at the end of the decay volume.  Diagnostic muon monitor systems may be similar to flux monitors, but access a smaller fraction of the muon spectrum.  They must, however, be available ``online'' during neutrino running, which poses different constraints in terms of simplicity of construction and radiation tolerance.

\begin{figure}[t]
\vskip 0.cm
  \centering
  \includegraphics[width=3in]{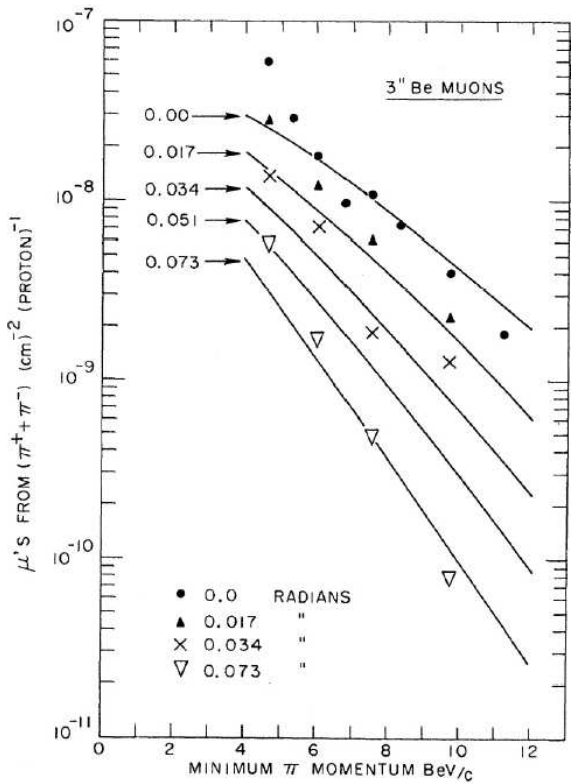}
\vskip 0.cm
  \caption{Emulsion measurements of muon fluxes in the steel shielding in the BNL neutrino experiment \cite{burns1965}.  The curves are fits using the CKP model\cite{ckp}.}
  \label{fig:burns1965-fig3}
\end{figure}

While the first neutrino experiments at BNL and CERN did not in any way measure their neutrino flux, the follow-on experimental run at BNL\cite{burns1965} did so, and was the first attempt to measure a neutrino flux using the tertiary muons.  Emulsions were placed in seven ``probe holes'' in the steel shielding at the end of the decay region, each probe hole containing up to four emulsions at different transverse distances to the beam axis.  Since the experiment had no focusing of the neutrino parents, this measurement, in the limit of no multiple scattering in the steel, should help corroborate their neutrino flux calculation.  The experimenters report an error $(20-30)\%$ from these measurements, due to the incomplete phase space sampled by the emulsions in the steel and the inability to go below a certain threshold (steel thickness) for fear of backgrounds from upstream hadrons.  Their flux data is shown in Figure~\ref{fig:burns1965-fig3}.  This tuning was done using a short run with a 3'' Be target to allow comparison with the existing thin-target hadroproduction data and with a 12'' Be target which was used during the neutrino run.

\begin{figure}[t]
\vskip 0.cm
  \centering
  \includegraphics[width=2.1in]{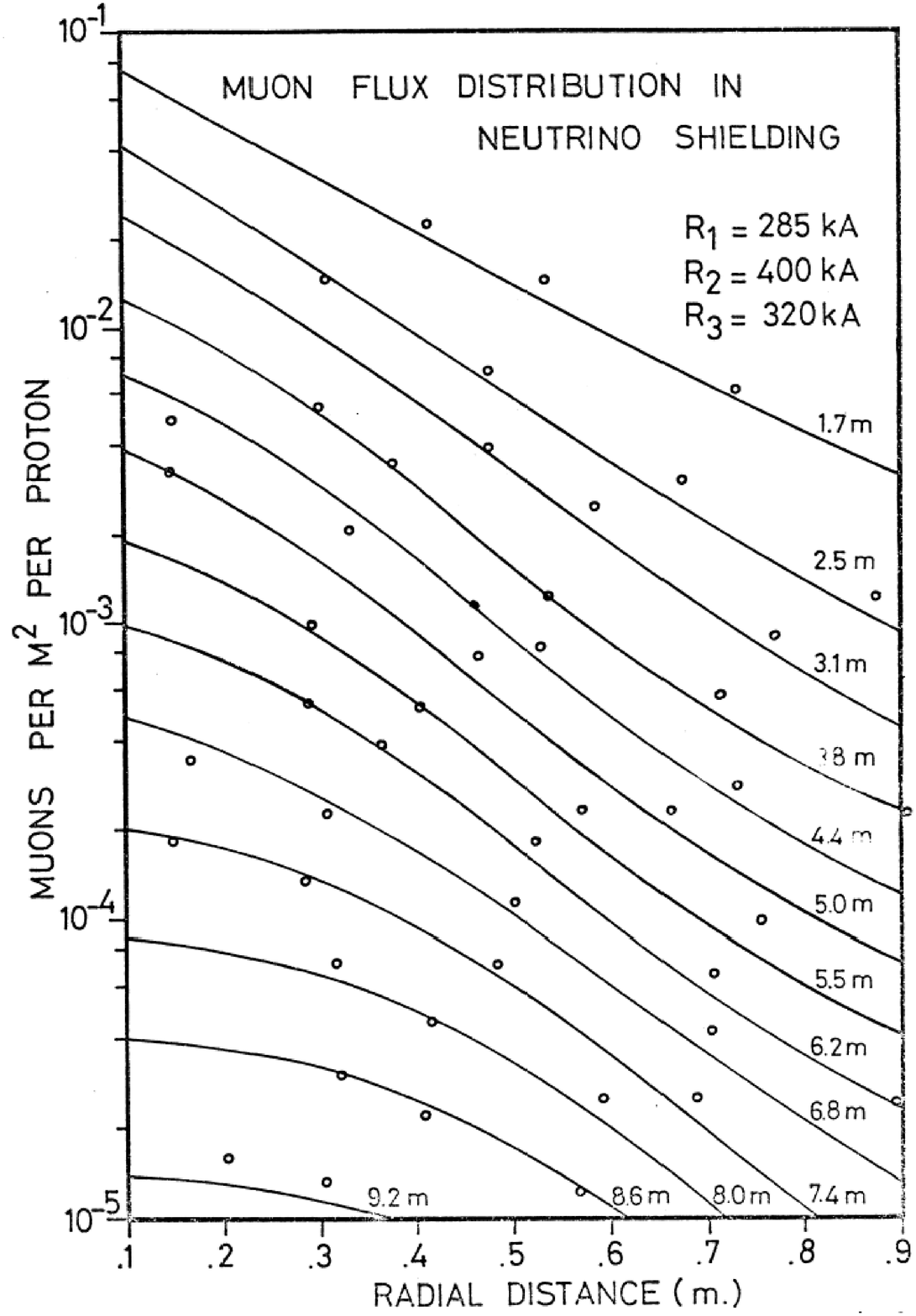}
  \includegraphics[width=2.1in]{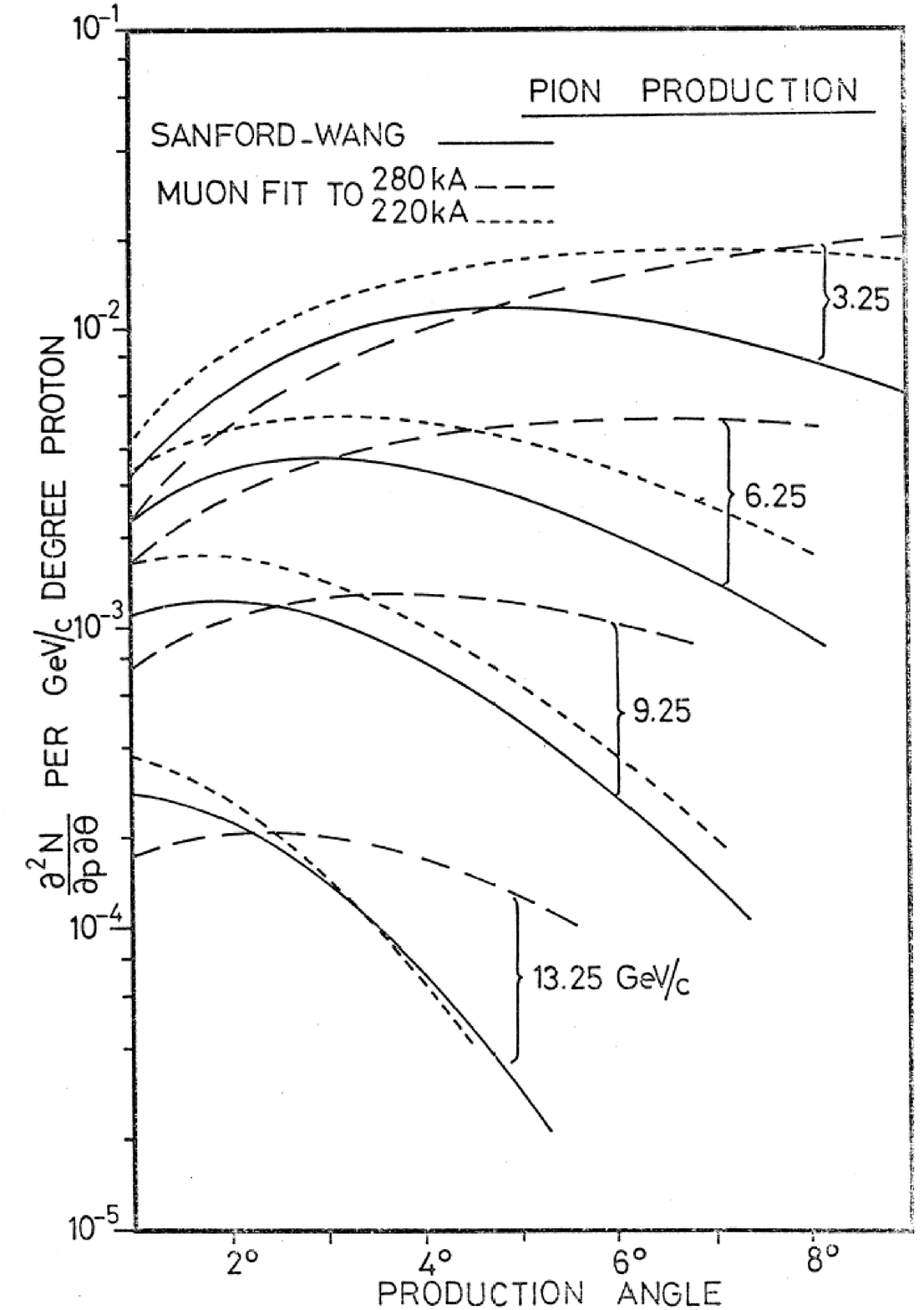}
  \includegraphics[width=2.1in]{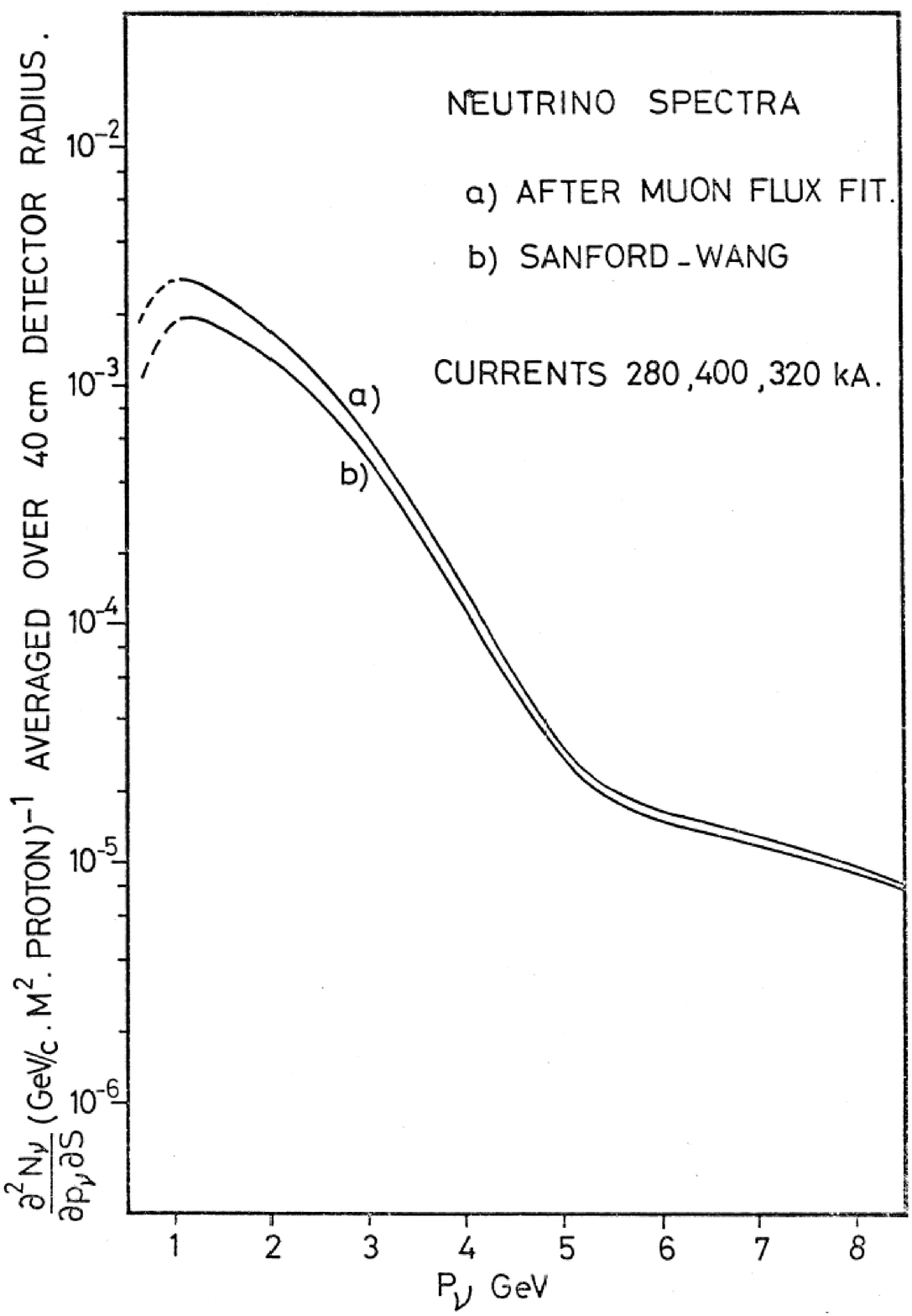}
\vskip 0.cm
  \caption{Demonstration of the muon flux-fitting procedure performed at CERN to determine the neutrino spectrum.  (left) muon fluxes at several lateral locations transverse to the beam axis and at several longitudinal depths in the muon filter.  (right) Sanford-Wang\cite{sanfordwang1967} parameterization of pion yields from the target $d^2N/dpd\Omega$ before and after the muon fit, assuming all target particles are created in primary interactions and the nominal Sanford-Wang $K/\pi$ ratio. (right) Neutrino flux at the bubble chamber before and after the fit.  Figures taken from \cite{wachsmuth1969}.}
  \label{fig:wachsmuth1969-figs}
\end{figure} 

CERN subsequently measured its neutrino spectrum using muon system measurements \cite{bloess1971}.  There are multiple challenges:  (1) they re-parameterized Sanford-Wang from this fit, so have to hope thick target effects simply scale the flux, not modify it; (2) focusing effects have minimal effect on the spectrum; (3) muons from $K$ decays only easily distinguished at large lateral offsets from beam axis, so have to assume $\pi/K$ ratio from external `beam survey' data \footnote{A similar measurement of the neutrino flux was attempted using the muon system at the IHEP beam \cite{anikeev1996}, and at the higher energies there the $K/\pi$ ratio uncertainty seems to be a bigger effect.}; (4) the first few meters of shielding have large hadron shower content, so one must only measure the flux above some threshold or perform a significant subtraction (CERN claims 50\% for the first data point at 1.7~m, 12\% in the second, and 6\% in the third -- these numbers were confirmed by measurements subsequently made with a W beam plug after the target, which demonstrated drops in roughly these proportions); (5) Both $\mu^+$ (proportional to $\nu_\mu$ flux) and $\mu^-$ (proportional to the $\overline{\nu}_\mu$ flux, which CERN claims is only 0.2\% of their flux); (6) the measured flux has to be corrected by $\sim$6\% for $\delta$-rays (the correction was obtained using emulsions placed on the chambers.  A `beam survey' of Ref. \cite{allaby1970} became available after this work, and agreed in neutrino flux prediction to within 10-15\%, which was their stated uncertainty.\footnote{Unfortunately for present-day beams, this method may not be easily applicable at high-intensity neutrino beams.  Higher intensities have required thicker beam stops to absorb hadronic showers from the remnant proton beam, and the neutrino energies (hence muon momenta) of interest have decreased for long-baseline neutrino oscillation searches.}

\begin{figure}[t]
\vskip -0cm
  \centering
  \includegraphics[height=3.1in]{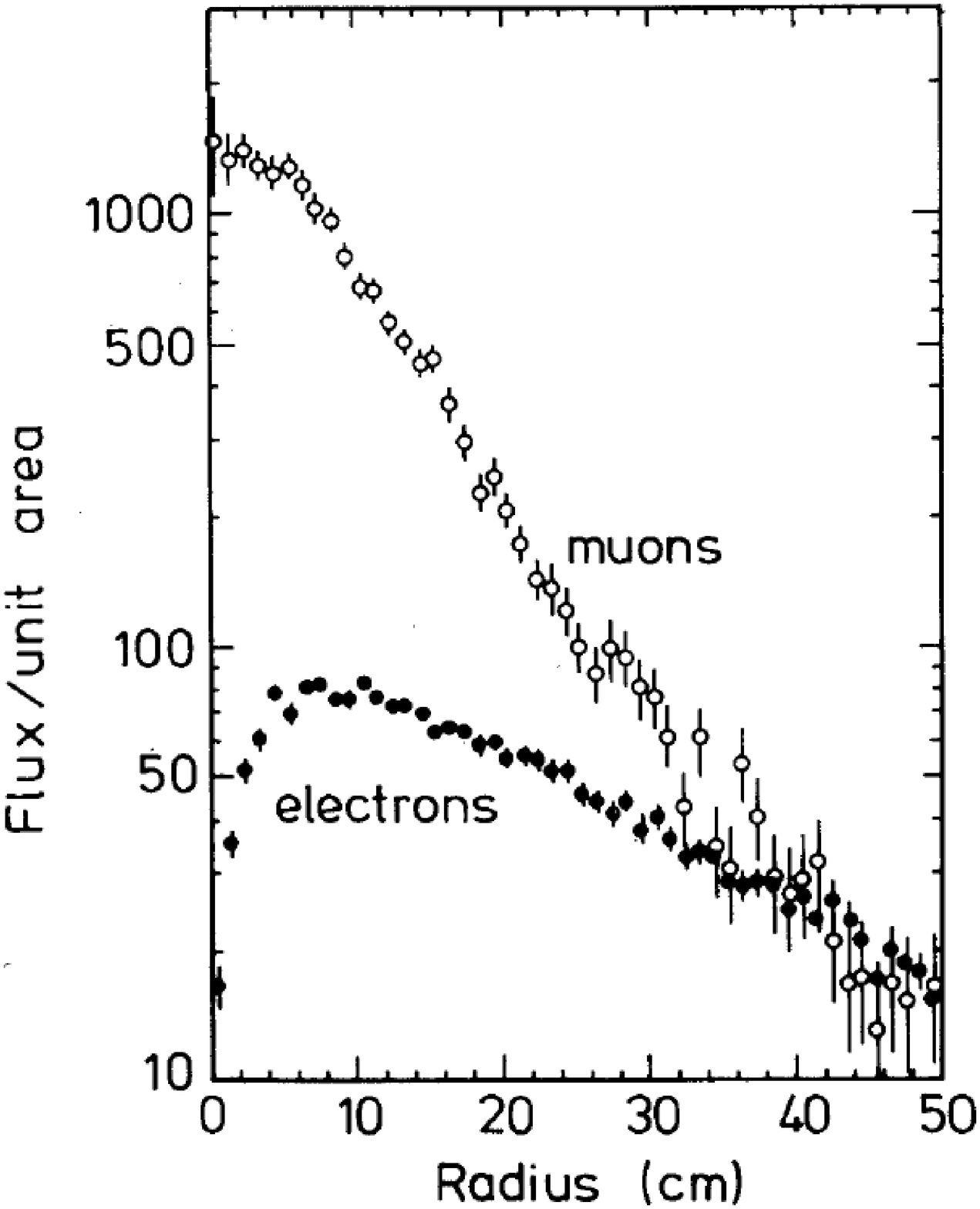}
  \includegraphics[height=3.1in]{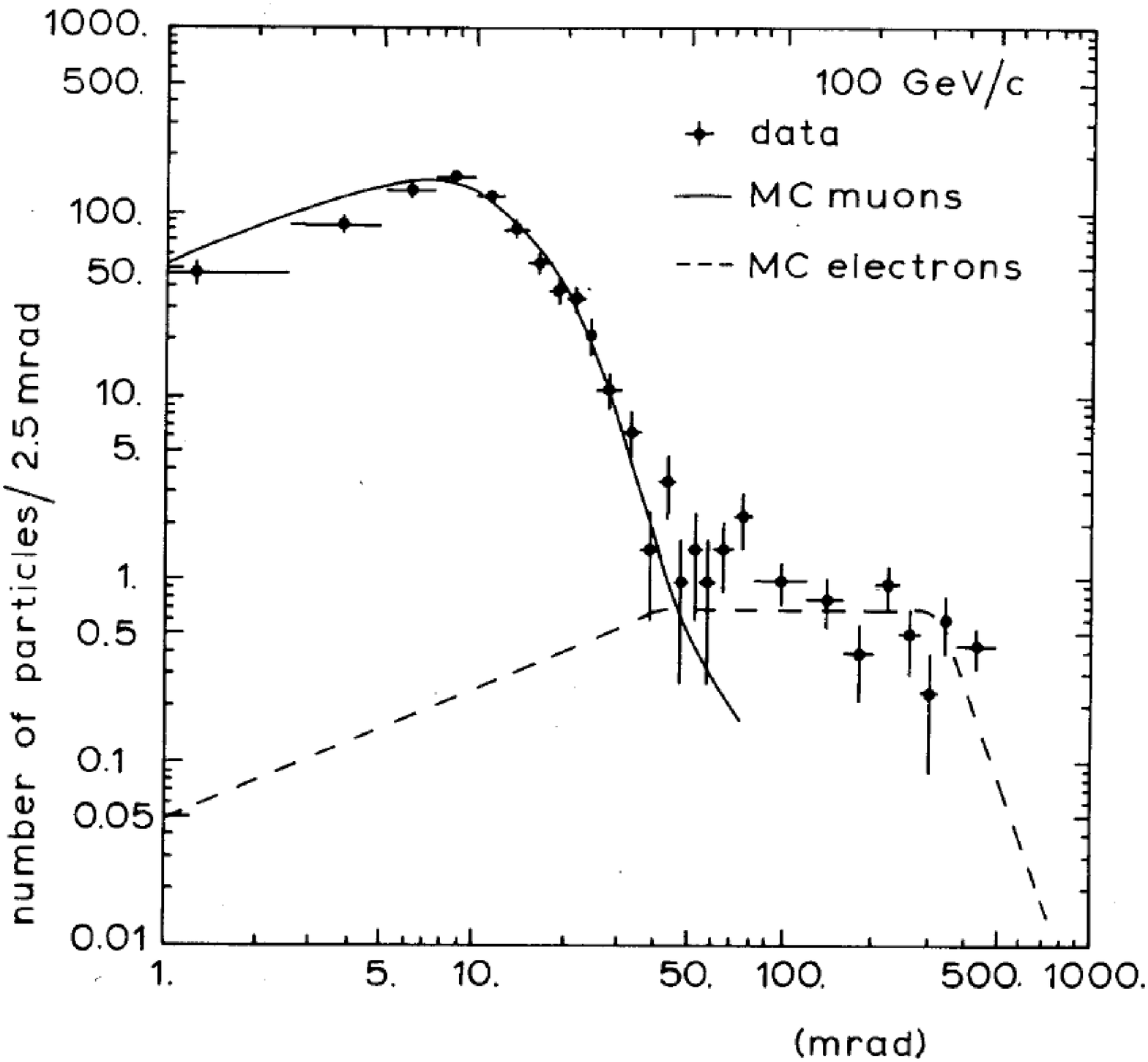}
  \caption{(left) Monte Carlo calculation of the transverse distribution of muons and knock-on electrons in one of the muon pits at the CERN WANF beam.  (right) Measured track angle in an emulsion sample placed in a 100~GeV/c muon beam, showing tracks at zero degrees (beam muons) and wide-angle tracks (knock-on electrons).  Taken from \cite{abt1985}.}
  \label{fig:abt1985-figures}
\end{figure}

The large flux of $\delta$ rays produced in the muon filter shielding makes flux measurements in a wide band beam quite challenging  \cite{heijne1979,abt1985}.  Charge-integrating detectors like solid state detectors or ionization chambers will measure a combination of muon signal and $\delta$-ray background.  As shown in the left plot of Figure~\ref{fig:abt1985-figures}, the electron component can be a significant fraction of the signal and has a different lateral shape than the muon beam due to multiple scattering of the electrons.  CERN employed a series of emulsions to count the tracks as well as the scattering angle of each track (straight-through tracks are presumably from beam muons while wide-angle tracks are from knock-on electrons), thereby obtaining a $\mu/e$ ``correction factor'' by which future muon measurements are adjusted.  Thus, absolute flux measurements are limited to 3-5\% due to the counting in emulsions and the ability to separate electrons from muons.

The technologies employed for muon monitors vary, though they must become increasingly simple and radiation-hard as the intensities increase and the ability to access these remote muon pits is reduced.  Ionization chambers were used in horn-focused wide-band beams such as the CERN PS neutrino beam \cite{bloess1971,pattison1969}, the BNL neutrino beam \cite{chi1989},the IHEP-Serpukhov beam \cite{bugorsky1977}, the KEK neutrino beam \cite{hill2001,ahn2006}, the Fermilab NuTeV beam \cite{yu1998}, the Fermilab NuMI beam \cite{kopp2006b}, and the CERN CNGS \cite{elsener2005}.  Solid state muon detectors were used in the CERN West Area Neutrino Facility (WANF) beam line off the SPS \cite{cavallari1978,heijne1980,heijne1983}.  Maximum muon rates ranged from $5\times10^5$/cm$^2$/spill \cite{bloess1971} to $5\times10^7$/cm$^2$/spill \cite{kopp2006b}.  ANL \cite{derrick1969}, CERN \cite{bloess1971}, and Serpukhov \cite{baratov1977f} had plastic scintillators in the downstream portion muon shield, where particle fluxes are lower $\sim$1/cm$^2$/spill).

Muon monitor systems have done real diagnostic work:  the CERN muon chambers at the PS neutrino line detected a flux asymmetry which was eventually traced to a magnetic field asymmetry in the horn \cite{dusseux1972}.  Also at CERN, a misalignment of some of the target hall components was detected at the WANF line:  they could achieve higher muon fluxes if they readjusted target and horn positions using motorize mounts \cite{casagrande1996}.  At NuTeV \cite{yu1998}, the monitors demonstrated the alignment of the neutrino beam, necessary for the desired precision in $\sin^2\theta_W$.  The NuMI secondary beam monitors detected misalignment of the target which, if uncorrected, could have been catastrophic:  the proton beam was initially aimed toward the target's edge \cite{zwaska2006}, which can cause stress on the target fins.  The NuMI chambers also diagnosed a failure in the target which caused it to fill with cooling water \cite{kopp2006b}.  For long-baseline experiments, the muon monitors are essential to demonstrate that the neutrino beam points to the remote neutrino detector within the required accuracy, as demonstrated at KEK \cite{ahn2006}.  

\begin{figure}[t]
\vskip -0.cm
  \centering
  \includegraphics[width=3.5in,angle=0]{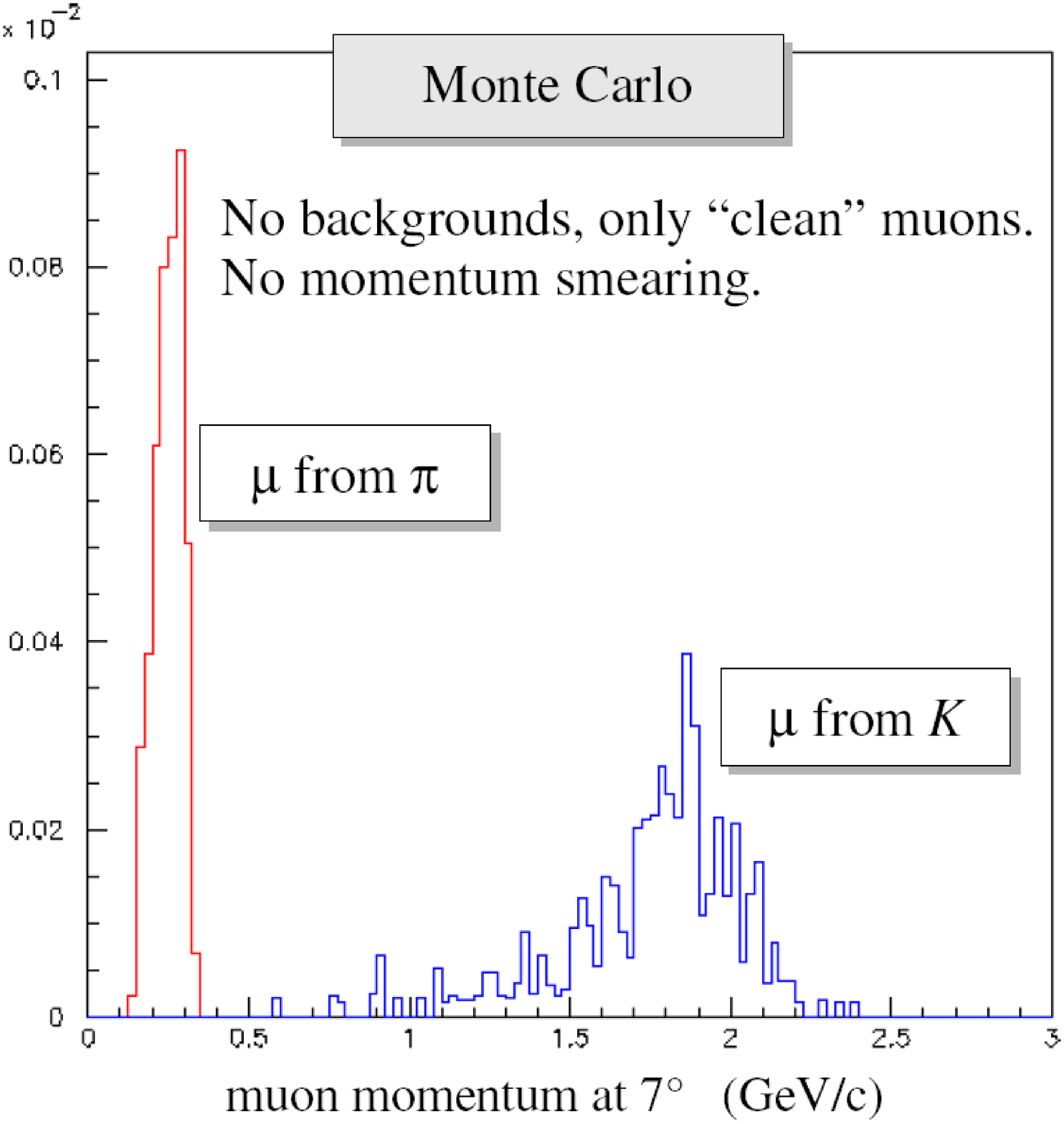}
\vskip -.5cm
  \caption{Muon momenta seen 7$^\circ$ off-axis from the MiniBooNE decay pipe from $\pi$ and $K$ decays.  Taken from \cite{hart2005}. }
  \label{fig:hart2005}
\end{figure} 

The MiniBooNE beam has a spectrometer system, the ``Little Muon Counter'' (LMC), placed 7$^\circ$ off-axis to its decay pipe (see Figure~\ref{fig:brice2004-dump}).  The spectrometer has an upstream collimator to define the $7^\circ$ angle of muons exiting the decay volume, and a small tracking system to measure muon momenta exiting the collimator.  The experiment, which is designed to search for $\nu_\mu\rightarrow\nu_e$ oscillations, benefits from direct measurements of intrinsic $\nu_e$ content in the beam, such as originates from $\mu$ decays or $K_{e3}$ decays.  While the former are constrained by measurements of the $\nu_\mu$ flux in the MiniBooNE detector, the $K_{e3}$ decays require separate knowledge about the $K/\pi$ ratio off the MiniBooNE target.  Figure~\ref{fig:hart2005} shows the result of a Monte Carlo calculation \cite{hart2005}, in the absence of backgrounds from neutrons, conversions, or resolution effects from scattering in the collimator, of what would be expected in the LMC spectrometer.

\newpage
\section{Two-Detector Experiments}
\label{two-det}

\begin{figure}[t]
\vskip -0.cm
  \centering
  \includegraphics[width=4.in,angle=0]{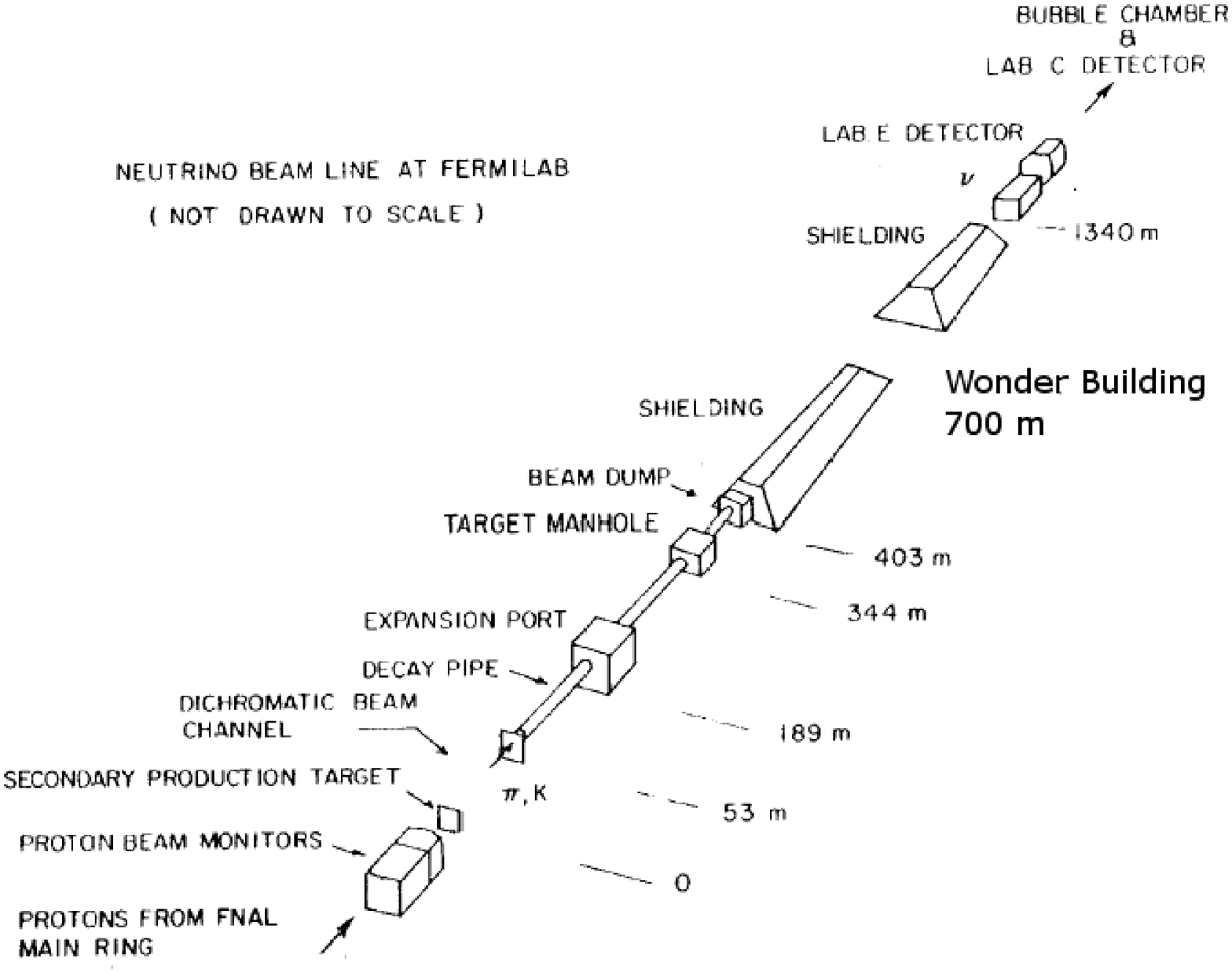}
\vskip -.5cm
  \caption{View of the Fermilab Neutrino Line in 1983, taken from \cite{blair1983}.  The dichromatic beam is directed at detectors at 700~m, 1100~m, and 1500~m.}
  \label{fig:blair1983-fig1}
\end{figure}

\begin{figure}[t]
\vskip -2.cm
  \centering
  \includegraphics[width=4.in,angle=270]{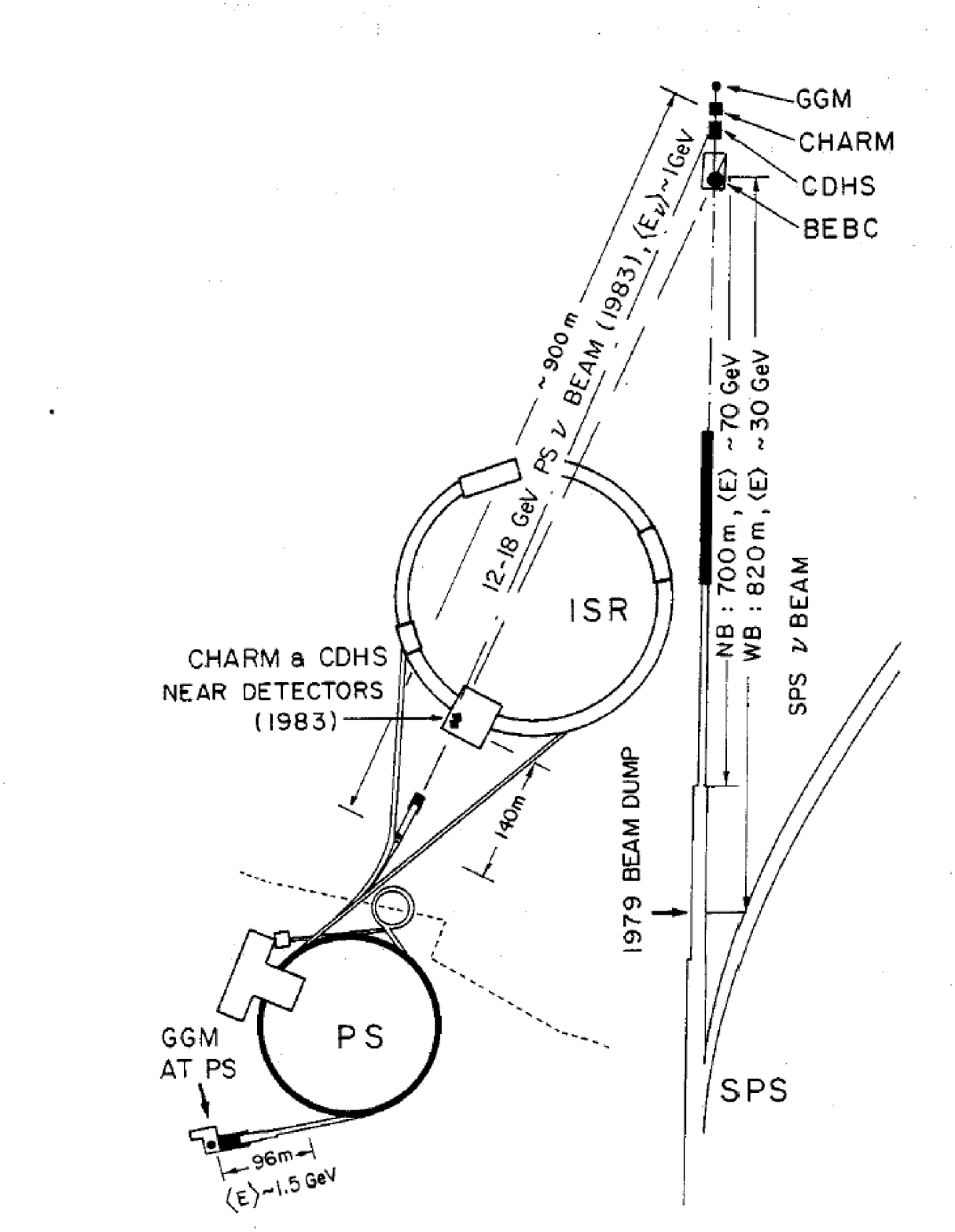}
\vskip -.5cm
  \caption{Plan view of the CERN PS neutrino beam two-detector experiments, taken from \cite{bergsma1988}.  A bare-target beam taken from the PS is delivered to the CHARM and CDHS experiments.}
  \label{fig:wachsmuth1982-fig1}
\end{figure} 

Two-detector experiments were pioneered at Fermilab (see Figure~\ref{fig:blair1983-fig1}) and CERN (see Figure~\ref{fig:wachsmuth1982-fig1}) for the purpose of studying neutrino oscillations.  Downstream of the neutrino beam, a ``near'' detector measures directly the energy spectrum of neutrinos from the beam.  A second detector measures the energy spectrum of neutrinos which have propagated for some time interval.  The distance to the second detector is presumably long compared to the time to arrive at the first detector.  Deviations between the two energy spectra may be used to infer the presence of neutrino oscillations, which manifest themselves as the disappearance of the $\nu_\mu$ beam \cite{stockdale1984,bergsma1988,ahn2006,michael2006}, or the appearance of a different neutrino flavor in the $\nu_\mu$ beam (eg:  \cite{ahn2004}).  The direct measurement of the flux in the first detector greatly reduces the need to calculate the beam spectrum, improving the experiments' sensitivity.  
More recently, long-baseline experiments have searched for oscillations across distances of order 100's of km, as pioneered at KEK;  such has required use of accurate GPS to locate the two separated detectors \cite{noumi1997}.  

\subsection{Calculating the Extrapolated Beam Flux}

\begin{figure}[t]
\vskip -0.cm
  \centering
  \includegraphics[width=5.2in]{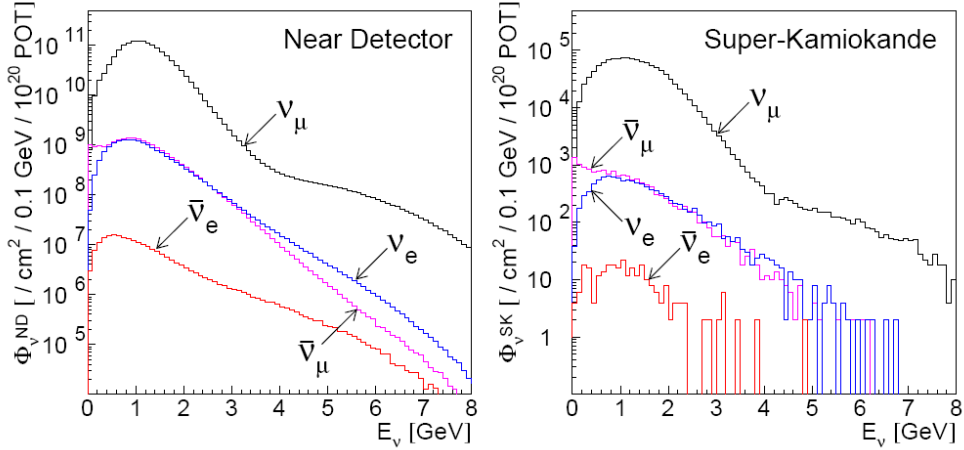}
\vskip -.0cm
  \caption{Energy spectra of neutrinos from the KEK beam at $z=300$~m and at $z=250$~km.  Taken from \cite{ahn2006}. }
  \label{fig:ahn2006-fig6}
\end{figure}

Even in the absence of oscillations, the energy spectra in the two detectors are are different, and the differences must be calculated so that a ``near-to-far'' extrapolation of that spectrum observed in the near detector can be computed.  A comparison of the expected energy spectra in the KEK beam at $z_{\mbox{near}}=300$~m (``near detector'') and at $z_{\mbox{far}}=250$~km (``far detector'') is shown in Figure~\ref{fig:ahn2006-fig6}.  The spectra are different, not simply scaleable by the ratio of the two solid angles subtended by the detectors, $z_{\mbox{near}}^2/z^2_{\mbox{far}}$.  What follows is a description of the near-to-far extrapolation technique.

If the beam were a point source, then the prediction could be estimated by
\begin{center}
$N_{\mbox{far}} =\mathcal{R}$$_{FN} N_{\mbox{near}}$
\end{center}
where the extrapolation factor $\mathcal{R}$$_{FN} = Z_{near}^2/Z_{far}^2$ is just the ratio of solid angles subtended by the two detectors.  Considering that the beam is an extended source, one could weight this extrapolation factor by the pion lifetime along the length of the decay tunnel:
\begin{equation}
\mathcal{R}_{FN} = \frac
{\int_{z\sim 0}^{L}  \frac {e^{-\frac{0.43m_{\pi}z}{E_{\nu}c\tau}}} {(Z_F-z)^2} dz}
{\int_{z\sim 0}^{L}  \frac {e^{-\frac{0.43m_{\pi}z}{E_{\nu}c\tau}}} {(Z_N-z)^2} dz}
\label{rfneqn}
\end{equation}
where the integral is over the length $L$ of the decay tunnel and the substitution $E_{\pi}\sim E_{\nu}/0.43$ has been made (c.f. Equation~\ref{eq:enu-vs-epi}).  The fact that $Z_F>>Z_N$ reduces the integral in the numerator to $\sim1/Z_F^2$, like a point source, while the integral in the denominator reflects the more complicated ``line source'' of neutrinos seen by the near detector. 

\begin{figure}[b]
\vskip -0.cm
  \centering
  \includegraphics[width=6in]{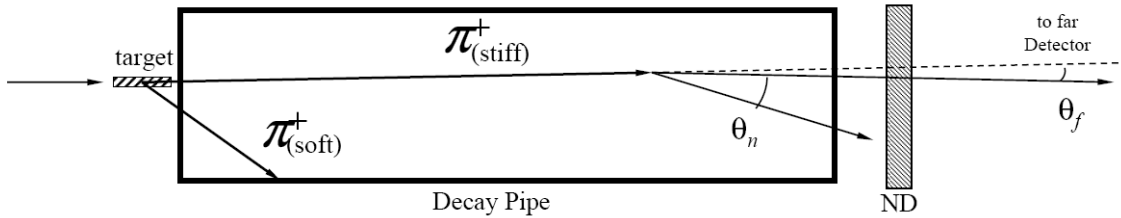}
\vskip -.0cm
  \caption{Demonstration of the solid angle differences in a two-detector neutrino experiment.  Not to scale.}
  \label{fig:near-far-angles}
\end{figure}

\begin{figure}[t]
\vskip -0.cm
  \centering
  \includegraphics[width=4.in]{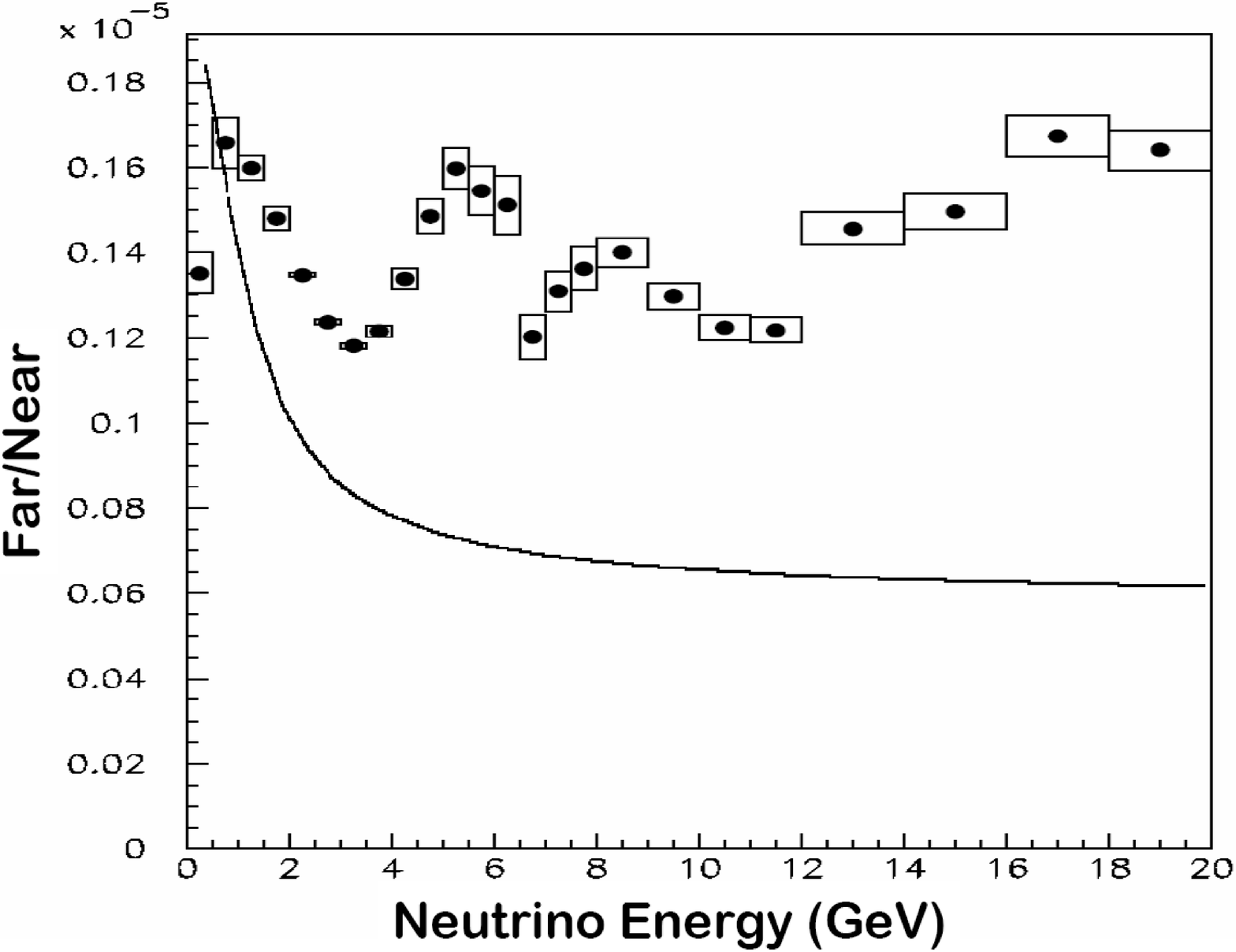}
\vskip -.0cm
  \caption{Far-over-near ratio for the NuMI LE beam and the two detector locations at $z=1040$~m and $z=735.4$~km.  The points are calculated with a Monte Carlo simulation of the beam line, the box sizes correspond to the uncertainties in the calculation due to the hadron production model assumed, and the curve is the idealized calculation using only the pion lifetime of Equation~\ref{rfneqn}.}
  \label{fig:far-near-calculation}
\end{figure}

Equation~\ref{rfneqn} is a simplification, since not all decaying pions produce neutrinos within the finite acceptances of the detectors and not all pions are able to decay before interacting along the decay pipe walls.  Furthermore, the significant acceptance differences between the near and far detectors are a function of the energy of the pion (hence neutrino), as is indicated schematically in Figure~\ref{fig:near-far-angles}.  Fast pions tend to live longer and decay downstream in the decay pipe, closer to the near detector.  A variety of pion decay angles $\theta$ will result in a neutrino which strikes the transversely large near detector, while only $\theta\approx0^\circ$ decays result in a neutrino arriving at the far detector, affecting the neutrino energy at each detector in Equation~\ref{eq:enu-vs-epi}.  Additionally, pions of different momenta enter the decay volume with different angular divergences.  For an unfocused beam, such as used in the CERN experiments, the pion angle leaving the target is $\theta\approx2/\gamma=2m_\pi/E_\pi$.  Such differences in angles of entry in the decay pipe result (for low-energy pions) in wider-angle decays to reach the neutrino detectors, and also greater likelihood of striking the decay volume walls.  Even for a horn-focused beam, pions of different momenta will have varying divergences depending upon the exact tune of the focusing system, and some pions will be unfocused due to the zero-field inner-apertures (``necks'') of the horns, as shown in Figure~\ref{fig:5-beams}.  

The calculation of $\mathcal{R}_{\mbox{FN}}$ requires a Monte Carlo, as shown in Figure~\ref{fig:far-near-calculation} for the NuMI beam.  The Monte Carlo prediction departs sharply from the idealized curve of Equation~\ref{rfneqn}, particularly for very high energy neutrinos (which come from unfocused parent mesons), but also at two other distinct values near $E_\nu\approx4.5$~GeV and $E_\nu\approx8$~GeV.  Inspection of Figure~\ref{fig:5-beams} reveals that these energies are the transition points at which pions no longer pass through horn 1 or horn 2, respectively, of the NuMI focusing system.  Changes in focusing, or the absence of focusing, causes the departure from the idealized curve.  Uncertainties in the extrapolated far detector flux are typically at the level of $(2-5)$\%, well below the uncertainty in direct prediction of the flux (which range from $(20-100)$\% just from hadron production alone), as discussed in Section~\ref{uncert}.  Thus, the two-detector experiment serves to reduce the uncertainty in the prediction of the ``far detector's'' energy spectrum.

\begin{figure}[t]
\vskip -0.cm
  \centering
  \includegraphics[width=5.5in]{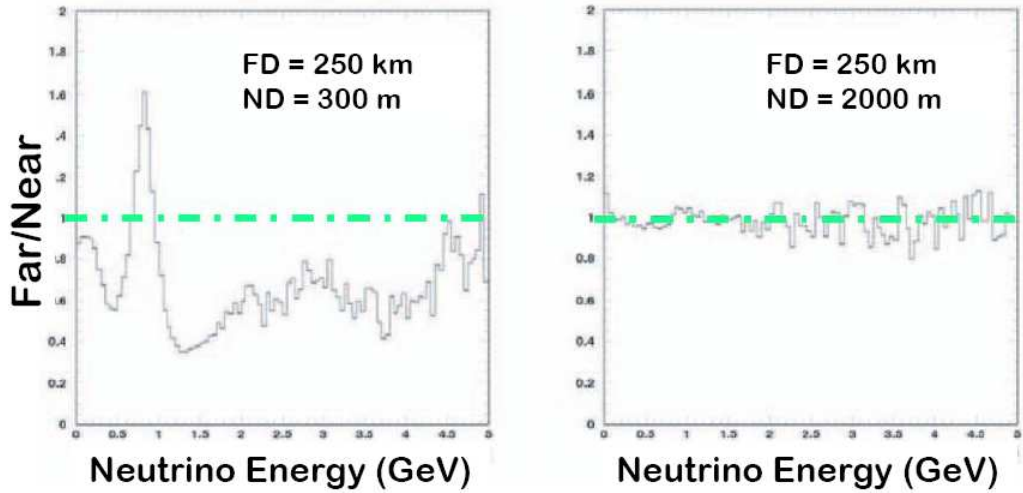}
\vskip -.0cm
  \caption{F/N ratio for the Tokai-to-Kamiokande experiment in Japan.  The ``far detector`` is Super Kamiokande at $z=250$~km.  The ``near detector'' is planned to be at $z=300$~m, though a proposal has been made to place an additional ``near detector'' at $z=2000$~m.  Taken from \cite{kearns2005}. }
  \label{fig:kearns2005-fig3}
\end{figure}

\begin{figure}[b]
\vskip -0.cm
  \centering
  \includegraphics[width=5.5in]{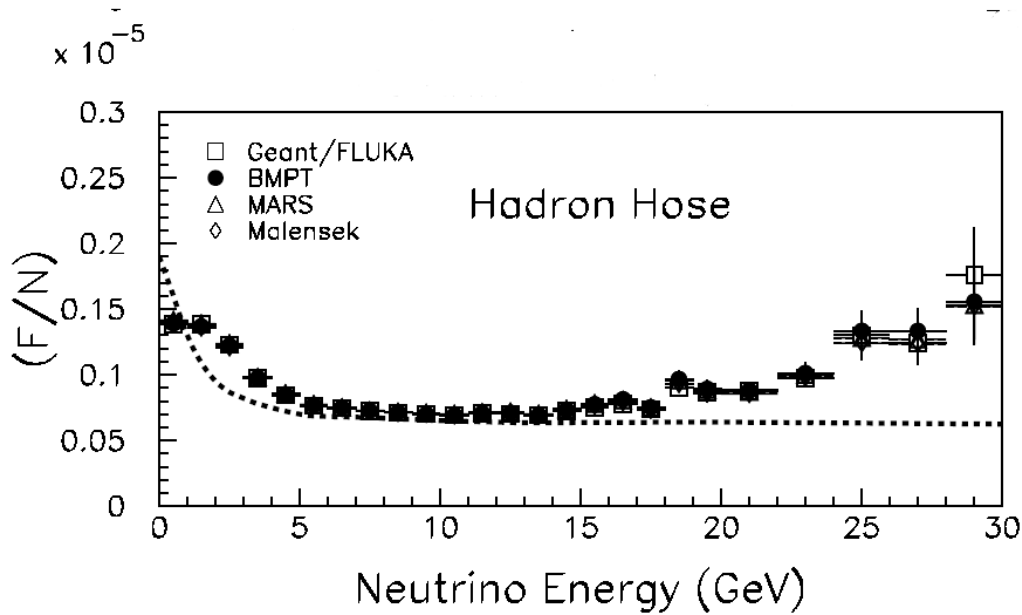}
\vskip -.0cm
  \caption{F/N for the NuMI beam with the addition of the proposed ``hadron hose'' focusing in the decay tunnel.  The curve is the idealized calculation using only the pion lifetime of Equation~\ref{rfneqn}.  The hadron hose draws pions away from the decay pipe walls, allowing them to decay and thereby following the pion lifetime curve of Equation~\ref{rfneqn}.  Taken from \cite{hylen2003}.}
  \label{fig:hylen2003-figxx}
\end{figure}

Two proposals have been developed to make the far and near detector spectra more similar, requiring less sophisticated calculation of $\mathcal{R}_{\mbox{FN}}$.  The T2K collaboration \cite{kearns2005}, noting that the structure in $\mathcal{R}_{\mbox{FN}}$ results from the large angular acceptance of the near detector, proposed placing a near detector further from their beam line, at $z=2000$~m, large compared to the 130~m long beam line.  The NuMI hadron hose proposal \cite{hylen2003} (see Section~\ref{hose}) was developed in part as a means to remove the angular correlations between the near and far detectors, thereby the energy difference resulting from Equation~\ref{eq:enu-vs-epi}.  The spiraling orbits in the hose field randomize the decay angles to the 2 detectors, as well as keep particles away from the decay pipe walls so they can continue to follow the pion lifetime curve of Equation~\ref{rfneqn} \cite{kopp2000}.  The results from these two ideas are shown in Figures~\ref{fig:kearns2005-fig3} and ~\ref{fig:hylen2003-figxx}, respectively.  Both proposals are quite effective in removing some of the complexity of the near-to-far extrapolation calculation.  

\clearpage
\newpage

\clearpage
\newpage

\clearpage
\newpage

\begin{figure}[t]
\vskip -0.cm
  \centering
  \includegraphics[width=3in]{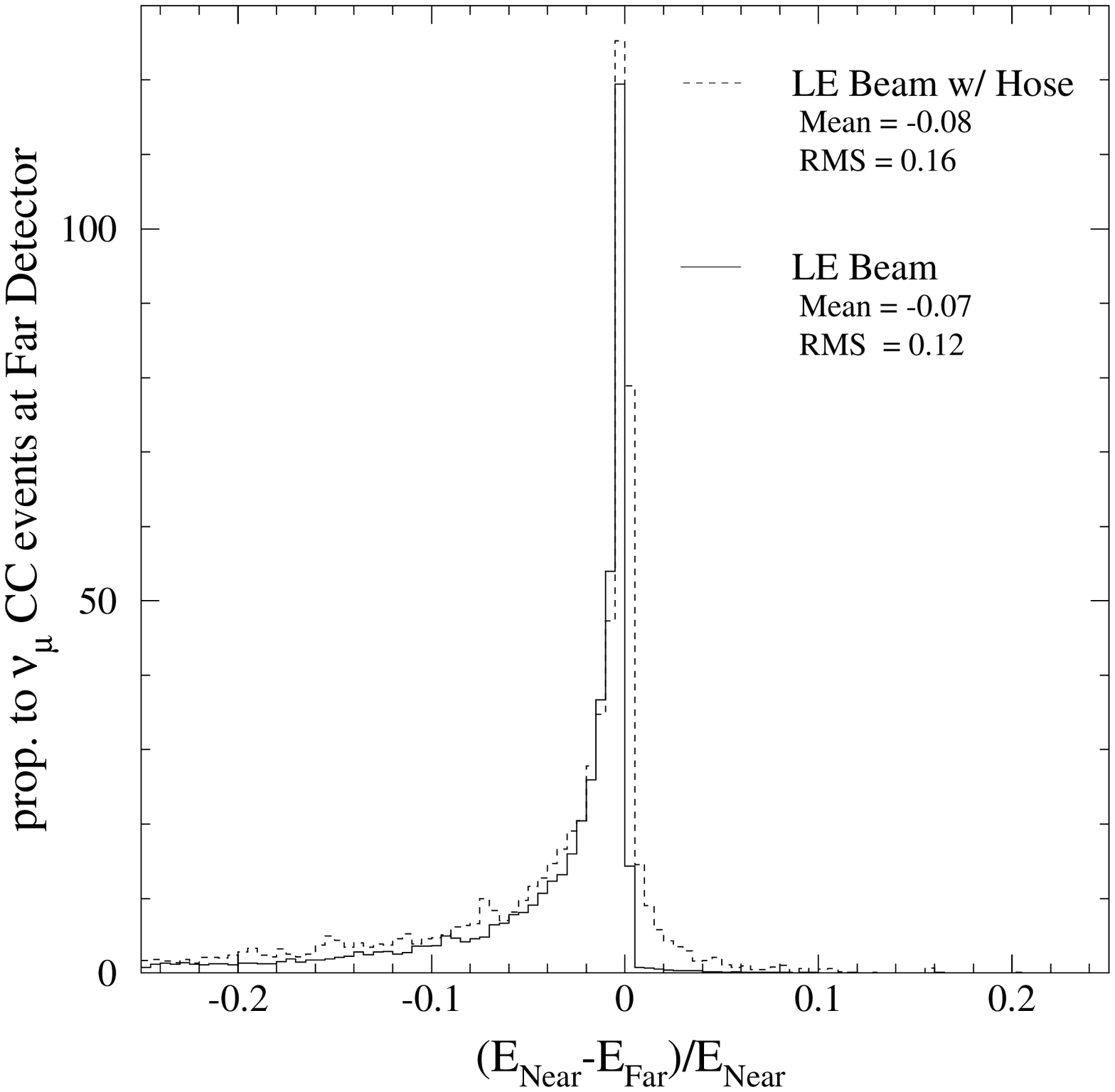}
  \includegraphics[width=3in]{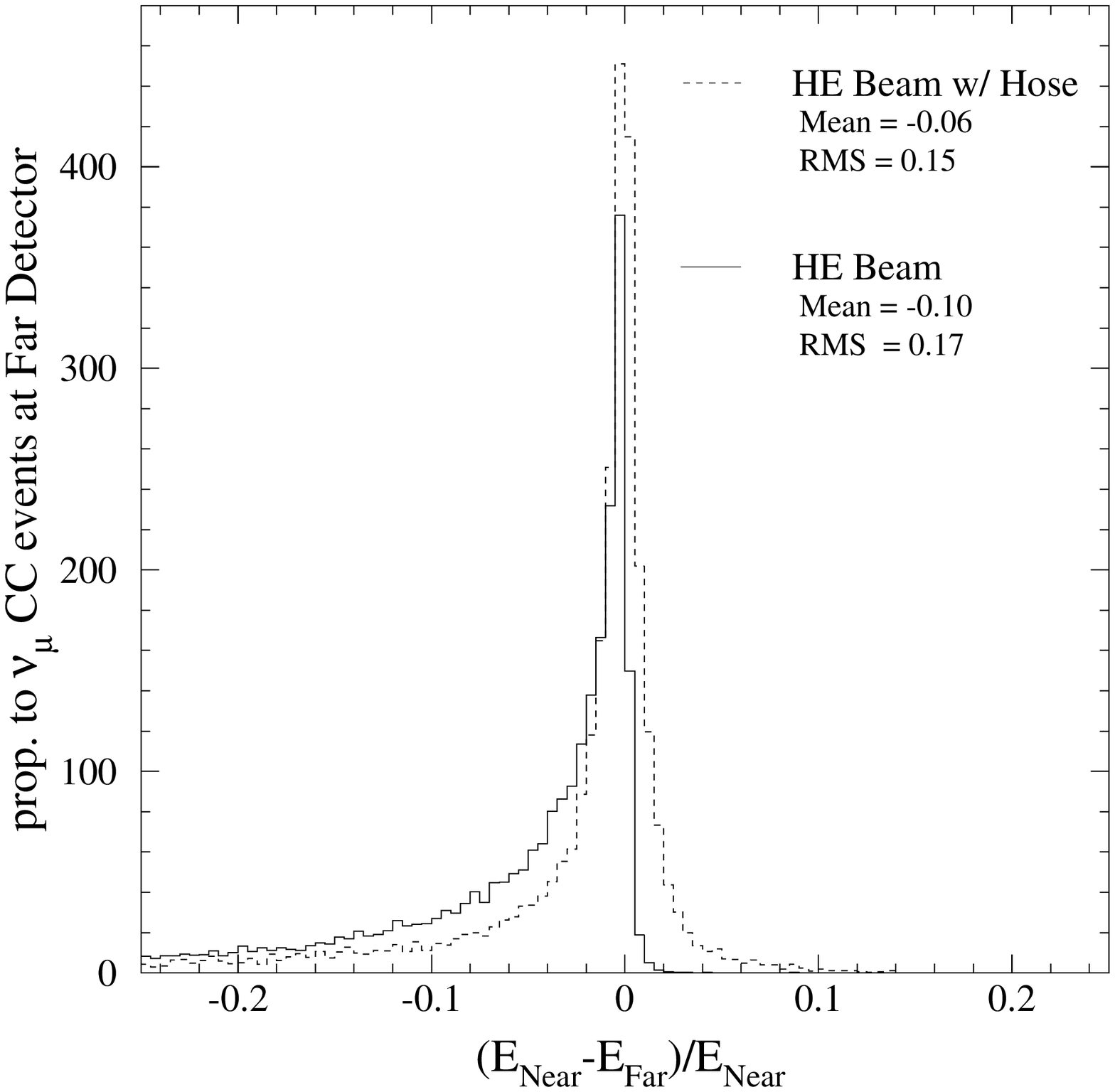}
\vskip -.0cm
  \caption{The difference in the energy of the neutrino that a given pion would emit toward the far detector
to that it would emit toward the near detector in NuMI/MINOS in the low-energy (left) and high-energy (right) beams, normalized to the near detector energy.  Curves are shown with and without the focusing of the Hadron Hose proposal.  Taken from \cite{hylen2003}.}
  \label{fig:hylen2003-fig5}
\end{figure} 

\begin{figure}[t]
\vskip -.25cm
  \centering
  \includegraphics[width=5.8in]{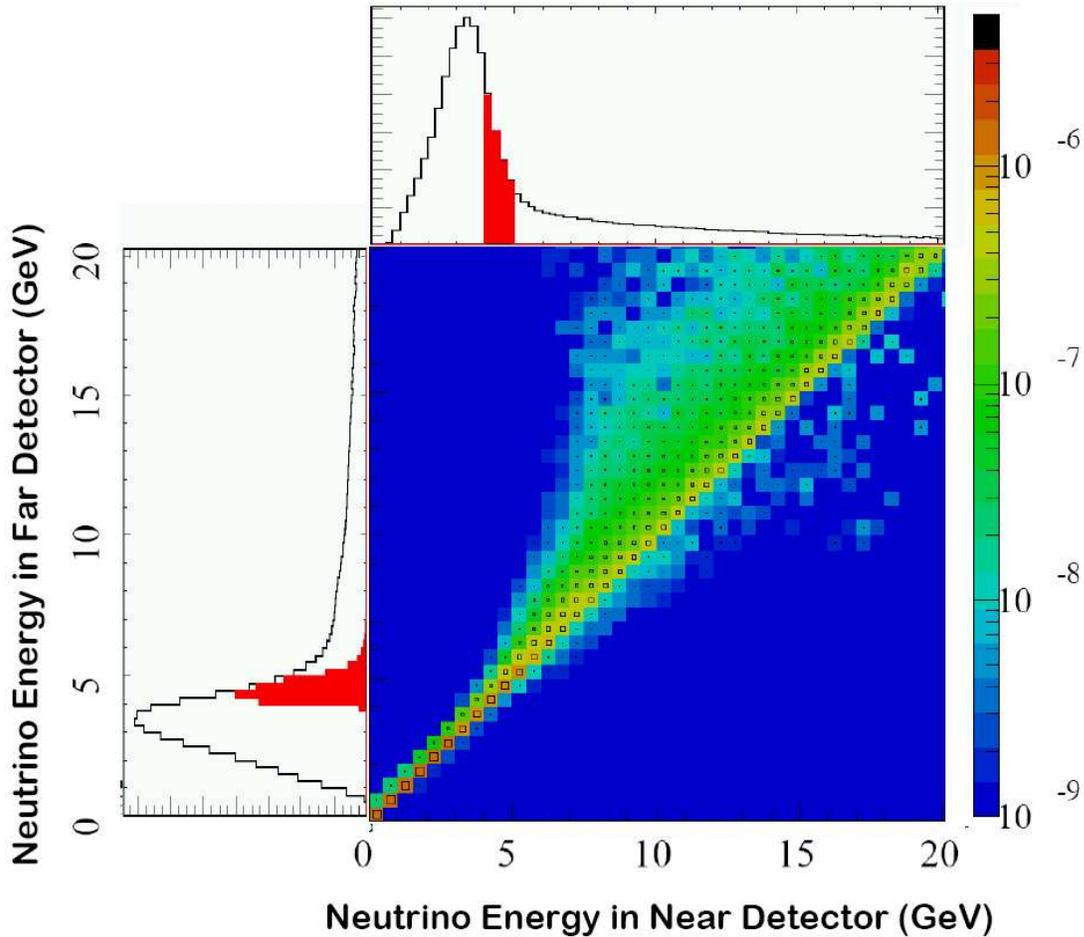}
\vskip -.5cm
  \caption{Graphic representation of the matrix method $M_{ij}$ for predicting the FD flux from the ND flux of \cite{szleper2001}.  The 2-dimensional color plot gives the values of $M_{ij}$ in Equation~\ref{eq:para-matrix}, the scale for which is at the right.  The histograms show energy spectra in the ND and FD.  Figure courtesy M. Messier.}
  \label{fig:messier-matrix}
\end{figure} 

That the decay angles to the ND and FD are substantially different is demonstrated in Figure~\ref{fig:hylen2003-fig5}.  In this figure is shown, for each pion focused in the NuMI beam, the ratio of the neutrino energy created by that pion if the neutrino strikes the center of the near detector at $z=1040$~m to the energy of the neutrino created by that same pion if it strikes the far detector at $z=735.4$~km.  Because of the wide array of decay angles which can strike the near detector, particularly for high-energy particles decaying at the downstream end of the decay pipe, the energy spectrum in the near detector is systematically lower in the ND than the FD.  With the inclusion of the Hadron Hose, the ratio of energy spectra at the two detectors becomes more symmetric.  

The systematic difference in decay angles motivated another technique \cite{szleper2001} employed for analysis of data from the NuMI beam \cite{michael2006} which permits some correction for the near-far difference in energy spectra.  Noting that a pion decay which gives rise to an energy $E^{ND}_i$ in the ND actually gives rise to a neutrino energy $E^{FD}_j$ in the FD, a matrix was developed which describes the relationship between the energies $E^{ND}_i$ and $E^{FD}_j$:
\begin{equation}
\phi^{FD}_j = \Sigma_i M_{ij} \phi^{ND}_i
\label{eq:para-matrix}
\end{equation}
where $\phi^{FD}_j$ is the flux of neutrinos in the $j$th energy bin in the FD and $\phi^{ND}_i$ is the flux of neutrinos in the $i$th energy bin in the ND.  Equation~\ref{eq:para-matrix} provides a more accurate prediction of the flux of neutrinos in the FD given a measurement of the $\phi^{ND}_i$ in the ND.  Conceptually, the matrix is similar to a point-spread function:  a particular flux of neutrinos bin of energy $i$ in the ND contributes to several bins $j$ in the FD.  

The matrix elements $M_{ij}$ is shown graphically in Figure~\ref{fig:messier-matrix}.  As noted in the scale, the values of $M_{ij}$ are typically of order $(z_{near}/z_{far})^2=(1/735.4)^2$.  The histograms show energy spectra in the ND and FD.  As expected, the matrix departs from a diagonal matrix at large neutrino energies, which arise from high-momentum pions decaying closer to the ND.  Summation of the rows of $M_{ij}$ ({\it i.e.} summation in $j$) would give the $i$th element of the far-over-near ratio $\mathcal{R}_{\mbox{FN}}^i$. 

Also shown in Figure~\ref{fig:messier-matrix} are the energy spectra calculated in the ND and FD.  For illustration, a set of neutrino energies are highlighted in the ND spectrum, and the corresponding neutrino energies from the same meson decays are indicated in the FD.  A single neutrino energy measured in the ND must, because of decay kinematics, contribute to the flux prediction for several energy bins in the the FD.  Such is the advantage of the matrix method.

\subsection{Systematic Uncertainties}
\label{uncert}

This section is not intended to be a full list of all systematic uncertainties for all two-detector experiments, nor of all possible techniques which can be used to limit them.  Surely these will change over time.  Rather, the figures shown here are meant to demonstrate where systematics appear in the spectrum and what kind of information is helpful in limiting them.  The NuMI beam will be used as a basis for discussion.

\begin{figure}[t]
\vskip -.5cm
  \centering
  \includegraphics[width=6.5in,height=4.in]{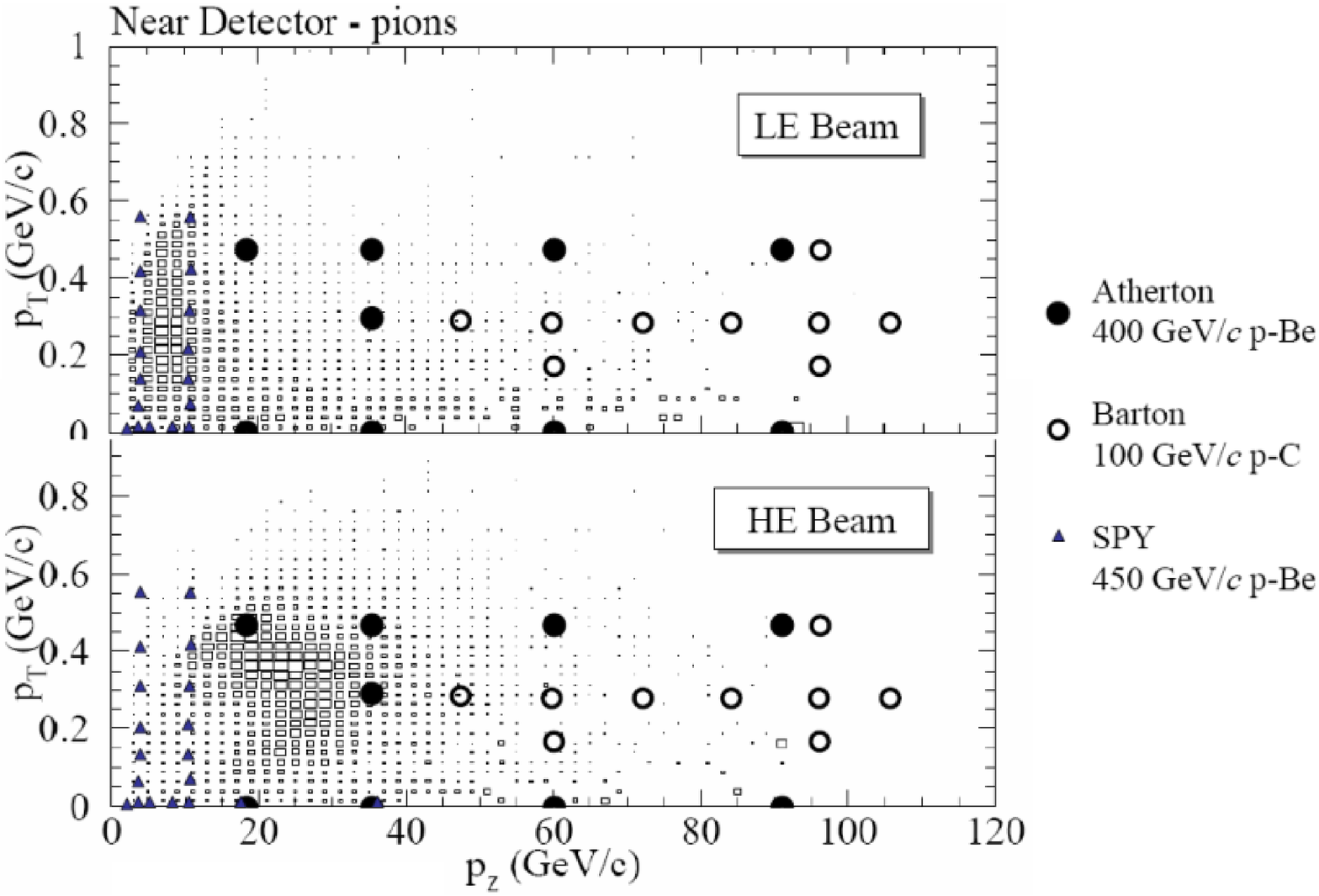}
\vskip -.5 cm
  \caption{Density plot showing the $p_T$ and $p_z$ of pions which yield neutrinos in the MINOS near detector.  The box size show the probability of producing such a pion in a $p+$C collision at 120~GeV/$c$ multiplied by the probability for it to go through the beam line and result in a neutrino interaction in the MINOS detector.  Upper plot:  NuMI LE beam; Lower plot:  NuMI HE beam (see Figure~\ref{fig:compare-parents}).  The symbols indicate previous hadron production measurements by Barton\cite{barton1983}, Atherton \cite{atherton1980}, and NA56/SPY \cite{ambrosini1999}.  Forthcoming measurements from NA49\cite{alt2006b} and MIPP \cite{messier2005} will cover much more of the $(x_F,p_T)$ plane.}
  \label{fig:pt-pz-numi}
\end{figure} 

\subsubsection{Hadron Production Uncertainties}

Hadron production data may not exist for a neutrino experiment's target material, thickness, beam energy, or relevant portion of $(x_F,p_T)$ phase space.  Prior to 2006, such was the case for NuMI (two-interaction length Carbon target, struck by $p_0$=120~GeV/$c$ protons).  

Figure~\ref{fig:pt-pz-numi} shows the $(x_F,p_T)$ of $\pi^+$ which contribute to the NuMI $\nu_\mu$ energy spectrum in either the LE or HE beam configurations.  As can be seen, the LE beam configuration focuses pions of $p_z\sim$10~GeV/$c$ with $p_T\sim250$~MeV/$c$.  The tail at higher $p_z$ and low $p_T$ comes from unfocused pions which pass through the field-free necks of the horns.  The HE beam focuses pions of $p_z\sim$30~GeV/$c$.

\begin{figure}[t]
\vskip -0.5cm
  \centering
  \includegraphics[width=3.2in]{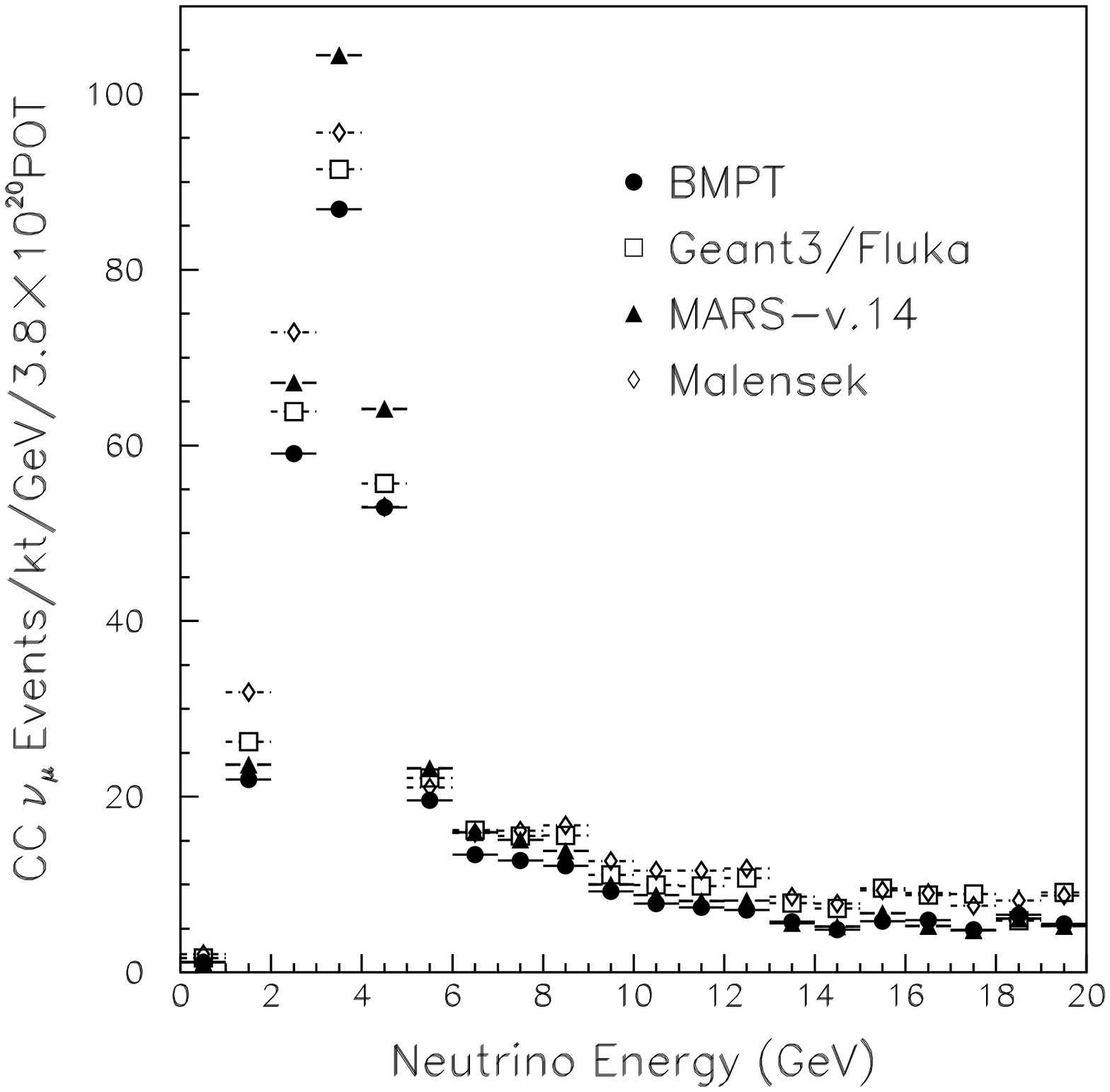}
  \includegraphics[width=3.2in]{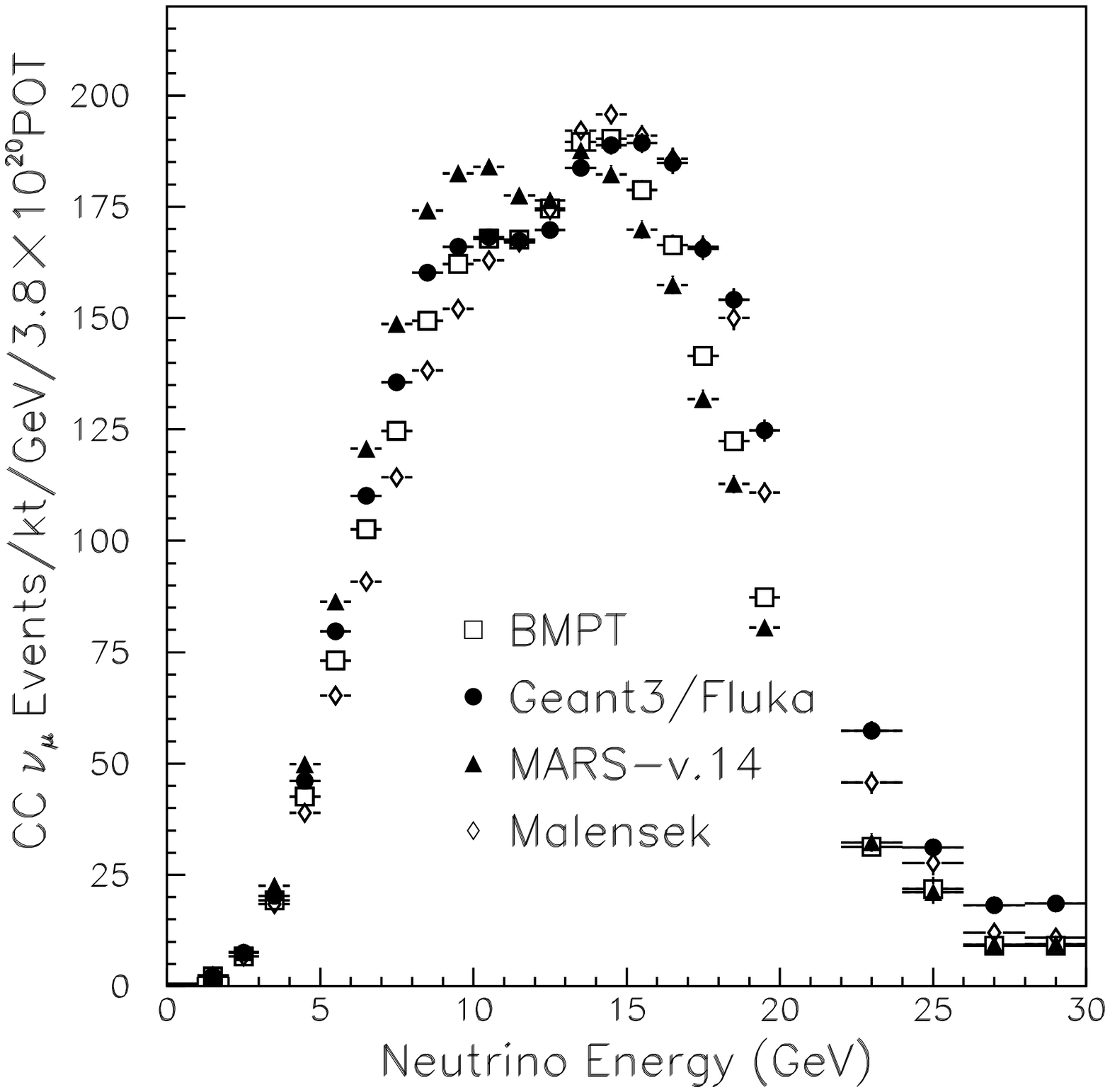}
\vskip 0.cm
  \caption{Energy spectrum of neutrinos interacting on Iron at $z=735.4$~km in the NuMI beam calculated in the LE beam configuration (left plot) and the HE beam configuration (right plot).  Hadron production models from Geant3/Fluka \cite{gfluka}, Malensek \cite{malensek}, BMPT \cite{bmpt}, and MARS-v.14 \cite{mars} are compared.}
  \label{fig:compare-numi-models}
\end{figure} 

In the figure is also shown the data points acquired by high-energy hadron production experiments.  Only Barton's data is on Carbon, and none are at the correct beam momentum or are a thick target measurement.  Less constraint is available for HE focusing region, and this must be entrusted to hadron production models to extrapolate.  No data is available for the high energy tail of either of these beams at $p_T\approx0$.  Inspection of Figure~\ref{fig:compare-numi-models} shows that these regions with little experimental constraint have sizeable model dependence of the flux calculations.  The variation amongst various models' predictions is $\sim$20\% in the LE beam and $\sim30$\% in the HE beam.

For a two detector experiment, the uncertainty of relevance is in the far/near ratio, since the near detector flux is measured directly and is used to estimate the far detector flux.  Further, the F/N uncertainties, in general smaller than for the direct flux calculation, are only large near the ``edges'' of the focusing system.  If all particles were perfectly focused, then hadron production uncertainties would amount to only a rate uncertainty in either detector.  It is the residual divergence of the beam that results in near-far differences, and such arise only at the limits of the focusing or in the high energy tail, where there is no focusing.  Thus, in Figure~\ref{fig:far-near-calculation} the largest far/near uncertainties (indicated by the vertical sizes of the boxes) appear at $E_\nu=6$~GeV, $E_\nu=8$~GeV, or above $E_\nu>12$~GeV.  Beamlines with complete focusing for all energies would therefore be anticipated to have reduced model dependence of the F/N calculation, such as the Hadron Hose in Figure~\ref{fig:hylen2003-figxx}.

\begin{figure}[t]
  \begin{centering}
\vskip -.4cm
    \includegraphics[width=2.in]{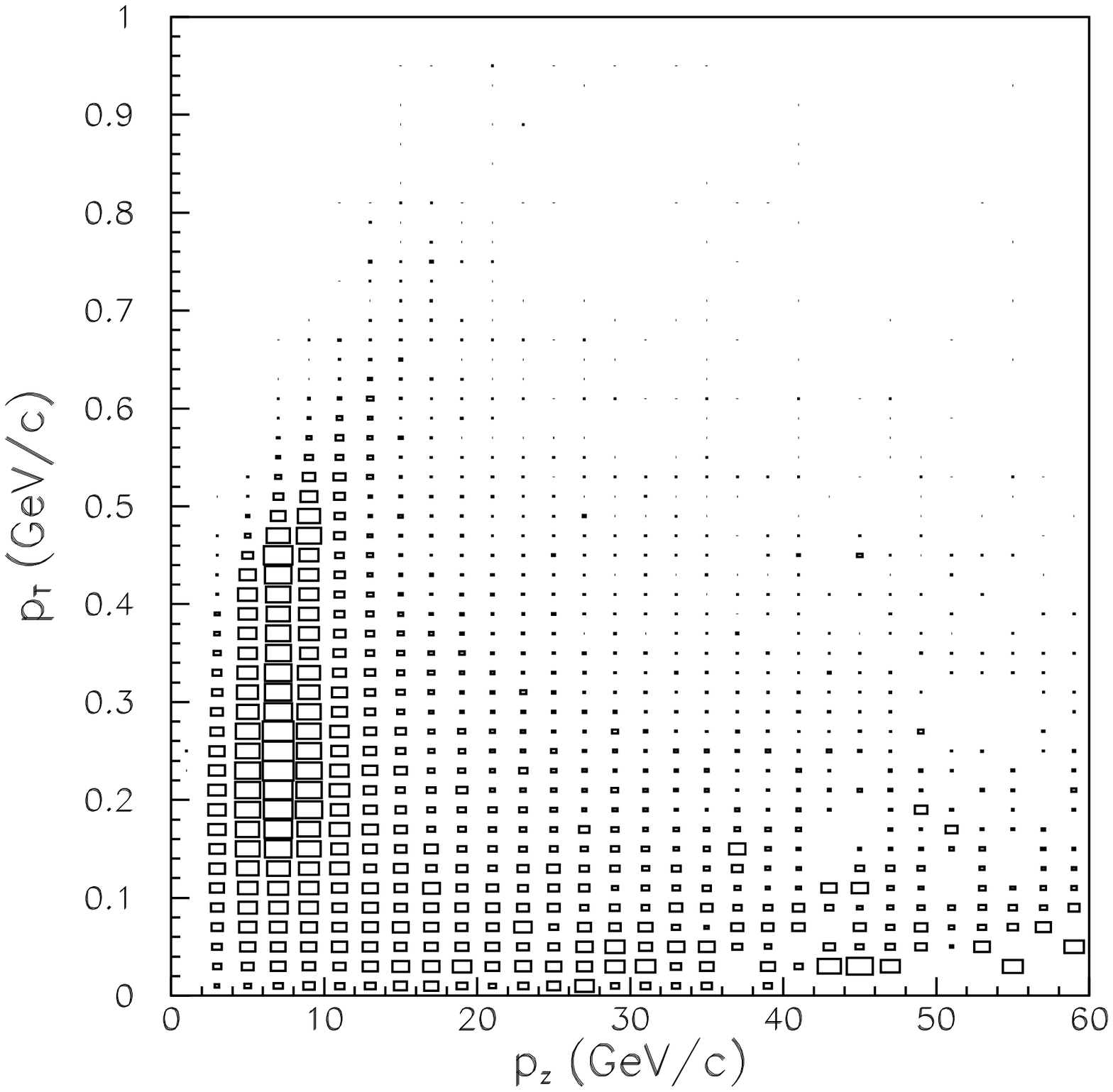}
    \includegraphics[width=2.in]{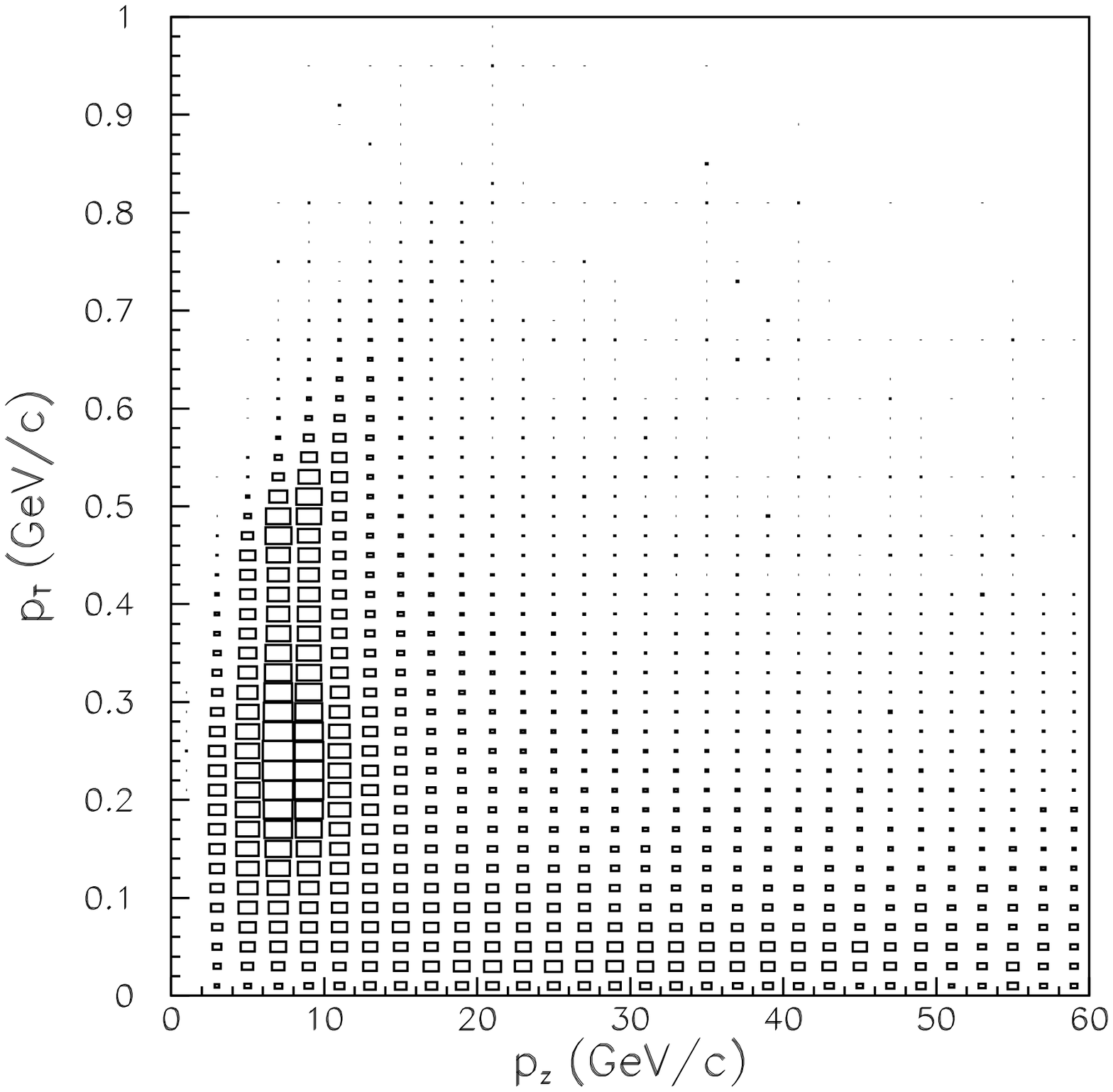}
    \includegraphics[width=2.in]{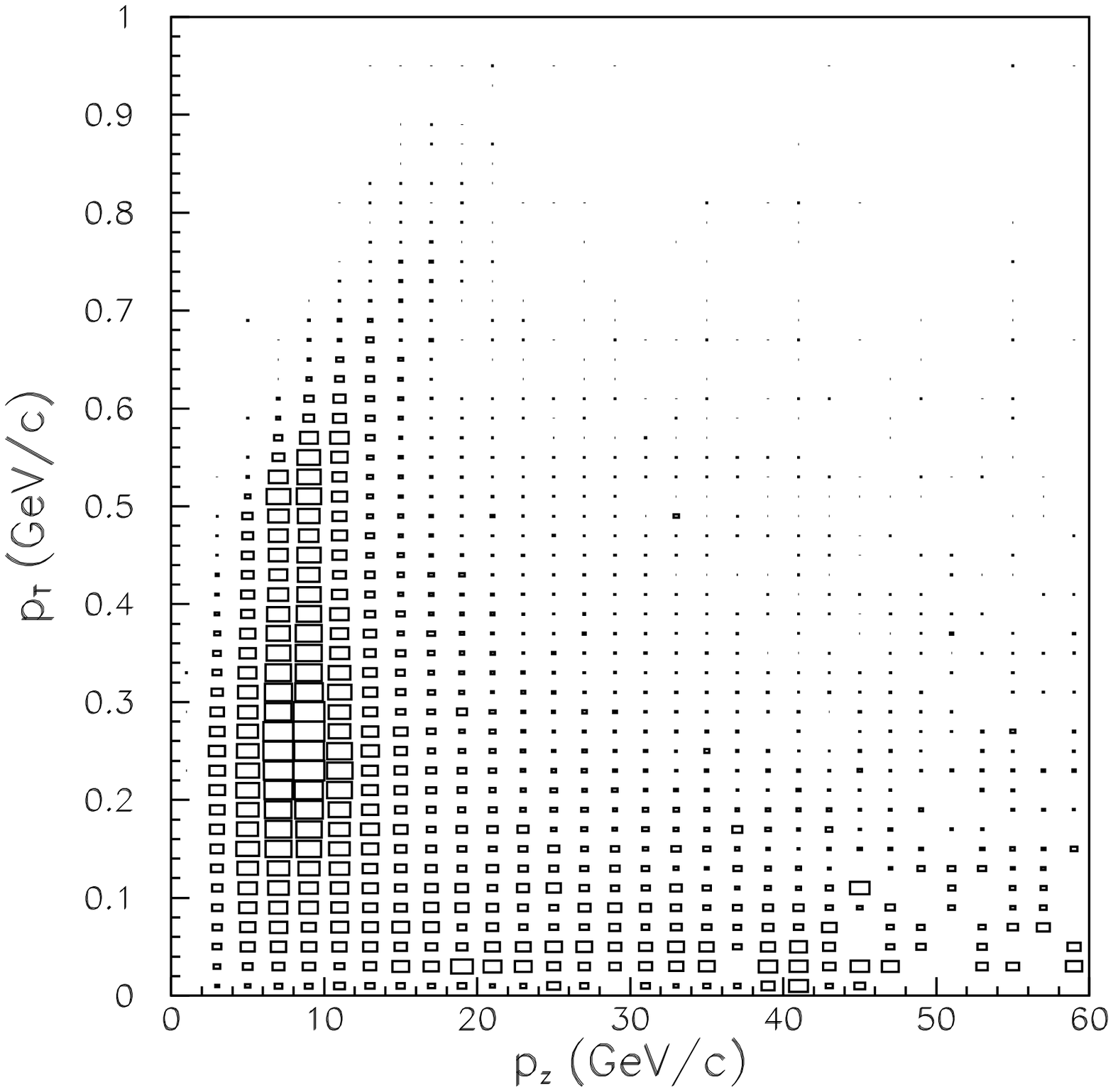}
    \includegraphics[width=2.in]{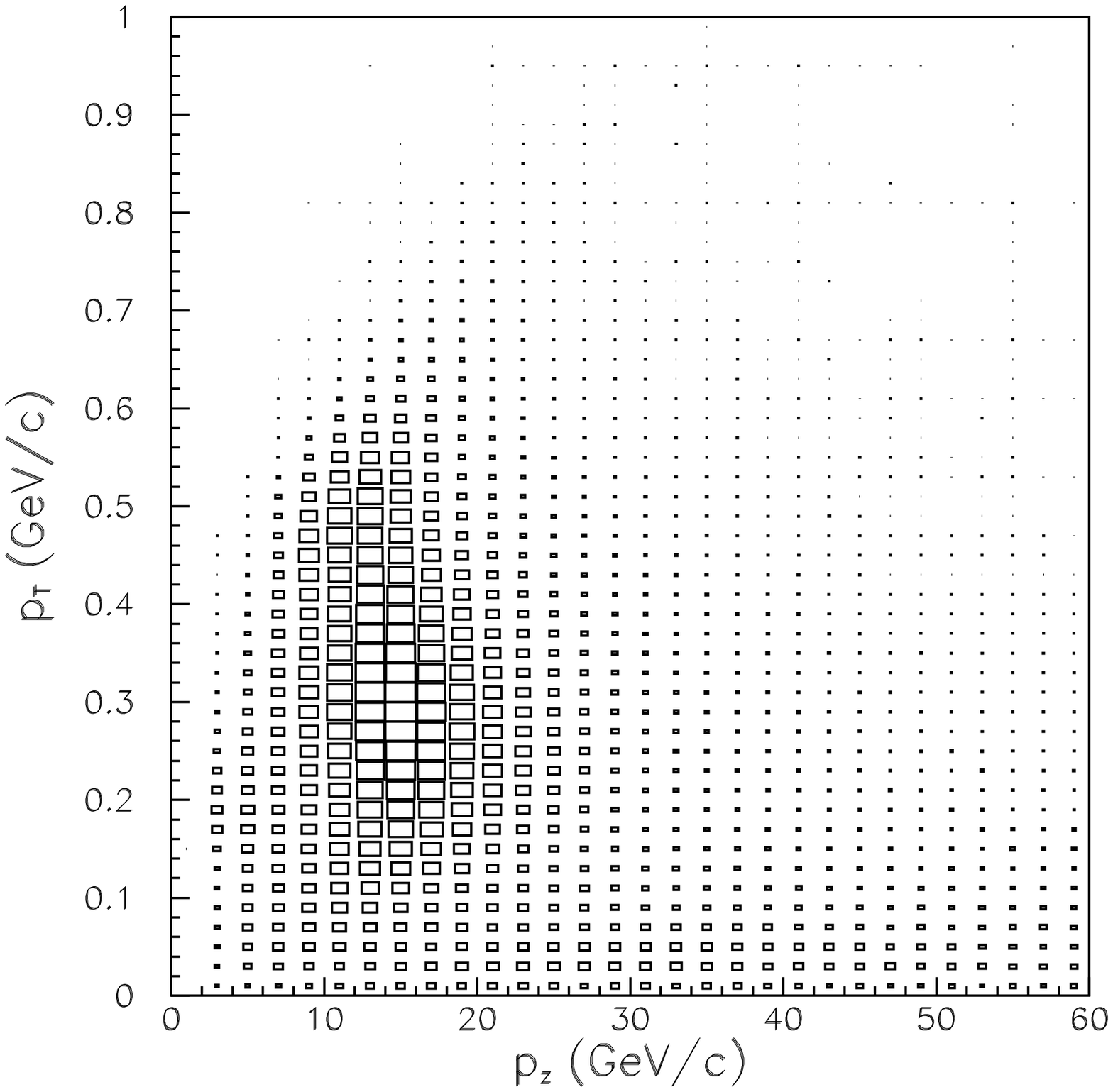}
    \includegraphics[width=2.in]{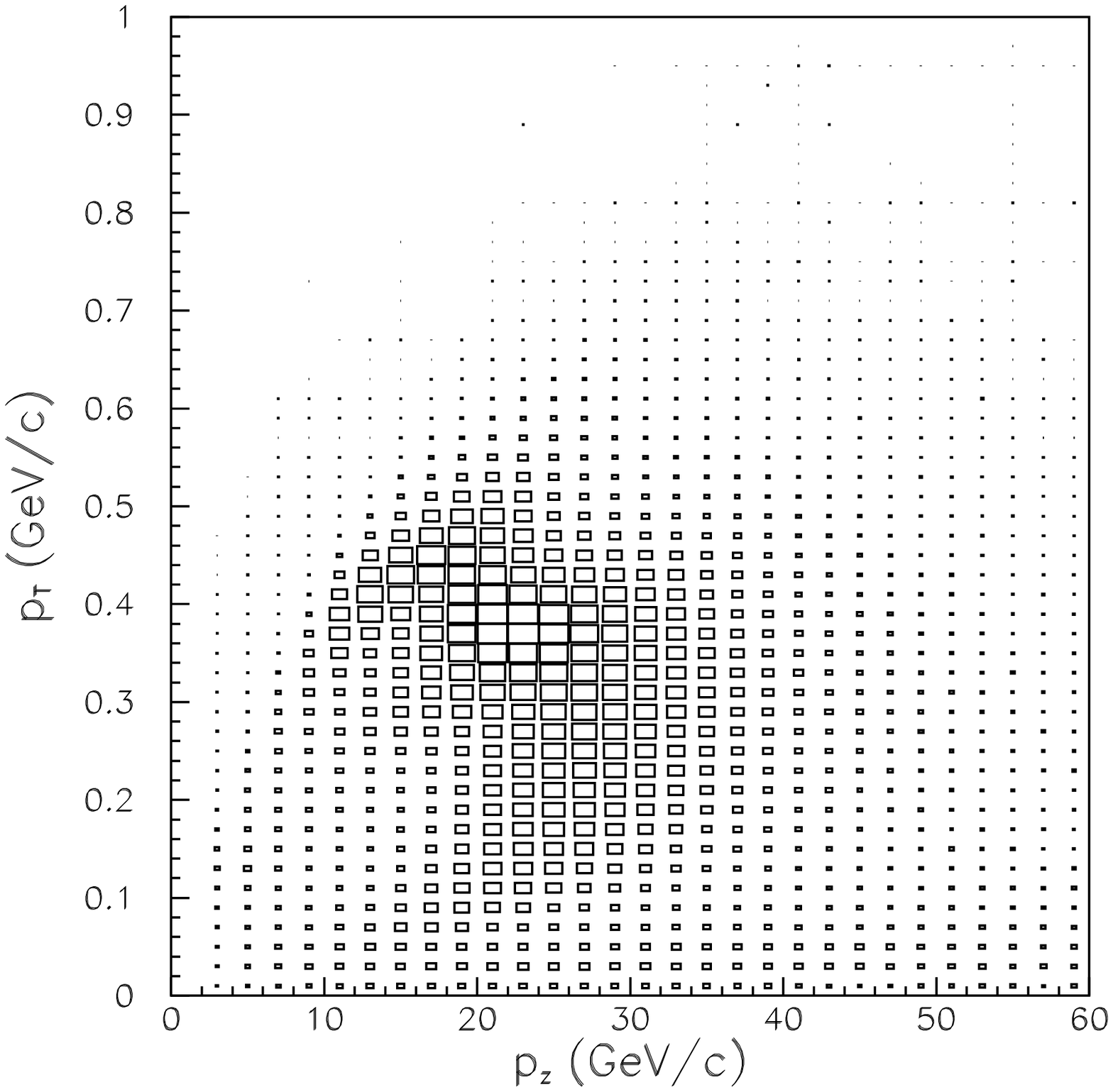}
    \includegraphics[width=2.in]{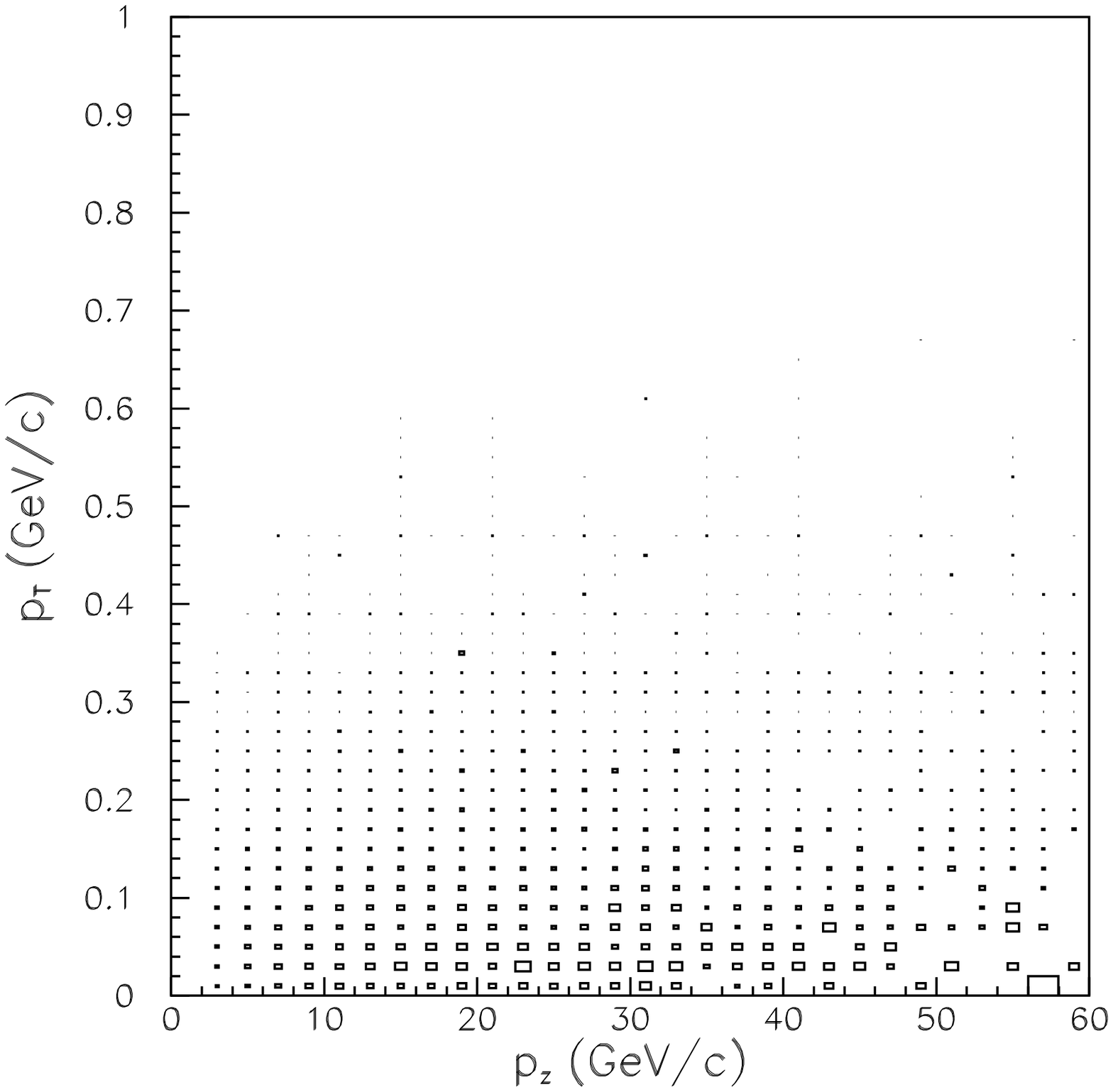}
      \caption{Distribution of $p_T$ and $x_F$ of $\pi^+$ from the NuMI target that contribute to the the charged-current event rate at the first detector at $z=1040$~m.  The box sizes are proportional to the probability of the pion resulting in a CC interaction in the ND.  The 6 plots correspond to the 6 beam configurations which have been run:  LE10/170kA (top left), LE10/185kA (top middle), LE10/200kA (top right), LE100/200kA (bottom left), LE250/200kA (bottom middle), LE10/0kA (bottom right) -- where each configuration is designated by the number of centimeters by which the target is upstream of the horn and by the current in the horns. As is evident, each beam configuration samples different region of $(x_F,p_T)$.  Taken from \cite{pavlovic2007}.
      \label{fig:ptxf_allbeams}}
  \end{centering}
\end{figure}

For NuMI, it was possible to tune the Monte Carlo hadron production model in $(x_F,p_T)$ to better agree with the near detector energy spectrum by virtue of simultaneously fitting data accumulated in several beam energy configurations \cite{michael2006}, lowering the uncertainty in the prediction of the far detector spectrum.  Similar tuning was used by neutrino experiments at BNL \cite{ahrens1986}, NOMAD \cite{astier2003} and NuTeV \cite{mcfarland1998,zeller2001}, but the NuMI capability to vary target position as well as horn current provides additional information.  In brief, variation of the horn current changes the $p_T$ kick received by particles through the horn (hence the $\langle p_T\rangle$ of the focusing), while the target position dictates the mean $x_F$ being focused, as discussed in Section~\ref{wbb}.  Several focusing uncertainties depend only on $p_z$ or $x_F$ of focused particles, as discussed in the next section, so the ability to control both $x_F$ and $p_T$ helps disentangle hadron production and other focusing effects.  Figure~\ref{fig:ptxf_allbeams} shows the $(p_T,x_F)$ sampled by several NuMI beam configurations.



\begin{figure}[t]
\vskip -.6cm
  \centering
  \includegraphics[width=3.2in]{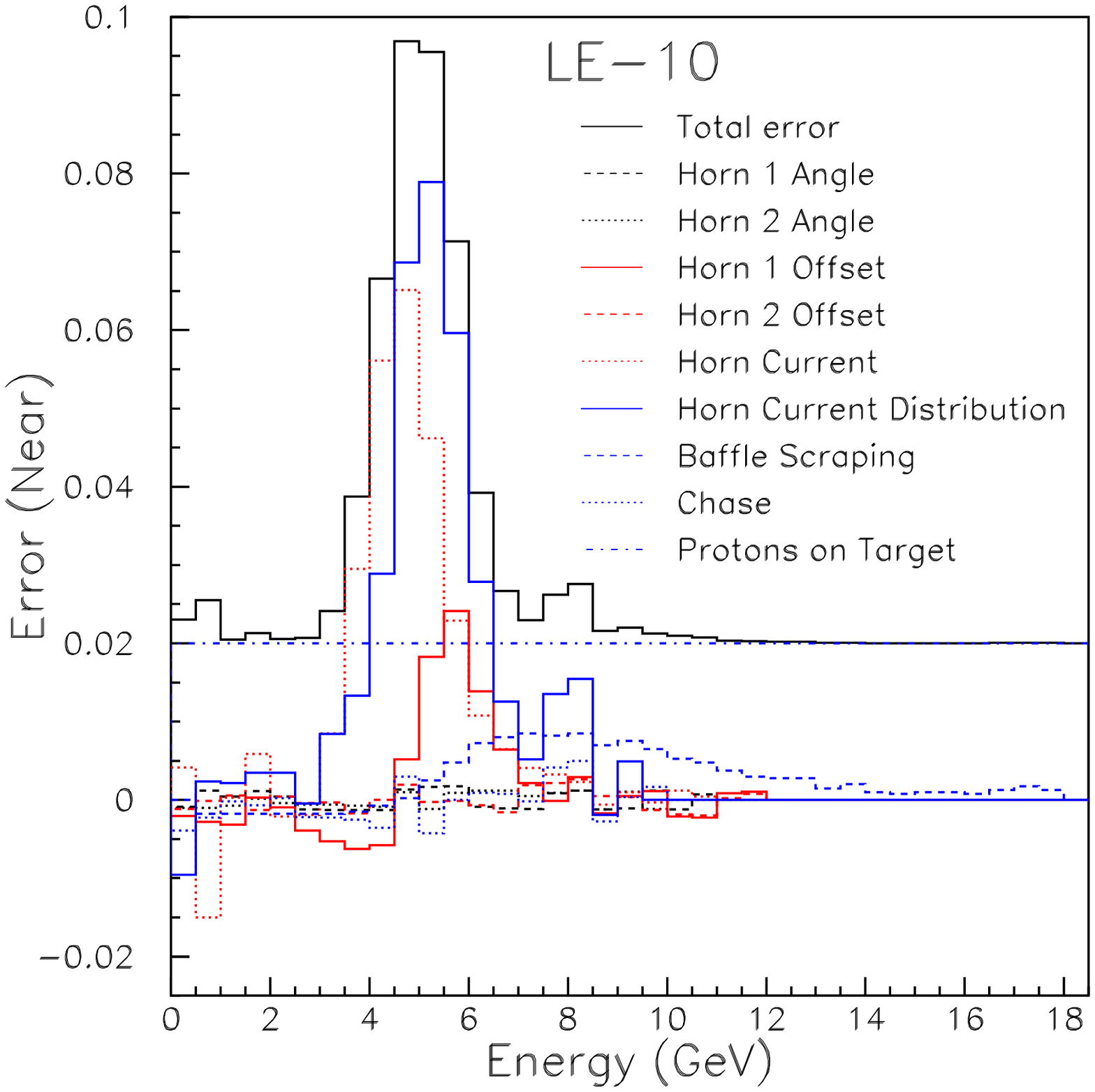}
  \includegraphics[width=3.2in]{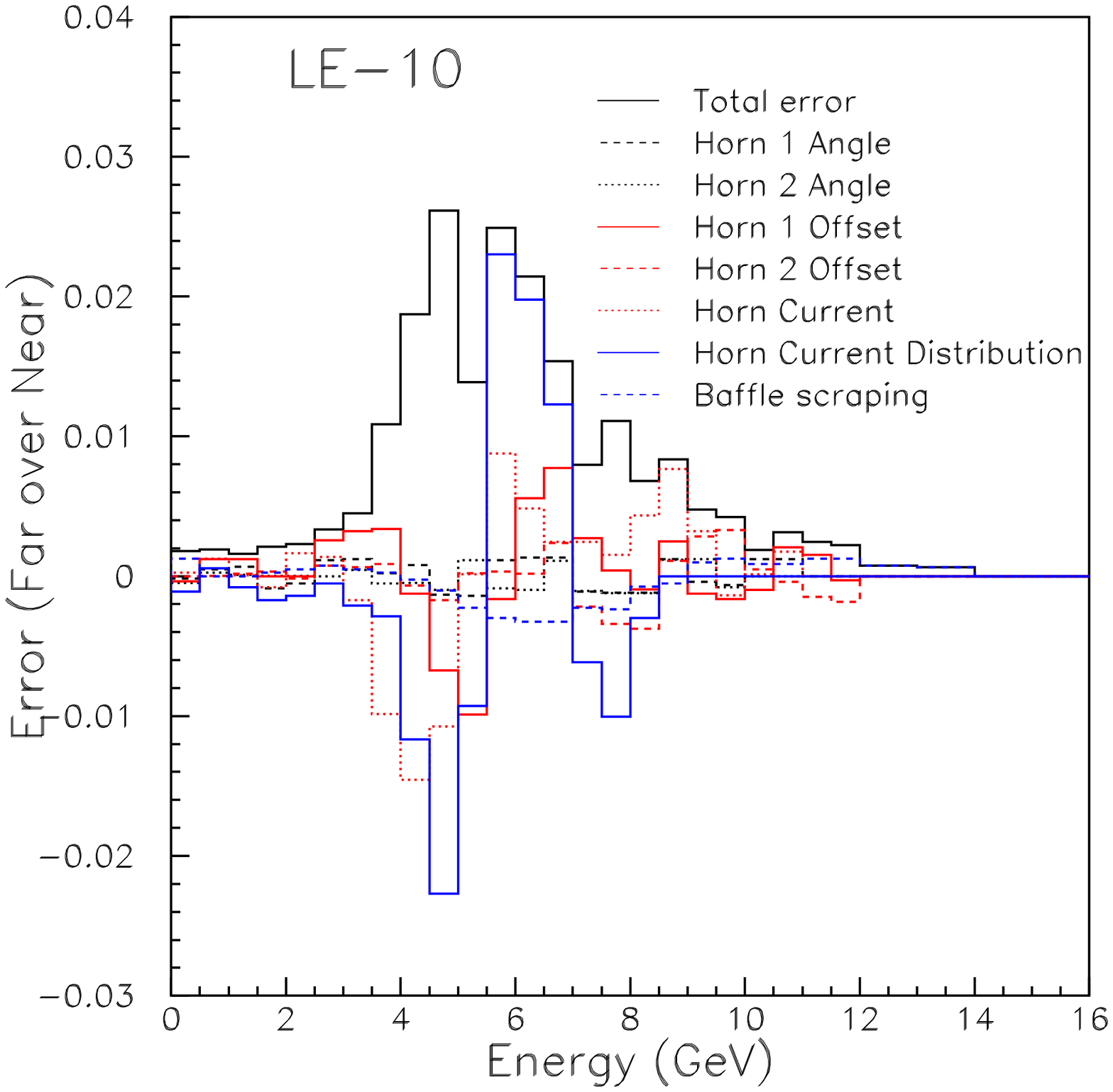}
\vskip 0.cm
  \caption{Uncertainties in the near detector spectrum (left plot) and in the far/near ratio (right plot) from focusing effects, estimated for the NuMI LE beam.  Taken from \cite{pavlovic2007}.}
  \label{fig:uncert-lebeam}
\end{figure}

\subsubsection{Focusing Uncertainties}

Several common systematic uncertainties manifest themselves in the detector flux prediction or the far/near calculation.  Figure~\ref{fig:uncert-lebeam} shows some of the larger uncertainties for the NuMI LE beam configuration:

{\bf Number of protons on target:}  two precision toroids, each calibrated by precision current sources, were tracked throughout the run against each other and the toroid reading the current in the accelerator.  A precision of $<$2\% was achieved.  

{\bf Proton beam halo} scraping on upstream collimating baffles can yield pions from a different target location than the nominal target location, and these pions can enter the focusing system, contributing high-energy neutrinos.  Proton halo was measured by primary beam instrumentation to be $<0.2$\% in magnitude.

The {\bf absolute current in the horns} was calibrated by precision toroids around the stripline to $\pm0.5\%$.  

The current pulse in the horn is $\sim1$~msec, at which the {\bf skin depth} in Aluminum is $\delta=7$~mm, larger than the 3~mm thickness of the horn conductor.  The distribution of the current in the conductors is therefore not a simple exponential, and the uncertainty on this distribution affects strongly those pions which graze and have considerable pathlength through the conductor material.

{\bf Misalignments} of the horns result in smaller-angle pions, not normally intercepted by the horns, receiving some focusing.  Such changes the spectrum for those pions at exactly the angles near the horn necks.

\section{Summary}

The present article summarized elements of particle production and focusing to derive conventional neutrino beams.  With 40~years of innovative advances, neutrino beams have gone from exploratory devices whose flux and composition was known little better than natural neutrino sources to instruments able to predict neutrino fluxes at the experiments at the few percent level.  Further, the power of neutrino beams has risen from a few detected neutrino interactions per week to a few detected interactions per second.  

In the coming years, new facilities will push the technological challenges of neutrino beams further.  Studies of neutrino oscillations across long baselines of hundreds of kilometers will be conducted at the CERN CNGS, Fermilab NuMI, and JPARC-nu facilities.  Probing rare transitions such as $\nu_\mu\rightarrow\nu_e$ oscillations, CP violation in the lepton sector, or the appearance of the $\nu_\tau$ in a $\nu_\mu$ beam will require sophisticated focusing systems capable of coping with Megawatts of protons delivered to these facilities over several years.  Further, precision cross-section experiments being conducted at the Fermilab-MiniBooNE or NuMI beam lines or the JPARC-nu beam line will demand accurate demonstration of the neutrino fluxes and composition.  

With many exciting fields of physics to be probed using the neutrino, we may look forward to continuing advances in beam line systems to realize these goals.

\vskip .25cm
\begin{large}{\bf Acknowledgements}\end{large}
\vskip .25cm
I thank L. Loiacono, J. Ma, R. Miyamoto and Z. Pavlovic assistance researching this paper.  V. Garkusha, M. Kordosky, K. Lang, A. Marchionni, A. Para and P. Vahle offered helpful critique of this manuscript.  It's a pleasure to acknowledge years of stimulating collaborations with colleagues on the NuMI facility and the MINOS experiment collaboration.  I thank D. Casper and D. Harris for the opportunity to lecture on conventional neutrino beams at the 2006 NuFact Summer School, out of which these notes grew.


\appendix
\section{Kinematic Relations}
\label{kinematics}
This appendix reproduces several useful kinematic formulae relating to the energy and decay distribution of neutrino daughters from pion and kaon parents.  More thorough review of relativistic kinematics can be found in \cite{hagedorn1963}, and numerous formulae relevant for neutrino beams can be found in \cite{ramm1963}.

\begin{figure}[t]
\vskip 0.cm
  \centering
  \includegraphics[width=4.5in]{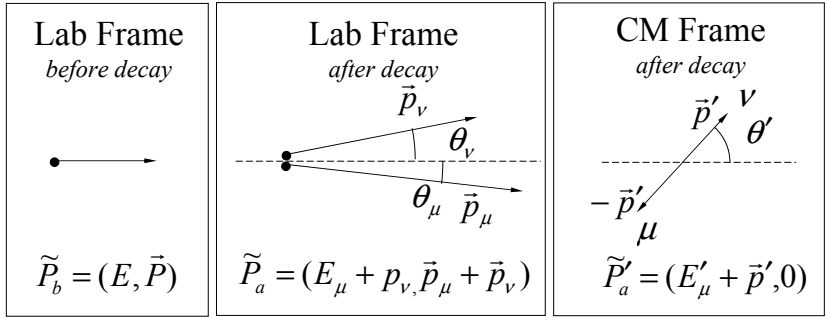}
\vskip 0.cm
  \caption{View of a parent $\pi$ or $K$ meson decay in the laboratory and center-of-mass (CM) frames, defining the momenta, energies, and angles of the parent and daughter particles.}
  \label{fig:reference-frames}
\end{figure}

Figure~\ref{fig:reference-frames} defines several momentum vectors and directions for the daughters from a $\pi$ or $K$ decay.  Being a two-body decay into a muon and neutrino, the momenta $p^\prime$ of the daughters in the center-of-mass frame can be calculated as:
\begin{equation}
p^\prime=\frac{M}{2}\left(1-\frac{m_\mu^2}{M^2}\right)
\label{eq:cm-momentum}
\end{equation}
where $M$ is the mass of the $\pi$ or $K$ parent and $m_\mu$ is the muon mass.  For the daughters in $\pi$ ($K$) decay, $p^\prime=29.8$~MeV (235.6~MeV).  The $\pi/K$ being spin zero, the angular distribution of the decay daughters is isotropic in the CM frame,
\begin{equation}
\frac{dP}{d\Omega^\prime}=\frac{1}{4\pi}
\label{eq:ang-cm}
\end{equation}

\begin{figure}[t]
\vskip 0.cm
  \centering
  \includegraphics[width=5in]{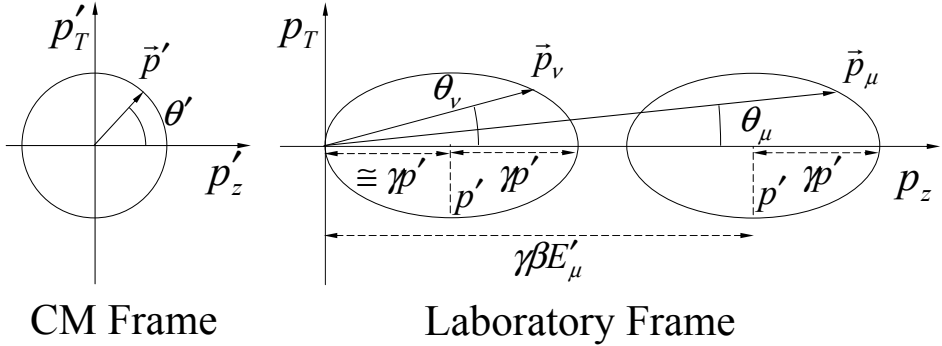}
\vskip 0.cm
  \caption{Lorentz transformation of a momentum 3-vector $p^\prime$ from the CM frame to the laboratory frame.  In the CM frame, $p^\prime$ lies on a circle, while in the lab it is required to lie on an ellipse (see Equation~\ref{eq:lab-ellipse}).  For $\beta\approx1$ parents as sketched above, the neutrino ellipse approximately is tangent to the $p_T$ axis and the muon ellipse is shifted to the right.  In $\pi\rightarrow\mu\nu$ decays, as sketched above, the muon energy in the lab alwa}
  \label{fig:lorentz-boost-ellipses}
\end{figure}

Transformation of the daughter momenta to the lab is done by the Lorentz boost:
\begin{equation}
E=\gamma(E^\prime+\beta p_z^\prime)
\label{eq:boost-e}
\end{equation}
\begin{equation}
p_z=\gamma(p_z^\prime+\beta E^\prime)
\label{eq:boost-pz}
\end{equation}
\begin{equation}
p_T=p_T^\prime
\label{eq:boost-pt}
\end{equation}
where $\gamma=E/M$, $\beta=(1-1/\gamma^2)^{1/2}$, and $E$ and $M$ are the energy of the parent meson in the lab and its mass, respectively.  The fact that $p_T^{\prime 2}+p_z^{\prime 2}=p^{\prime 2}$ means that the daughter momentum vectors lie on a circle in the CM frame (see Figure~\ref{fig:lorentz-boost-ellipses}), while this relation, upon substitution of Equations~\ref{eq:boost-pz} and \ref{eq:boost-pt} yield that the daughter momentum vectors lie on an ellipse in the laboratory frame:
\begin{equation}
\frac{(p_z-\beta\gamma E^\prime)^2}{\gamma^2 p^{\prime 2}} + \frac{p_T^2}{p^{\prime 2}}=1
\label{eq:lab-ellipse}
\end{equation}
As indicated in Figure~\ref{fig:lorentz-boost-ellipses}, the ellipse for the neutrino momentum vector in the lab approximately intercepts the origin for very relativistic parents ($\beta\approx1$) since $E^\prime=p^\prime$ for the neutrino.  Because of the large muon energy $E_\mu^\prime$ (=109~MeV or 258~MeV for $\pi$ and $K$ decays, respectively), the ellipse for the muon is shifted to the right.  As indicated in Figure~\ref{fig:lorentz-boost-ellipses}, the neutrino momentum in the lab ranges from 0 to $p_\nu^{max}=2\gamma p^\prime=(1-\frac{m_\mu^2}{M^2}) E$ (when looking at all possible decay angles), which is 0.43$E$ for $\pi\rightarrow\mu\nu$ decays and 0.96$E$ for $K\rightarrow\mu\nu$ decays.  The muon momentum ranges from 0.57$E$ to $E$ in $\pi\rightarrow\mu\nu$ decays and 0.04$E$ to $E$ in $K\rightarrow\mu\nu$ decays.

Transformation of the daughter angles is found from Equations~\ref{eq:boost-pz} and \ref{eq:boost-pt}, noting that $p_T=p\sin\theta$, $p_T^\prime=p^\prime\sin\theta^\prime$, $p_z^\prime=p^\prime\cos\theta^\prime$, and $p_z=p\cos\theta$:
\begin{equation}
\gamma\tan\theta=\frac{\sin\theta^\prime}{\cos\theta^\prime+(\beta/\beta^\prime)}
\label{eq:boost-theta}
\end{equation}
where $\beta^\prime=p^\prime/E^\prime$ is the daughter velocity in the CM frame (=1 for the neutrino and =0.28 or 0.91 for the muon in $\pi$ or $K$ decays).  The maximum decay angle in the lab arises from $\theta^\prime=\pi/2$, yielding 
$\tan\theta_{\mbox{max}}=\beta^\prime/\gamma\beta$.
For the neutrinos from relativistic parents, this reduces to $\theta_\nu^{\mbox{max}}\sim1/\gamma$, for the muons $\theta_\mu^{\mbox{max}}\sim\beta^\prime/\gamma$.  Thus the muons are more forward-boosted.

The angular distribution of neutrinos in the lab frame is found from Equations~\ref{eq:ang-cm} and \ref{eq:boost-theta}.  The angular distribution in the lab is found from
$dP/d\Omega=(dP/d\Omega^\prime)(d\Omega^\prime/d\Omega)=(dP/d\Omega^\prime)(d\theta^\prime/d\theta)(\sin\theta^\prime/\sin\theta)$.
For neutrinos from very relativistic parents with $\beta\approx1$, Equation~\ref{eq:boost-theta} can be inverted to give $\cos\theta^\prime\approx(1-\gamma^2\tan^2\theta)/(1+\gamma^2\tan^2\theta)$, and
\begin{equation}
\frac{dP}{d\Omega}\approx\frac{1}{4\pi}\frac{4\gamma^2(1+\tan^2\theta)^{3/2}}{(1+\gamma^2\tan^2\theta)^2}
\label{eq:angular-spread}
\end{equation}
which reduces to Equation~\ref{eq:flux-angle} in the limit that $\theta<<1$.

The neutrino energy in the lab is found from Equation~\ref{eq:boost-e}.  For $\beta\approx1$, again using $\cos\theta^\prime\approx(1-\gamma^2\tan^2\theta)/(1+\gamma^2\tan^2\theta)$ the lab energy is
\begin{equation}
E_\nu\approx
\frac{\left(1-\frac{m_\mu^2}{M^2}\right) E}{1+\gamma^2\tan^2\theta_\nu}
\label{eq:enu-lab}
\end{equation}
which reduces to Equation~\ref{eq:enu-vs-epi} in the limit that $\theta_\nu<<1$.  Equations~\ref{eq:enu-lab} and \ref{eq:angular-spread} combine to show that the energy distribution of the neutrinos, $dP/dE_\nu=(dP/d\Omega_\nu)(d\Omega_\nu/dE_\nu)$, is constant (which averages over all decay angles $\theta_\nu$).

\clearpage
\newpage
\begin{large}
\begin{center}
Problem Set
\end{center}
\end{large}

\begin{enumerate}

\item  Assume a ``half-horn'' with an inner-conductor which is partially tapered and at $z=f$ is perpendicular to the $z$ axis, as in Figure~\ref{fig:problem-1-figure}.  We will calculate the trajectory of a particle of momentum $p$ through such a horn but not resort to the ``thin lens'' approximation that the pathlength $s$ through the horn is small compared to the distance $z_{in}$ from the particle source (target) to the horn.  
\begin{enumerate}
\item Show that, if the particle enters the horn at a coordinate $(z_{\mbox{in}},r_{\mbox{in}})$, its trajectory through the horn is given by
$$r=r_{\mbox{in}}exp[A_0(\cos\theta-\cos\theta_{\mbox{in}})]$$
$$z-z_{\mbox{in}}=A_0r_{\mbox{in}}\int_{\theta=0}^{\theta_{\mbox{in}}}d\theta\cos\theta exp[A_0(\cos\theta-\cos\theta_{\mbox{in}}))]$$
where $r_{\mbox{in}}/z_{\mbox{in}}=\tan\theta_{\mbox{in}}$ and $A_0\equiv\frac{2\pi}{\mu_0}\frac{p}{I}=p(\mbox{GeV}/c)/(6\times10^{-5} I\mbox{(kA)})$.  The above are the equations integrated numerically for several values of $\theta_{\mbox{in}}$ in Figure~\ref{fig:dohm1975-fig3}.
\item If the particle angles entering this horn are small, namely $\theta_{\mbox{in}}^2<<1$ and $A_0^2\theta_{\mbox{in}}^4<<1$, show that the entrance points of the particles into the horn should follow:  $$(x_{\mbox{in}}-\frac{5}{8}f)^2 + \frac{3}{4}A_0y_{\mbox{in}}^2 = (\frac{3}{8}f)^2$$
which is an ellipse with major and minor axes given by $a=\frac{3}{8}f$, $b=2a/(3A_0)^{\frac{1}{2}}$.
\item  Show that if instead one requires a tighter ``small-angle'' approximation $A_0\theta_{\mbox{in}}^2<<1$, one obtains the condition of a parabola, $$\lambda (f-x_{\mbox{in}})=y_{\mbox{in}}^2.$$
What is the coefficient $\lambda$?  The above form is for a parabolic lens.
\begin{figure}[h]
\vskip -0.5cm
  \centering
  \includegraphics[width=2.5in]{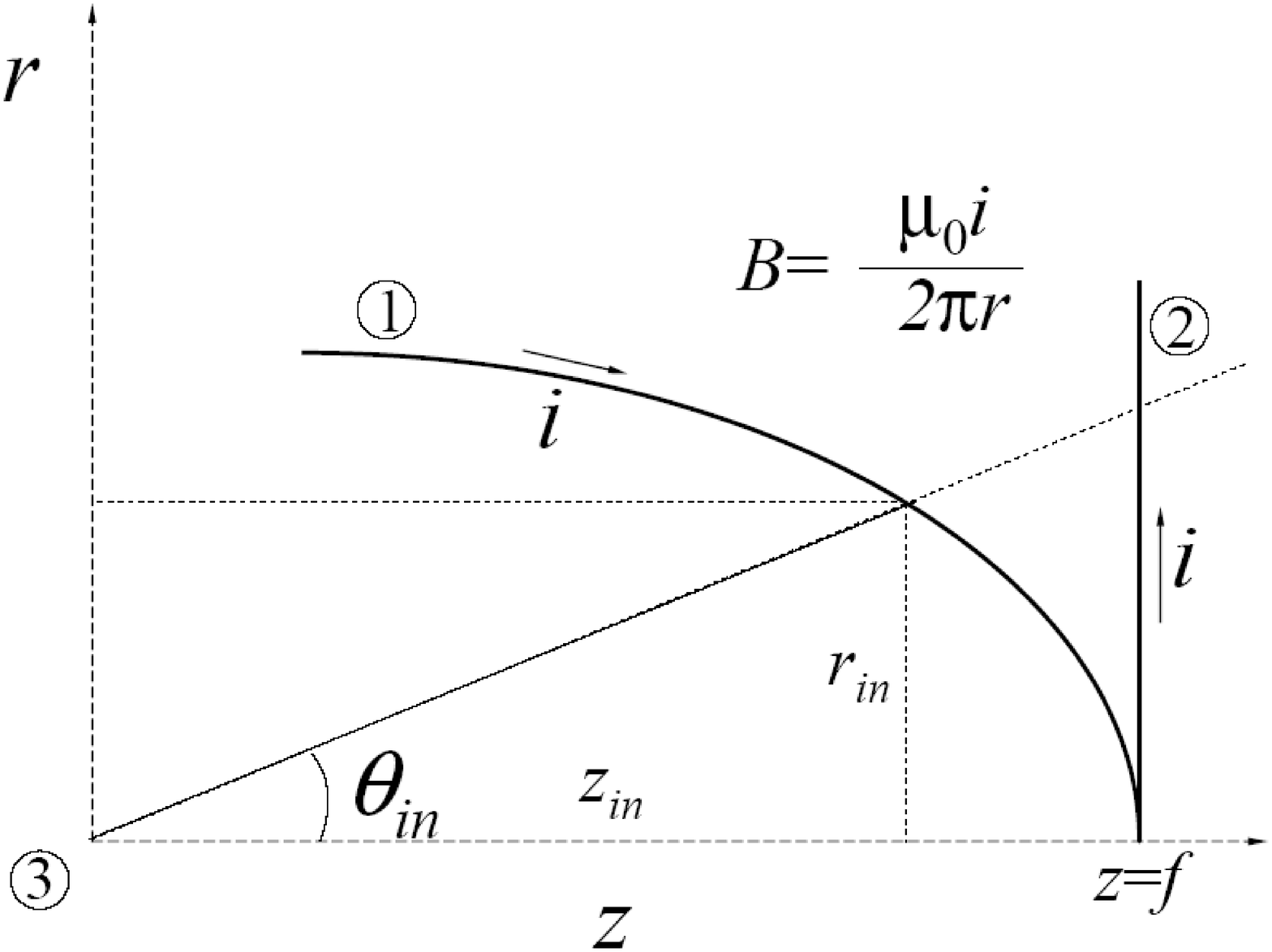}
  \includegraphics[width=2.5in]{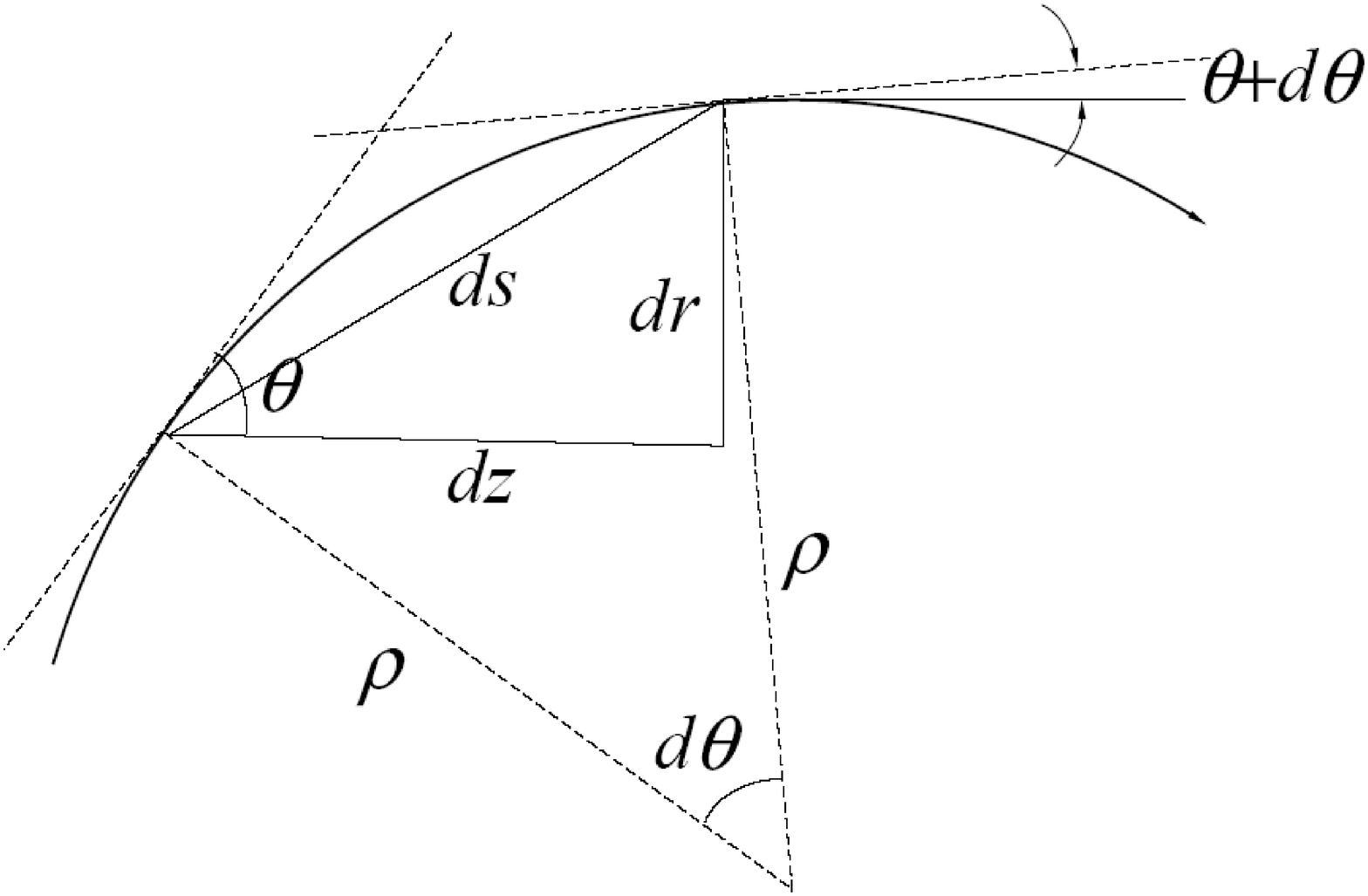}
\vskip -.5cm
  \caption{Schematic of a particle trajectory in a lens not assumed to be ``thin.''  1.-upstream, tapered end of inner conductor, 2.-downstream end of inner conductor, which points perpendicular to the $z$ axis, 3.-source of particles (target) at $(z,r)=(0,0)$.  }
  \label{fig:problem-1-figure}
\end{figure} 
\end{enumerate}

\begin{figure}[t]
\vskip -0.5cm
  \centering
    \includegraphics[width=3.in]{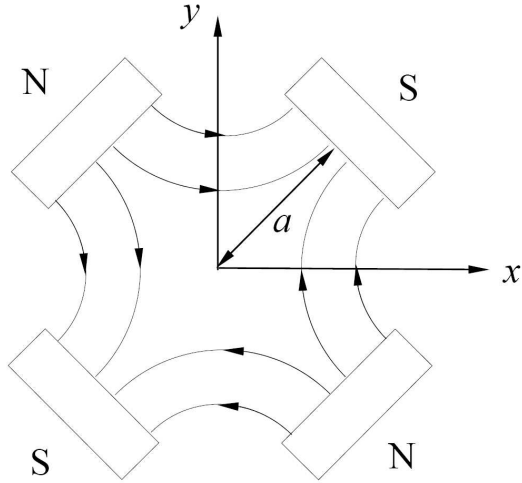}
\vskip -.5cm
  \caption{End-view of a quadrupole magnet of aperture $a$.}
  \label{fig:problem-2-figure}
\end{figure} 

\item  This problem concerns the wide-band focusing properties of quadrupole magnets.  

\begin{enumerate}
  \item  Show that the focal length of a quadrupole of length $L$, aperture $a$, and maximum field $B_0$ (in Tesla) at the pole tip, acting on a momentum $p$ (in GeV/$c$) is $$f=\frac{pa}{0.3B_0L}.$$
    The magnetic potential of a quadrupole is conveniently expressed as $\phi=\frac{B_0}{a}xy$.
  \item  Show that the net focal length of a pair of identical quadrupoles, one focusing and the next defocusing, and separated by a distance $d$, is $$f^*=\frac{p^2a^2}{0.09B_0^2L^2d}.$$  It is most practical to use transfer matrices which act on vectors $(x~~x^\prime)$, and first derive the transfer matrix for a drift space of length $d$ and the focusing or defocusing quadrupole.  The three matrices multiplied together gives the net transfer matrix, which contains the focal length.  Assume the thin lens approximation.
  \item  Prove Equation~\ref{eq:foci}
  \item  Prove Equation~\ref{eq:mom-bite} in the limit that $a<<R$ and $u<<D$.
\end{enumerate}

\item Suppose you are an experimenter on the MiniBooNE team.  Following the example of \cite{danby1962}, you are about to place a temporary beam dump (see Figure~\ref{fig:brice2004-dump}) in the decay tunnel to stop a number of mesons and thereby test your beam flux prediction.  By what fraction should the $\nu_{\mu}$ flux change?  The $\nu_e$ flux?  Assume that half of $\nu_e$'s come from $K$ decay and half from $\mu$ decay for this experiment.  Important information:  the MiniBooNE decay pipe is 50~m, the temporary dump is at 25~m, and the $\nu_\mu$ focusing peak is $\sim$800~MeV.  

\item  This problem concerns the distribution of neutrino daughters from pi/$K$ decay.  Assume that the meson parent is travelling with $\beta\approx$1.    
\begin{enumerate}
  \item  Assuming the decay neutrinos are isotropic in the CM frame ($dP/d\Omega^\prime=1/4\pi$=constant), show that in the laboratory frame 
            $$dP/d\Omega=(dP/d\Omega^\prime)\times (d\Omega^\prime/d\Omega) = \frac{1}{4\pi}\frac{4\gamma^2(1+\tan^2\theta)^{3/2}}{1+\gamma^2\tan^2\theta)^2},$$
    where $\gamma=(1-\beta^2)^{-1/2}$ is the boost factor for the pion/kaon parent meson.  In the small-angle limit $\tan\theta<<1$ this formula reduces to Equation~\ref{eq:flux-angle}.
  \item  Use your result above to show that the energy distribution in the lab of neutrinos from pion or kaon decays is flat in neutrino energy, {\it i.e.}
            $$dP/dE_\nu = (dP/d\Omega)(d\Omega/dE_\nu)=\mbox{constant}$$
    What is the constant?
\end{enumerate}


\end{enumerate}

\end{document}